\documentclass[10pt,journal,compsoc]{IEEEtran}
\usepackage{balance}  %

\usepackage[linesnumbered, boxed, vlined, figure]{algorithm2e} %

\usepackage{multirow}
\usepackage{graphicx}
\usepackage{amsmath}
\usepackage[labelformat=simple]{subcaption}

\usepackage{latexsym} %

\usepackage{url}
\usepackage[cal=dutchcal]{mathalfa}

\renewcommand{\baselinestretch}{0.991}
\SetKwInput{Algorithm}{Algorithm}
\SetKwInput{Function}{Function}
\SetKwInput{Result}{Result}
\SetKwProg{Fn}{Function}{}{}
\makeatletter
\newcommand{\nosemic}{\renewcommand{\@endalgocfline}{\relax}}%
\newcommand{\dosemic}{\renewcommand{\@endalgocfline}{\algocf@endline}}%
\let\oldnl\nl%
\newcommand{\nonl}{\renewcommand{\nl}{\let\nl\oldnl}}%
\makeatother

\usepackage{tikz}
\usetikzlibrary{decorations.pathreplacing,calc}
\newcommand{\tikzmark}[1]{\tikz[overlay,remember picture] \node (#1) {};}

\newcommand*{\AddNote}[4]{%
	\begin{tikzpicture}[overlay, remember picture]
	\draw [decoration={brace,amplitude=0.5em, mirror, raise=10pt},decorate,thick, black, xshift=50pt,yshift=0pt]
	($(#3)!(#1.north)!($(#3)-(0,1)$)$) --  
	($(#3)!(#2.south)!($(#3)-(0,1)$)$)
	node [align=center, text width=2.5cm, pos=0.5, anchor=south, rotate=90, yshift=12pt] {#4};
	\end{tikzpicture}
}%

\graphicspath{
	{plots/ldbc/}
	{plots/ldbc/SOCIAL-PUB10M-SUB5K-S25-R25/}
	{plots/nyc/NYC-PUB1M-SUB5K-S25-R25/}
	{plots/biogrid/BIOGRID-PUB1500K-SUB5K-S25-R25/}
	{plots/ldbc/ldbc-p1m-s10k/}
	{plots/nyc/NYC-PUB100K-SUB5K-S25-R25/}
	{plots/ldbc/SOCIAL-PUB100K-SUB5K-S10-R25/}
	{plots/ldbc/SOCIAL-PUB100K-SUB5K-R25-VAR_SELECT/}
	{plots/ldbc/SOCIAL-PUB100K-SUB5K-S25-R25/}
	{plots/biogrid/BIOGRID-PUB100K-SUB5K-S25-R25/}
	{plots/ldbc/SOCIAL-PUB1M-SUB5K-S25-R25/}
	{plots/ldbc/SOCIAL-PUB100K-SUB5K-S25-R25-VARYING1000-5000/}}

\newcommand{\Inv}{\textsc{Inv}}
\newcommand{\InvCache}{\textsc{Inv+}}

\newcommand{\InvIncr}{\textsc{Inc}}
\newcommand{\InvIncrCache}{\textsc{Inc+}}

\newcommand{\Tree}{\textsc{TriC}}
\newcommand{\TreeCache}{\textsc{TriC+}}

\newcommand{\Neo}{Neo4j}

\newcommand{\edgeInd}{\mathit{edgeInd}}
\newcommand{\sourceInd}{\mathit{sourceInd}}
\newcommand{\targetInd}{\mathit{targetInd}}
\newcommand{\queryInd}{\mathit{queryInd}}

\newcommand{\rootIndex}{\mathit{rootInd}}

\newcommand{\edge}{\mathit{edge}}
\newcommand{\cPaths}{\mathit{CP}}
\newcommand{\mv}{\mathit{matV}}
\newcommand{\parent}{\mathit{prnt}}
\newcommand{\children}{\mathit{chn}}
\newcommand{\select}{\mathcal{\sigma}}
\newcommand{\overlap}{\mathcal{o}}
\newcommand{\len}{\mathcal{l}}

\newcommand{\markM}{\mathit{mark\_Matched}}
\newcommand{\rt}{\mathit{root}}
\newcommand{\search}{\mathit{depthFirstSearch}}

\usepackage{enumitem}

\newcounter{qcounter}

\newcommand{\spacecut}[1]{}

\usepackage{times}

\newenvironment{mylisting}[1] {\vspace*{0ex} \begin{list}{\arabic{qcounter}.}{
			\usecounter{qcounter}
			\setlength{\labelwidth}{2ex}
			\setlength{\labelsep}{1ex}
			\setlength{\leftmargin}{3.5ex}
			\setlength{\rightmargin}{0ex}
			\setlength{\itemsep}{0.6ex}
			\setlength{\parsep}{0.5ex}
		} #1 } {\end{list}}

\newcommand{\ctitle}[1]{\vspace{1ex}\noindent\textbf{#1}}
\newcommand{\cititle}[1]{\vspace{1ex}\noindent\textit{#1}}

\usepackage{float}
\usepackage{xcolor}

\newtheorem{definition}{Definition}
\newtheorem{example}{Example}

\makeatletter%
\begin{document}
	
	\title{Efficient Continuous Multi-Query Processing \\ over Graph Streams}

		\author{Lefteris~Zervakis, Vinay~Setty, Christos~Tryfonopoulos, Katja~Hose%
		\IEEEcompsocitemizethanks{
			\color{black}{
			\IEEEcompsocthanksitem
			L. Zervakis, V. Setty, and K. Hose are with the Department of Computer Science, Aalborg University, Aalborg, Denmark.\protect\\
			E-mail: \{lefteris,vinay,khose\}@cs.aau.dk
			\IEEEcompsocthanksitem
			L. Zervakis and C. Tryfonopoulos are with the Department of Informatics and Telecommunications, University of the Peloponnese, Tripolis, Greece.\protect\\
			E-mail: \{zervakis,trifon\}@uop.gr
			\IEEEcompsocthanksitem
			V. Setty is also with the Department of Electrical Engineering and Computer Science, University of Stavanger, Stavanger, Norway.\protect\\
			E-mail: vinay.j.setty@uis.no
			\IEEEcompsocthanksitem
			}
		}
	}

	\IEEEtitleabstractindextext{
\begin{abstract}

Graphs are ubiquitous and ever-present data structures that have a wide range of applications involving social networks, knowledge bases and biological interactions. The evolution of a graph in such scenarios can yield important insights about the nature and activities of the underlying network, which can then be utilized for applications such as news dissemination, network monitoring, and content curation. Capturing the continuous evolution of a graph can be achieved by long-standing sub-graph queries. Although, for many applications this can only be achieved by a set of queries, state-of-the-art approaches focus on a single query scenario. 
In this paper, we therefore introduce the notion of continuous multi-query \textcolor{black}{processing} over graph streams and discuss its application to a number of use cases. 
To this end, we designed and developed a novel algorithmic solution for efficient multi-query evaluation against a stream of graph updates and experimentally demonstrated its applicability. Our results against two baseline approaches using real-world, as well as synthetic datasets, confirm a two orders of magnitude improvement of the proposed solution. 

\end{abstract} 		
	}
	\maketitle

\IEEEraisesectionheading{\section{Introduction}\label{sec:intro}}

\IEEEPARstart{I}{n} recent years, graphs have emerged as prevalent data structures to model information networks in several domains such as social networks, knowledge bases, communication networks, biological networks and the World Wide Web. These graphs are \emph{massive} in scale and \emph{evolve} constantly due to frequent updates. For example, Facebook has over 1.4 billion daily active users who generate over 500K posts/comments and four million likes every minute resulting in massive updates to the Facebook social graph\footnote{Facebook quarterly update \url{http://bit.ly/2BIM30d}}. 

To gain meaningful and up-to-date insights in such frequently updated graphs, it is essential to be able to monitor and detect continuous patterns of interest. There are several applications from a variety of domains that may benefit from such monitoring. In social networks, such applications may involve targeted advertising, spam detection~\cite{BoshmafMBR11, Wang10}, and fake news propagation monitoring based on specific patterns~\cite{SongGCW14, web:fakeNews:Neo4J}. Similarly, other applications like (i) protein interaction patterns in biological networks~\cite{dip-DB,UniProt}, (ii) traffic monitoring in transportation networks, (iii) attack detection (e.g., distributed denial of service attacks in computer networks), (iv) question answering in knowledge graphs~\cite{c-sparql}, and (v) reasoning over RDF graphs~\cite{streamReasonWeb} may also benefit from such pattern detection.

\begin{figure}[!t]
	\begin{subfigure}[t]{0.5\linewidth}
		\centering
		\includegraphics[scale=.4]{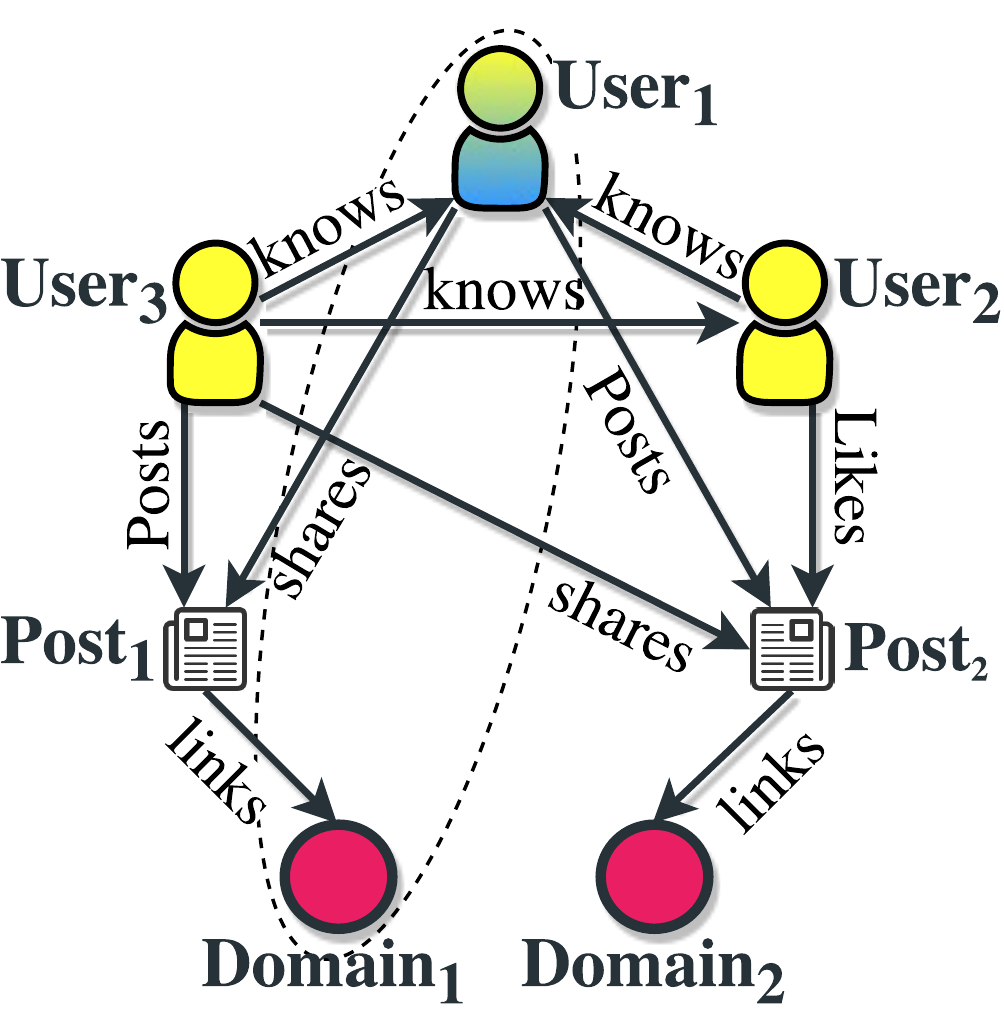}	
		\\
		(a) 
	\end{subfigure}%
	\begin{subfigure}[t]{0.5\linewidth}
		\centering
		\includegraphics[scale=.4]{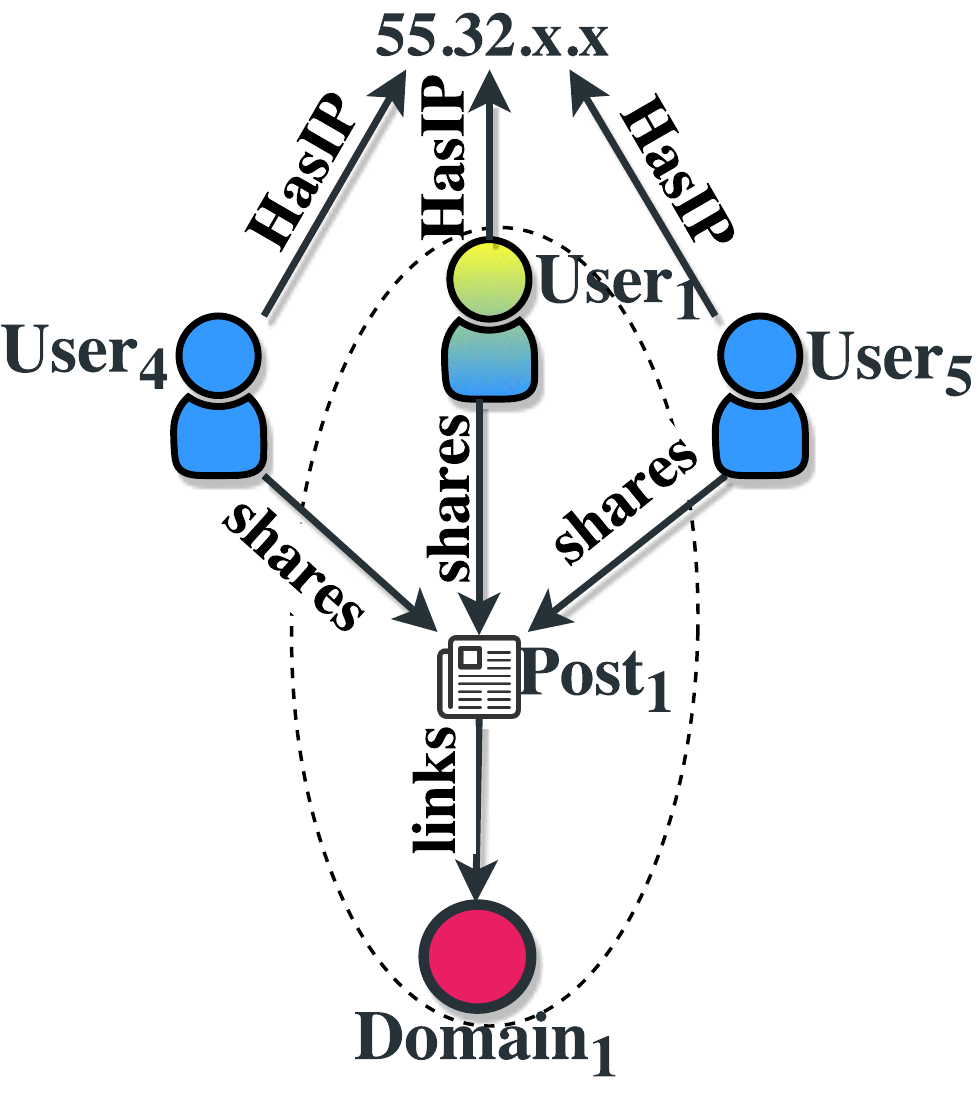}
		\\
		(b)	
	\end{subfigure}%
	
	\caption{Spam detection: Users sharing and liking content with links to flagged domains. (a) A clique of users who know each other, and (b) Users sharing the same IP address.}
	\label{fig:motivation:intro}
\end{figure}

For the applications mentioned above it is necessary to express the required patterns as \emph{continuous sub-graph queries} over (one or many) streams of graph updates and appropriately \emph{notify} the subscribed users for any patterns that match their subscription. Detecting these query patterns is fundamentally a sub-graph isomorphism problem which is known to be NP-complete due to the exponential search space resulting from all possible sub-graphs \cite{shasha-pods02,closuretree2006ICDE}. The typical solution to address this issue is to pre-materialize the necessary sub-graph views for the queries and perform exploratory joins \cite{sun2012efficient}; an expensive operation even for a single query in a static setting. 

These applications deal with graph streams in such a setup that is often essential to be able to support hundreds or thousands of continuous queries simultaneously. This leads to several challenges that require: (i) quickly detecting the affected queries for each update, (ii) maintaining a large number of materialized views, and (iii) avoiding the expensive join and explore approach  for large sets of queries.

To better illustrate the remarks above, consider the application of spam detection in social networks. Fig.~\ref{fig:motivation:intro} shows an example of two graph patterns that may emerge from malicious user activities, i.e., users posting links to domains that have been flagged as fraudulent. Notice that malicious behavior could be caused either because a group of users that know each other share and like each other's posts containing content from a flagged domain (Fig.~\ref{fig:motivation:intro}(a)), or because the group of users shared the same flagged post several times from the same IP. Even though these two queries are fundamentally different and produce different matching patterns, they share a common sub-graph pattern, i.e.,  ``User1$\xrightarrow{\text{shares}}$Post1$\xrightarrow{\text{links}}$Domain1''. If these two queries are evaluated independently, all the 
computations 
for processing the common pattern have to be executed twice. 
However, by identifying  
common patterns in query sets,  
we can amortize the costs of processing and answering them.

One simple approach to avoid processing all the (continuous) queries upon receiving a graph update is to index the query graphs using an inverted-index at the granularity of edges. While this approach may help us quickly detect all the affected queries for a given graph update, we still need to perform several exploratory joins to answer the affected queries. For example, in Fig.~\ref{fig:motivation:intro}, we would need to join and explore the edges matching the pattern ``User1$\xrightarrow{\text{Shares}}$Post1~and~Post1$\xrightarrow{\text{Links}}$Domain1'' upon each update to process the two queries. On the contrary, if we first identify the maximal sub-graph patterns shared among the queries instead, we can minimize the number of operations necessary to answer the queries. 
Therefore, a solution which groups queries based on their shared patterns would be expected to deliver significant performance gains. To the best of our knowledge, none of the existing works provide a solution that exploits common patterns for continuous multi-query answering.

In this paper, we address this gap by proposing a novel \emph{algorithmic solution}, coined \Tree~(\textsc{Tri}e-based \textsc{C}lustering) to index and cluster continuous graph queries. In \Tree, we first decompose queries into a set of directed paths such that each vertex in the query graph pattern belongs to at least one path (path covering problem \cite{Diestel_Book05}). However, obtaining such paths leads to redundant query edges and vertices in the paths; this is undesirable since it affects the performance of the query processing. Therefore, we are interested in finding paths which are shared among different queries, with minimal duplication of vertices. 
The paths obtained are then indexed using `tries' that allow us to minimize query answering time by (i) quickly identifying the affected queries, (ii) sharing materialized views between common patterns, and (iii) efficiently ordering the joins between materialized views affected from the update.

\begin{figure}[!ht]
	\centering
	\resizebox{0.5\textwidth}{!}{
\begingroup
  \makeatletter
  \providecommand\color[2][]{%
    \GenericError{(gnuplot) \space\space\space\@spaces}{%
      Package color not loaded in conjunction with
      terminal option `colourtext'%
    }{See the gnuplot documentation for explanation.%
    }{Either use 'blacktext' in gnuplot or load the package
      color.sty in LaTeX.}%
    \renewcommand\color[2][]{}%
  }%
  \providecommand\includegraphics[2][]{%
    \GenericError{(gnuplot) \space\space\space\@spaces}{%
      Package graphicx or graphics not loaded%
    }{See the gnuplot documentation for explanation.%
    }{The gnuplot epslatex terminal needs graphicx.sty or graphics.sty.}%
    \renewcommand\includegraphics[2][]{}%
  }%
  \providecommand\rotatebox[2]{#2}%
  \@ifundefined{ifGPcolor}{%
    \newif\ifGPcolor
    \GPcolortrue
  }{}%
  \@ifundefined{ifGPblacktext}{%
    \newif\ifGPblacktext
    \GPblacktexttrue
  }{}%
  \let\gplgaddtomacro\g@addto@macro
  \gdef\gplbacktext{}%
  \gdef\gplfronttext{}%
  \makeatother
  \ifGPblacktext
    \def\colorrgb#1{}%
    \def\colorgray#1{}%
  \else
    \ifGPcolor
      \def\colorrgb#1{\color[rgb]{#1}}%
      \def\colorgray#1{\color[gray]{#1}}%
      \expandafter\def\csname LTw\endcsname{\color{white}}%
      \expandafter\def\csname LTb\endcsname{\color{black}}%
      \expandafter\def\csname LTa\endcsname{\color{black}}%
      \expandafter\def\csname LT0\endcsname{\color[rgb]{1,0,0}}%
      \expandafter\def\csname LT1\endcsname{\color[rgb]{0,1,0}}%
      \expandafter\def\csname LT2\endcsname{\color[rgb]{0,0,1}}%
      \expandafter\def\csname LT3\endcsname{\color[rgb]{1,0,1}}%
      \expandafter\def\csname LT4\endcsname{\color[rgb]{0,1,1}}%
      \expandafter\def\csname LT5\endcsname{\color[rgb]{1,1,0}}%
      \expandafter\def\csname LT6\endcsname{\color[rgb]{0,0,0}}%
      \expandafter\def\csname LT7\endcsname{\color[rgb]{1,0.3,0}}%
      \expandafter\def\csname LT8\endcsname{\color[rgb]{0.5,0.5,0.5}}%
    \else
      \def\colorrgb#1{\color{black}}%
      \def\colorgray#1{\color[gray]{#1}}%
      \expandafter\def\csname LTw\endcsname{\color{white}}%
      \expandafter\def\csname LTb\endcsname{\color{black}}%
      \expandafter\def\csname LTa\endcsname{\color{black}}%
      \expandafter\def\csname LT0\endcsname{\color{black}}%
      \expandafter\def\csname LT1\endcsname{\color{black}}%
      \expandafter\def\csname LT2\endcsname{\color{black}}%
      \expandafter\def\csname LT3\endcsname{\color{black}}%
      \expandafter\def\csname LT4\endcsname{\color{black}}%
      \expandafter\def\csname LT5\endcsname{\color{black}}%
      \expandafter\def\csname LT6\endcsname{\color{black}}%
      \expandafter\def\csname LT7\endcsname{\color{black}}%
      \expandafter\def\csname LT8\endcsname{\color{black}}%
    \fi
  \fi
    \setlength{\unitlength}{0.0500bp}%
    \ifx\gptboxheight\undefined%
      \newlength{\gptboxheight}%
      \newlength{\gptboxwidth}%
      \newsavebox{\gptboxtext}%
    \fi%
    \setlength{\fboxrule}{0.5pt}%
    \setlength{\fboxsep}{1pt}%
\begin{picture}(7200.00,5040.00)%
    \gplgaddtomacro\gplbacktext{%
      \csname LTb\endcsname%
      \put(1342,704){\makebox(0,0)[r]{\strut{}0.00}}%
      \csname LTb\endcsname%
      \put(1342,1390){\makebox(0,0)[r]{\strut{}0.01}}%
      \csname LTb\endcsname%
      \put(1342,2076){\makebox(0,0)[r]{\strut{}0.10}}%
      \csname LTb\endcsname%
      \put(1342,2762){\makebox(0,0)[r]{\strut{}1.00}}%
      \csname LTb\endcsname%
      \put(1342,3447){\makebox(0,0)[r]{\strut{}10.00}}%
      \csname LTb\endcsname%
      \put(1342,4133){\makebox(0,0)[r]{\strut{}100.00}}%
      \csname LTb\endcsname%
      \put(1342,4819){\makebox(0,0)[r]{\strut{}1000.00}}%
      \csname LTb\endcsname%
      \put(2140,484){\makebox(0,0){\strut{}1}}%
      \csname LTb\endcsname%
      \put(3472,484){\makebox(0,0){\strut{}10}}%
      \csname LTb\endcsname%
      \put(4805,484){\makebox(0,0){\strut{}100}}%
      \csname LTb\endcsname%
      \put(6137,484){\makebox(0,0){\strut{}1000}}%
    }%
    \gplgaddtomacro\gplfronttext{%
      \csname LTb\endcsname%
      \put(198,2761){\rotatebox{-270}{\makebox(0,0){\strut{}Answering time (msec/update)}}}%
      \put(4138,154){\makebox(0,0){\strut{} Number of queries }}%
      \csname LTb\endcsname%
      \put(1606,4624){\makebox(0,0)[l]{\strut{}\Tree}}%
      \csname LTb\endcsname%
      \put(1606,4360){\makebox(0,0)[l]{\strut{}\Neo}}%
      \csname LTb\endcsname%
      \put(1606,4096){\makebox(0,0)[l]{\strut{}\InvIncr}}%
      \csname LTb\endcsname%
      \put(1606,3832){\makebox(0,0)[l]{\strut{}\Inv}}%
    }%
    \gplbacktext
    \put(0,0){\includegraphics{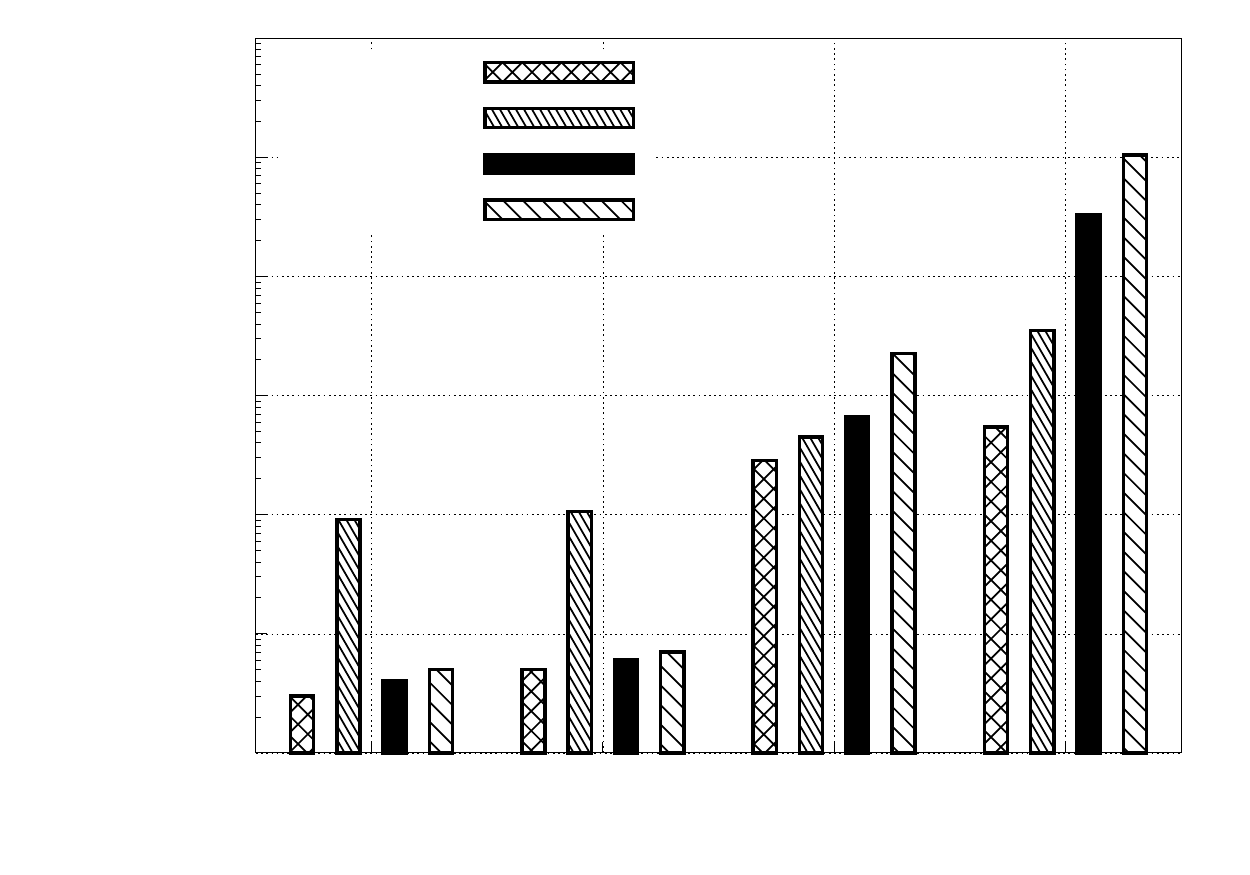}}%
    \gplfronttext
  \end{picture}%
\endgroup
 	}
	
	\caption{\textcolor{black}{Log-log plot comparing query answering time for the $ SNB $ dataset from the LDBC benchmark, when varying the number of queries exponentially.}}
	\label{fig:teaser}
\end{figure}

\color{black}
Fig.~\ref{fig:teaser} shows the potential for improvement in query answering time with our query clustering solution \Tree, for the LDBC graph benchmark \cite{ErlingALCGPPB15}. We can observe that \Tree~provides a speedup of two orders of magnitude in query answering time, 
compared to two advanced baselines using the ``inverted indexing technique" (\Inv, \InvIncr) and the graph database \Neo\ 
that do not exploit the common sub-graph patterns in the queries.
\color{black}

In summary, our \emph{contributions} are:
\begin{itemize}
	\item
	We formalize the problem of continuous multi-query answering over graph streams (Section~\ref{sec:dataModel}). 
	\item
	We propose a novel query graph clustering algorithm that is able to efficiently handle large numbers of continuous graph queries by resorting on (i) the decomposition of continuous query graphs to minimum covering paths and (ii) the utilization of tries for capturing the common parts of those paths (Section~\ref{sec:algTree}).
	\item
	\textcolor{black}{
	Since no prior work in the literature has considered continuous multi-query answering, we designed and developed two algorithmic solutions that utilize inverted indexes for the query answering. Additionally, we deploy and extend   \Neo~\cite{neo4j}, a well-established graph database solution, to support our proposed paradigm. To this end, the proposed solutions will serve as baselines  approaches during the experimental evaluation.  (Section ~\ref{sec:competitors}).
	}

	\item
	We experimentally evaluate the proposed solution using three different datasets from social networks, transportation, and biology domains, and compare the performance against the \textcolor{black}{three}  baselines. In this context, we show that our solution can achieve up to two orders of magnitude improvement in query processing time (Section~\ref{sec:exps}). 
\end{itemize}

\section{Related Work}
\label{sec:related}

\textcolor{black}{
Structural graph pattern search using graph isomorphism has been studied in the literature before \cite{shasha-pods02,closuretree2006ICDE}. In \cite{HanLL13}, the authors propose a solution that aims at reducing the search space for a single query graph. The solution identifies candidate regions in the graph that can contain query embeddings, while it is coupled with a neighborhood equivalence locating strategy to generate the necessary enumerations. In the same spirit \cite{ren2015exploiting} aims at reducing the search space in the graph by exploiting the syntactic similarities present on vertex relationships. \cite{ren2016multi} considers the sub-graph isomorphism problem when multiple queries are answered simultaneously. However, these techniques are designed for static graphs and are not suitable for processing continuous graph queries on evolving graphs. 
}

\begin{figure*}[!t]
	\begin{minipage}[b]{5cm}
		\centering
		\small{
			\begin{tabular}[b]{|l|}
				\hline
				\multicolumn{1}{|c|}{Update stream $ S $}\\
				\hline
				$ u_1 =$ ($ checksIn $ = ($ P_1 $, plc)) \\
				$ u_2 =$ ($ checksIn $ = ($ P_2 $, plc)) \\
				$ u_3 =$ ($ checksIn $ = ($ P_3 $, plc )) \\
				\hline
			\end{tabular}
		}\\ (a)
	\end{minipage}
	\begin{minipage}[b]{10cm}
		\centering
		\includegraphics[height=0.1\textheight, keepaspectratio]{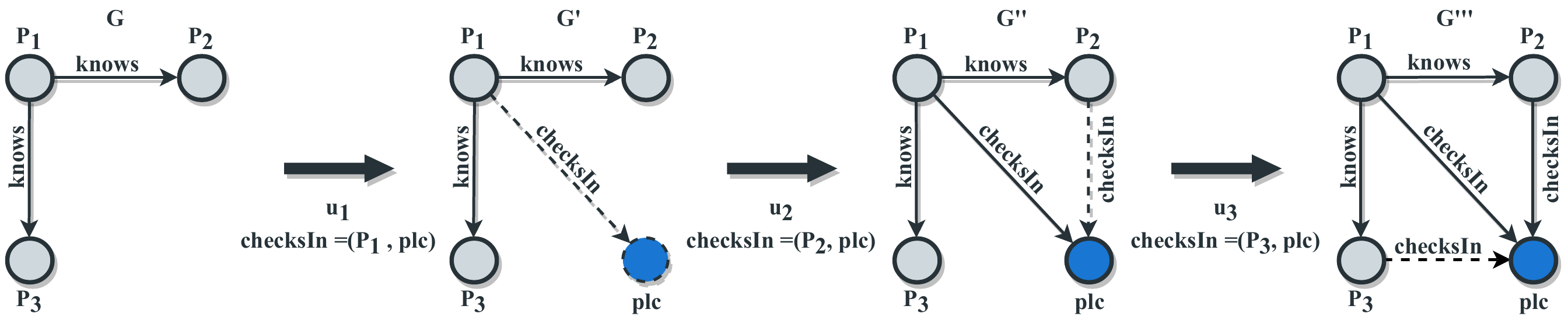}
		\\(b)
	\end{minipage}
	\caption{(a) An update stream $ S $ and (b) the evolution of graph $ G $ after inserting $ u_i \in S $.}
	\label{fig:graphEvolution}
\end{figure*}

Continuous sub-graph matching has been considered in \cite{wang-icde09} but the authors assume a static set of sub-graphs to be matched against update events, use approximate methods that yield false positives, and small (evolving) graphs. An extension to this work considers the problem of uncertain graph streams \cite{wang-tkde2010}, over wireless sensor networks and PPIs. 
The work in \cite{GillaniPL16} considers a setup of continuous graph pattern matching over knowledge graph streams. The proposed solution utilizes finite automatons to represent and answer the continuous queries. 
However, this approach can support a handful of queries,  since, each query is evaluated separately, while, it generates false positives due to the adopted sliding window technique. These solutions are not suitable for answering large number of continuous queries on graphs with high update rates.

 There are a few publish/subscribe solutions on ontology graphs proposed in \cite{GtoPSS-www05,Ontology-PubSub4}, but they are limited to the RDF data model. Distributed pub/sub middleware for graphs has recently been proposed in \cite{graps}, however, the main focus is on node constraints (attributes) while ignoring the graph structure.

In graph streams research; \cite{cgstream-cikm12,pan-cikm12} propose algorithms to identify correlated graphs from a graph stream. This differs from our setup since a sliding window that covers a number of batches of data is used, and the main focus is set on identifying subgraphs with high Pearson correlation coefficients. In \cite{gao2016toward}, the authors propose continuous pattern detection in graph streams with snapshot isolation. However, this solution considers only single queries at a time and the patterns detected are also approximate.%

The work in \cite{choudhury-multiRelational} provides an exact subgraph search algorithm that exploits the \emph{temporal} characteristics of representative queries for online news or social media monitoring. This algorithm exploits the structural and semantic characteristics of the graph through a specialized data structure. 
An extension of this work, considers continuous query answering with graph patterns over dynamic multi-relation graphs \cite{ChoudhuryHCAF15}. 
Finally, in ~\cite{sun2012efficient} the authors perform subgraph matching over a billion node graph by proposing graph exploration methods based on cloud technologies. 
While the aforementioned works are similar to the query evaluation scenario, the emphasis is on efficient search mechanisms, rather than continuous answering over streaming graph data.

\section{Data Model and Problem Definition}
\label{sec:dataModel}

In this section we outline the data (Section~\ref{sec:graphModel}) and query model (Section~\ref{sec:queryModel}) that our approach builds upon. 

\subsection{Graph Model}
\label{sec:graphModel}
In this paper, we use attribute graphs~\cite{book:cormen} (Definition~\ref{def:graph}), as our data model, as they are used natively in a wide variety of applications, such as social network graphs, traffic network graphs, and citation graphs. Datasets in other data models can be mapped to attribute graphs in a straightforward manner so that our approach can be applied to them as well. 

\begin{definition}
	An \emph{attribute graph} $G$ is defined as a directed labeled multigraph:
	\[G = (V, E, l_V, l_E, \Sigma_V, \Sigma_E)\] 
	where $V$ is the set of vertices and $E$ the set of edges. An edge $ e \in E $ is an ordered pair of vertices $ e : (s, t)$, where $s,t \in V $ represent source and target vertices.
	$l_V: V \rightarrow \Sigma_V $ and $l_E: E  \rightarrow \Sigma_E$ are labeling functions assigning labels to vertices and edges from the label sets $ \Sigma_V $ and $ \Sigma_E $.   
	\label{def:graph}
\end{definition}

For ease of presentation, we denote an edge $ e $ as $ e = (s, t) $, where $e$, $ s $  and $ t $ are the labels of the edge($ l_E(e) $), source vertex ($ l_V(s) $) and target vertex ($ l_V(t) $) respectively.

\spacecut{
Relational data, for instance, can be mapped to an attribute graph by creating a vertex for each tuple in a relation (using an artificial identifier as label based on the primary key). For each of the tuple's attribute values a vertex is created using the attribute value as label and connected to the vertex representing the tuple via edges labeled with the attribute names. Foreign key relationships lead to edges connecting vertices created for other relations. 

RDF~\cite{web:rdf} datasets consist of triples of the form (subject, predicate, object), where subjects always correspond to entities identified by URIs (blank nodes can be assigned URIs for the mapping) and objects can either be URIs or literals. The predicate (URI) describes the relationship between subject and object. 
An RDF dataset can be mapped to an attribute graph by creating a vertex for each URI (vertex label) that occurs at a subject or object position in any triple. For each triple with URIs at both subject and object position an edge connecting the corresponding vertices is created with the triple's predicate as label. For each triple with a literal at object position, a new vertex is created for the object and connected to the vertex representing the triple's subject with an edge using the triple's predicate as label. This basic process can be extended by rules handling special cases such as reification and triples with property URIs as subjects or objects. 

}

As our goal is to facilitate efficient continuous multi-query \textcolor{black}{processing} over graph streams, we also provide formal definitions for updates and graph streams (Definitions~\ref{def:update} and~\ref{def:stream}). 

\begin{definition}
	An update $ u_t$ on graph $ G $ is defined as an \emph{addition} $ (e) $ of an edge $ e $ at time $ t $. An addition leads to new edges between vertices and possibly the creation of new vertices.
	\label{def:update}
\end{definition}

\begin{definition}
	A \emph{graph stream} $ S =  (u_1, u_2, \ldots, u_t)$ of graph $ G $ is an \emph{ordered sequence} of updates. 
	\label{def:stream}
\end{definition}

Fig.~\ref{fig:graphEvolution}(a) presents an update stream $ S $ consisting of three graph updates $ u_1 $, $ u_2 $, and $ u_3 $ generated from social network events.
While, Fig.~\ref{fig:graphEvolution} (b) shows the initial state of graph $ G $ and its evolution after inserting sequentially the three updates.

\begin{figure}[!t]
	\centering
	\includegraphics[height=0.1\textheight,keepaspectratio]{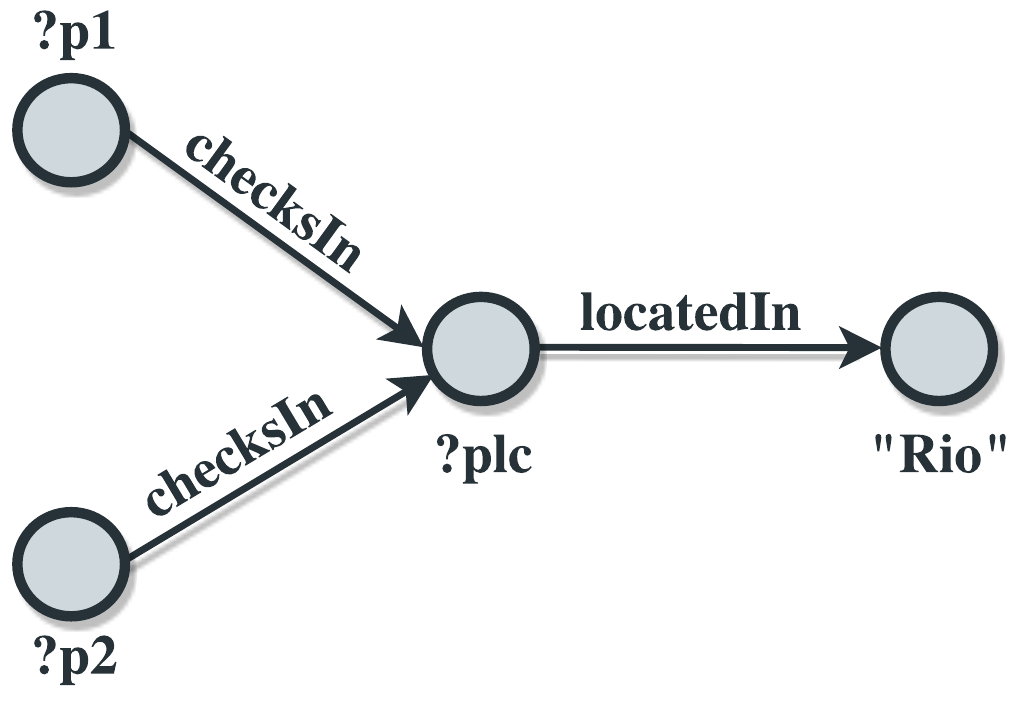}

	\caption{Example query graph pattern.}
	\label{fig:ex:checkIn1}	
\end{figure}

\subsection{Query Model}
\label{sec:queryModel}

For our query model we assume that users (or services operating on their behalf) are interested to learn when certain patterns emerge in an evolving graph. 
Definition~\ref{def:queryGraph} provides a formal definition of such \emph{query graph patterns} defining structural and attribute constraints. 

\begin{definition}
	A \emph{query graph pattern} $Q_i$ is defined as a directed labeled multigraph:
	\[Q_i = (V_{Q_i}, E_{Q_i}, vars, l_V, l_E, \Sigma_V, \Sigma_E)\]
	where $ V_{Q_i} $ is a set of vertices, $ E_{Q_i} $ a set of edges, and $vars$ a set of variables. %
	$l_V: V \rightarrow \{\Sigma_V \cup vars\} $ and $l_E: E  \rightarrow \Sigma_E$ are labeling functions assigning labels (and variables) to vertices and edges. 
	\label{def:queryGraph}
\end{definition}

Let us consider an example where a user wants to be notified when his friends visit places nearby. Fig.~\ref{fig:ex:checkIn1} shows the corresponding query graph pattern that will result in a user notification when two people check in at the same place/location in Rio. 

\begin{figure*}[!ht]
	\begin{minipage}[t]{0.6\linewidth}
		\centering
		\vspace{0pt}
		\includegraphics[scale=0.50]{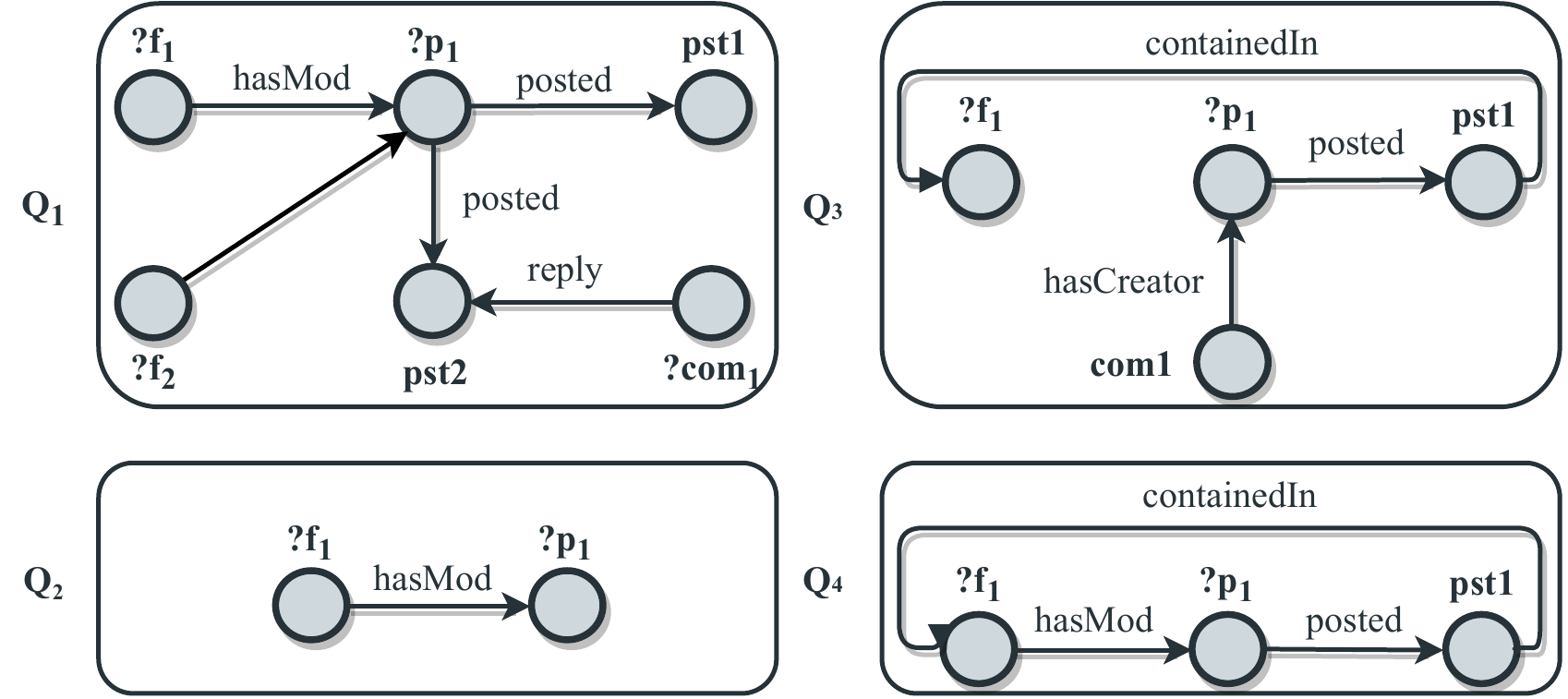}
		\\ \normalsize{(a)}
	\end{minipage}
	\begin{minipage}[t]{0.3\linewidth}
		\centering
		\vspace{0pt}
		\scriptsize{
			\begin{tabular}{|c|l|}
				\hline
				Query ID & Set of Covering Paths \\ 
				\hline
				\hline
				\multirow{3}{*}{$ Q_1 $} & $ P_1 = \{ ?var \xrightarrow{hasMod} ?var $ $ \xrightarrow{posted} ``pst 1"\} $ \\
				& $ P_2 = \{ ?var \xrightarrow{hasMod} ?var $ $ \xrightarrow{posted} ``pst 2"\} $ \\
				& $ P_3 = \{ ?var \xrightarrow{reply} ``pst 2"\} $ \\
				\hline
				$ Q_2 $ & $ P_1 = \{ ?var \xrightarrow{hasMod} ?var\}$ \\
				\hline
				\multirow{2}{*}{$ Q_3 $} & $ P_1 = \{ ``com 1" \xrightarrow{hasCreator} ?var \xrightarrow{posted} $ \\
				& $ ``pst 1" \xrightarrow{containedIn} ?var\} $ \\
				\hline
				\multirow{2}{*}{$ Q_4 $} & $ P_1 = \{ ?var \xrightarrow{hasMod} ?var \xrightarrow{posted} ``pst 1" $ \\
				& $ \xrightarrow{containedIn} ?var\} $ \\
				\hline
			\end{tabular}	
		} \\ \vspace{9pt} \normalsize{(b)}
	\end{minipage}
	\caption{\textcolor{black}{(a) Four query graph patterns that capture events generated inside a social network and (b) their covering paths.}}
	\label{fig:exampleQueries}
\end{figure*}

Based on the above definitions, let us now define the problem of \emph{multi-query \textcolor{black}{processing} over graph streams}.

	\textit{Problem Definition.} 
	Given a set of query graph patterns $Q_{DB} = \{Q_1, Q_2, \ldots, Q_k\}$, an initial attribute graph $G$, and a graph stream $S$ with continuous updates $u_t \in S$, the problem of \emph{multi-query \textcolor{black}{processing} over graph streams} consists of continuously identifying all satisfied query graph patterns $Q_i \in Q_{DB}$ when applying incoming updates. 

\ctitle{Query Set and Graph Modifications.}
\color{black}
A set of query graph patterns $ Q_{DB} $ is subject to modifications (i.e., additions and  deletions). In this work, we focus on streamlining the query indexing phase, while developing techniques that allow processing each incoming query graph pattern separately, thus supporting continuous additions in $ Q_{DB} $. In the same manner, a graph $ G $ is subject to edge additions and deletions, our main objective is to efficiently determine the queries satisfied by an edge addition. The proposed model does not require indexing the entire graph $ G $ and retains solely the necessary parts of $ G $ for the query answering. To this end, we do not further discuss deletions on $ Q_{DB} $ and $ G $, as we focus on providing high performance query answering algorithms.
\color{black}

\section{Trie-Based Clustering}
\label{sec:algTree}

To solve the problem defined in the previous section, we propose \Tree\ (\textsc{Tri}e-based \textsc{C}lustering).  
As motivated in Section \ref{sec:intro}, the \emph{key idea} behind \Tree\, lies in the fact that query graph patterns overlap in their structural and attribute restrictions. After identifying and indexing these shared characteristics (Section~\ref{sec:tree:indexing}), they can be exploited to batch-answer the indexed query set and in this way reduce query response time (Section~\ref{sec:tree:answering}). 

\subsection{Query Indexing Phase}
\label{sec:tree:indexing}

\Tree\ indexes each query graph pattern $ Q_i $ by applying the following two steps:
\begin{mylisting}
	\item 
	Transforming the original query graph pattern $ Q_i $ into a set of path conjuncts, that cover all vertices and edges of $ Q_i $, and  when combined can effectively re-compose $ Q_i $.
	
	\item 
	Indexing all paths in a trie-based structure along with unique query identifiers, while clustering all paths of all indexed queries by exploiting commonalities among them.
\end{mylisting}

\begin{figure}[!t]
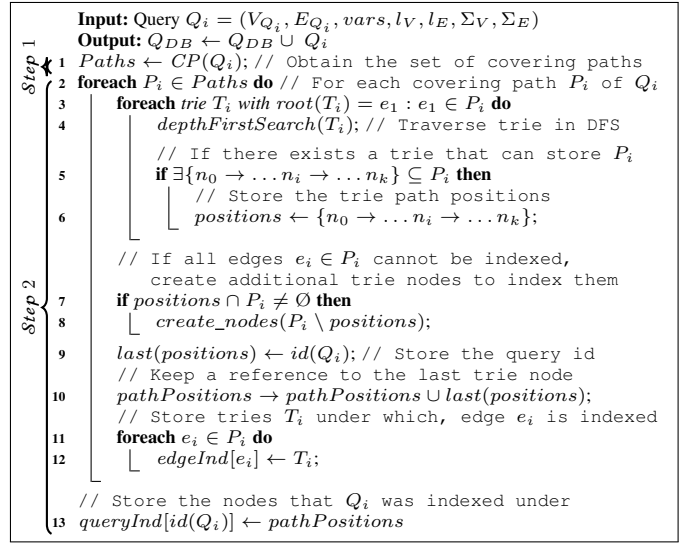

	\vspace{-15pt}
\IncMargin{1.2em}
\begin{algorithm}[H]
\scriptsize

	\KwIn{Query $Q_i = (V_{Q_i}, E_{Q_i}, vars, l_V, l_E, \Sigma_V, \Sigma_E)$} 
	\KwOut{$ Q_{DB} \leftarrow Q_{DB} \cup\ Q_i$}
	\tikzmark{left}\tikzmark{top1}\tikzmark{bottom1}$Paths  \leftarrow \cPaths(Q_i)$\tcp*[l]{Obtain the set of covering paths \hspace*{-5mm}}
	\tikzmark{top2}\ForEach(\tcp*[h]{For each covering path $P_i$ of $Q_i$\hspace*{-5mm}}){$P_i \in Paths$}{
		
		\ForEach{trie $T_i $ with $\rt(T_i) = e_1 : e_1 \in P_i$}{
			$\search(T_i)$\tcp*[l]{Traverse trie in DFS\hspace*{-5mm}}
			\BlankLine
			
			\tcp{If there exists a trie that can store $P_i$}
			\If{$\exists \{n_0 \rightarrow \ldots n_i \rightarrow \ldots n_k\} \subseteq P_i$}{
				\tcp{Store the trie path positions}
				$positions \leftarrow \{n_0 \rightarrow \ldots n_i \rightarrow \ldots n_k\}$\;
			}
		}

		\tcp{If all edges $e_i \in P_i$ cannot be indexed, create additional trie nodes to index them}
		\If{$positions \cap P_i \neq \O$}{
						$create\_nodes (P_i \setminus positions)$\;
		}
		$ last(positions) \leftarrow id(Q_i)$\tcp*[l]{Store the query id\hspace*{-5mm}}
		\tcp{Keep a reference to the last trie node}
		$pathPositions \rightarrow pathPositions \cup last(positions)$\; 

		\tcp{Store tries $T_i$ under which, edge $e_i$ is indexed\hspace*{-5mm}}
		\ForEach{$e_i \in P_i$}{
			$\edgeInd[e_i] \leftarrow T_i$\;
		}
	}

	\tcp{Store the nodes that $ Q_i $ was indexed under}
	$ \queryInd[id(Q_i)] \leftarrow pathPositions$ \tikzmark{bottom2}
	\AddNote{top1}{bottom1}{left}{$ \mathbcal{Step\ 1} $}
	\AddNote{top2}{bottom2}{left}{$ \mathbcal{Step\ 2} $}
	
\end{algorithm}
\DecMargin{1.2em} 	\vspace{-15pt}

	\caption{Query indexing phase of Algorithm~\Tree.}
	\label{pseudo:tree:indexing}
\end{figure}

In the following, we present each step of the query indexing phase of Algorithm~\Tree, give details about the data structures utilized and provide its pseudocode (Fig.~\ref{pseudo:tree:indexing}).

\ctitle{\underline{$ \mathbcal{Step\ 1:} $} Extracting the Covering Paths.}
In the first step of the query indexing process, Algorithm~\Tree\ decomposes a query graph pattern $ Q_i $ and  extracts a set of paths $ \cPaths(Q_i) $ (Fig.~\ref{pseudo:tree:indexing}, line 1). This set of paths, covers all vertices $ V \in Q_i $ and edges $ E \in Q_i $. 
At first, we give the definition of a path and subsequently define and discuss the covering path set problem.

\begin{definition}
	A \emph{path} $ P_i \in Q_i$ is defined as a list of vertices 
	$ P_i = \{ v_1 \xrightarrow{e_1} v_2 \xrightarrow{e_2} \dots v_k \xrightarrow{e_k} v_{k+1}\} $ where $ v_i \in Q_i $, such that two sequential vertices $ v_i, v_{i+1} \in P_i $ have exactly one edge $e_i \in Q_i$ connecting them, i.e., $ e_k = (v_k, v_{k+1}) $.
	
	\label{def:path}
\end{definition}

\begin{definition}
	\textcolor{black}{
	The \emph{covering paths} \cite{book:softwareTesting} $ CP $ of a query graph $ Q_i $ is defined as a set of paths $ CP(Q_i) = \{P_1, P_2, \ldots, P_k\}$ that cover all \emph{vertices} and \emph{edges} of $Q_i$. In more detail, we are interested in the least number of paths while ensuring that for every vertex $ v_i \in Q_i $ there is at least one path $ P_j $ that contains $ v_i $, i.e., $ \forall i \exists j : v_i \in P_j $, $ v_i \in Q_i $. In the same manner, for every edge $ e_i \in Q_i $ there is at least one path $ P_j $ that contains $ e_i $, i.e., $ \forall i \exists j : e_i \in P_j $.
	}
	
	\label{def:coverPath}
\end{definition}

\ctitle{Obtaining the Set of Covering Paths.}
\textcolor{black}{
The problem of obtaining a set of paths that covers all vertices and edges is a graph optimization problem that has been studied in literature \cite{book:softwareTesting, NtafosH79}. 
In our approach, we choose to solve the problem by applying a greedy algorithm, as follows: For all vertices $ v_i $ in the query graph $ Q_i $ execute a depth-first walk until a leaf vertex (no outgoing edge) of the graph is reached, or there is no new vertex to visit. Subsequently, repeat this step until all vertices and edges of the query graph $ Q_i $ have been visited at least once and a list of paths has been obtained. Finally, for each path in the obtained list, check if it is a sub-path of an already discovered path, and remove it from the list of covering paths. The end result of this procedure yields the \emph{set of covering paths}. 
}

\begin{example}
	In Fig.~\ref{fig:exampleQueries}(a) we present four query graph patterns.
	These query graph patterns capture activities of users inside a social network. By applying Definition~\ref{def:coverPath} on the four query graph patterns presented, Algorithm~\Tree\ extracts four sets of covering paths, presented in Fig.~\ref{fig:exampleQueries}(b).
\end{example}

Obtaining a set of paths serves two purposes: (a) it gives a less complex representation of the query graph that is easier to manage, index and cluster, as well as (b) it provides a streamlined approach on how to perform the materialization of the subgraphs that match a query graph pattern, i.e., the query answering during the evolution of the graph.

\ctitle{Materialization.} Each edge $ e_i $ that is present in the query set has a materialized view that corresponds to its $ \mv[e_i] $. The materialized view of $ e_i $ stores all the updates $ u_i $ that contain $ e_i $. In order to obtain the subgraphs that satisfy a query graph pattern $ Q_i $ all edges $ e_i \in Q_i $ must have a non-empty materialized view (i.e., $ \mv \neq \O $) and the materialized views should be joined as defined by the query graph pattern.

In essence, the query graph pattern determines the \emph{execution plan} of the query. However, given that a query pattern in itself is a graph there is a high number of possible execution plans available. A path $ P_i = \{ v_1 \xrightarrow{e_1} v_2 \xrightarrow{e_2} \dots v_k\} $ serves as a model that defines the order in which the materialization should be performed. Thus, starting from the source vertex $ v_1 \in P_i $ and joining all the materialized views from $ v_1 $ to the leaf vertex $ v_k \in P_i : |P| = k $ yields all the subgraphs that satisfy the path $ P_i $. After all paths $ P_i $ that belong in $ Q_i $ have been satisfied, a final join operation must be performed between all the paths. This join operation will produce the subgraphs that satisfy the query graph $ Q_i $. To achieve this path joining set, additional information is kept about the intersection of the paths $ P_i \in Q_i$. The intersection of two paths $ P_i $ and $ P_j $ are their common vertices.

\begin{figure}[!t]
	\centering
	\includegraphics[scale=0.50]{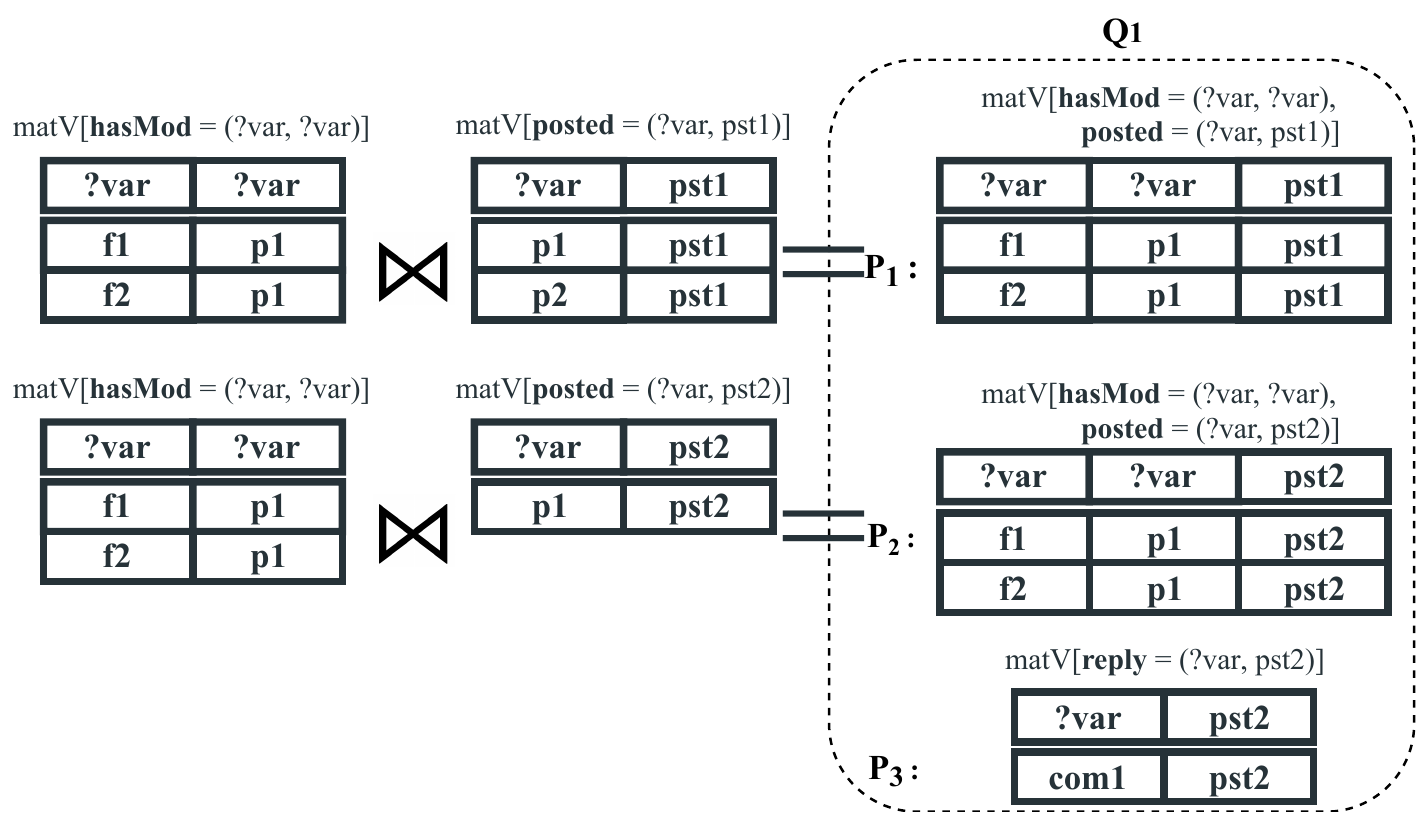}	
	
	\caption{\textcolor{black}{Materialized views of $ Q_1 $.}}
	\label{fig:pathJoin}
\end{figure}

\color{black}
\begin{example}
	Fig.~\ref{fig:pathJoin} presents some  possible materialized views that correspond to the covering paths of query graph $ Q_1 $ (Fig.~\ref{fig:exampleQueries} (b)). In order to locate all subgraphs that satisfy the structural and attribute restrictions posed by paths $ P_1 $, $ P_2 $ and $ P_3 $ their materialized views should be calculated. More specifically, path  $ P_1 = \{ ?var \xrightarrow{hasMod} ?var \xrightarrow{posted} ``pst1"\} $, is formulated by two edges, edges $ hasMod = (?var, ?var) $ and $ posted = (?var, pst1) $, thus, their materialized views $ \mv[hasMod = (?var, ?var)] $ and  $ \mv[posted = (?var, pst1)] $ must be joined. These two views contains all updates $ u_i $ that correspond to them, while the result of their join operation will be a new materialized view $ \mv[hasMod = (?var, ?var), posted = (?var, pst1)] $ as shown in Fig.~\ref{fig:pathJoin}. In a similar manner, the subgraphs that satisfy path $ P_2 $ are calculated, while $ P_3 $ that is formulated by a single edge does not require any join operations. Finally, in order to calculate the subgraphs that match $ Q_1 $ all materialized views that correspond to paths $ P_1 $, $ P_2 $ and $ P_3 $ must be joined.
\end{example}
\color{black}

\ctitle{\underline{$ \mathbcal{Step\ 2:} $} Indexing the Paths.}
Algorithm~\Tree\ proceeds into indexing all the paths, extracted in $ \mathcal{Step\ 1} $, into a trie-based data structure. For each path $ P_i \in \cPaths(Q_i) $, \Tree\ examines the forest for trie roots that can index the first edge $ e_1 \in P_i $ (Fig.~\ref{pseudo:tree:indexing}, lines $ 3 - 6 $).
To access the trie roots, \Tree\ utilizes a hash table (namely $ \rootIndex $) that indexes the values of the root-nodes (keys) and the references to the root nodes (values). 
If such trie $ T_i $ is located, $ T_i $ is traversed in a \emph{DFS} manner to determine in which sub-trie path $ P_i $ can be indexed (Fig.~\ref{pseudo:tree:indexing}, line $ 4 $). 
Thus, \Tree\ traverses the forest to locate an existing trie-path $ \{n_1 \rightarrow \ldots n_i \rightarrow \ldots n_k\} $ that can index the ordered set of edges $ \{e_1, \dots, e_k\} \in P_i $. 
If the discovered trie-path can index $ P_i $ partially (Fig.~\ref{pseudo:tree:indexing}, line $ 7 $), \Tree\ proceeds into creating a set of new nodes under $ n_k $ that can index the remaining edges (Fig.~\ref{pseudo:tree:indexing}, line $ 8 $). 
Finally,  the algorithm stores the identifier of $ Q_i $ at the last node of the trie path (Fig.~\ref{pseudo:tree:indexing}, line $ 9 $).

Algorithm~\Tree\ makes use of two additional data structures, namely $ \edgeInd $ and $ \queryInd $. The former data structure is a hash table that stores each edge $ e_i \in P_i$ (key) and a collection of trie roots $ T_i $ which index $ e_i $ as the hash table's value (Fig.~\ref{pseudo:tree:indexing}, lines $ 11 - 12 $). Finally, \Tree\ utilizes a matrix $ \queryInd $ that indexes the query identifier along side the set of nodes under which its covering paths $ P_i \in CP(Q_i) $ was indexed  (Fig.~\ref{pseudo:tree:indexing}, line $ 13 $).

\begin{figure*}[!t]
	
	\includegraphics[width=\linewidth]{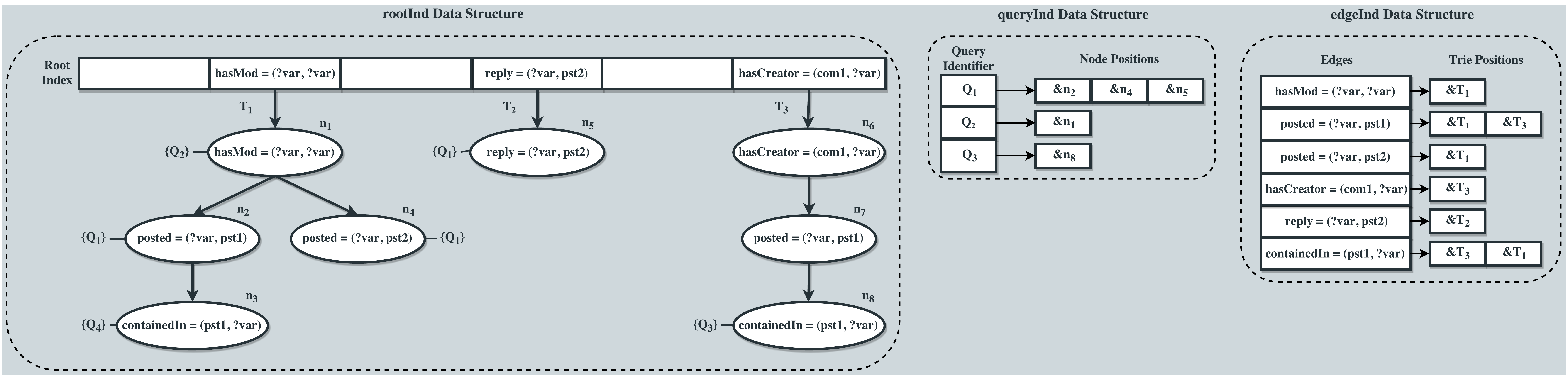}
	
	\caption{Data structures utilized by Algorithm~\Tree\ to cluster query graph patterns.}
	\label{fig:forest}
\end{figure*}

\begin{example}
	Fig.~\ref{fig:forest} presents an example of $ \rootIndex $, $ \queryInd $ and $ \edgeInd $ of Algorithm~\Tree\ when indexing the set of covering paths of Fig.~\ref{fig:exampleQueries} (b). Notice, that \Tree\ indexes paths $ P_1, P_2 \in Q_1 $, path $ P_1 \in Q_2 $ and path $ P_1 \in Q_4 $ under the same trie $ T_1 $, thus, clustering together their common structural restrictions (all the aforementioned paths) and their attribute restrictions. Additionally, note that the $ \queryInd $ data structure keeps references to the last node where each path $ P_i \in Q_i $ is stored, e.g. for $ Q_1 $ it keeps a set of node positions $\{ \&n_2, \&n_4, \&n_5 \}$ that correspond to its original paths $ P_1 $, $ P_2 $ and $ P_3 $ respectively. Finally, $ \edgeInd $ stores all the unique edges present in the path set of Fig.~\ref{fig:exampleQueries} (b), with references to the trie roots under which they are indexed, e.g. edge $ posted = (?var, pst1) $ that is present in $ P_1 \in Q_1 $, $ P_1 \in Q_3 $ and $ P_1 \in Q_4 $, is indexed under both tries $ T_1 $ and $ T_3 $, thus this information is stored in set $ \{\&T_1, \&T_3\} $.
\end{example}

\color{black}
The time complexity of Algorithm \Tree\ when indexing a path $ P_i $, where $ |P_i| = M $ edges and $ B $ the branching factor of the forest, is $\mathcal{O}(M*B)$, since \Tree\ uses a DFS strategy, with the maximum depth bound by the number of edges. Thus, for a new query graph pattern $Q_i$ with $ N $ covering paths, the total time complexity is $\mathcal{O}(N * M * B)$. Finally, the space complexity of Algorithm \Tree\ when indexing a query $ Q_i $ is $ \mathcal{O}(N * M) $, where $ M $ is the number of edges in a path and $ N $ the cardinality of $ Q_i $'s covering paths.
\color{black}

\ctitle{Variable Handling.}
A query graph pattern $ Q_i $ contains vertices that can either be literals (specific entities in the graph) identified by their label, or variables denoted with the generic label ``?var''. This approach is applied in order to alleviate restrictions posed by naming conventions and thus leverage on the common structural constraints of paths.

\begin{figure}[!t]
	\vspace{-15pt}
\IncMargin{1.2em}
\begin{algorithm}[H]
\scriptsize
	
	\KwIn{Update $u_i = (e_i): e_i = (s,t) $}
	\KwOut{Locate matched queries}

	\tikzmark{left}\tikzmark{top1}$affectedTries \leftarrow \edgeInd[e_i]$\tcp*[l]{Get affected tries}

	\ForEach{$T_i \in affectedTries$}{
		\ForEach(\tcp*[h]{Traverse $ T_i $ in DFS\hspace*{-5mm}}){node $n_i \in T_i$}{
			\If(\tcp*[h]{If current node indexes $e_i$\hspace*{-20mm}}){$\edge(n_c) = e_i$}{
				$fndPos \leftarrow n$ \tcp*[l]{Store the position}
				break \tcp*{Terminate the traversal}
			}
		}

		\tcp{Update $ \mv $s of $ fndPos $ and its children}
		 $ affectedQueries \leftarrow $ Trie Traversal \& Materialization ($ fndPos $)\hspace*{-5mm}\;
	}

	\ForEach{query $Q_i \in affectedQueries$}{
		$results \leftarrow \O$\;
		
		\ForEach(\tcp*[h]{For the covering paths of $Q_i$ \hspace*{-5mm}}){$P_i \in Q_i$}{
			$results \leftarrow results \Join  \mv[P_i]$\;
		}
		
		\If{$results \neq \O$}{
			\tikzmark{bottom1}\AddNote{top1}{bottom1}{left}{$ \mathbcal{Step\ 1} $}$\markM(Q_i)$\;
		}		
	}

\end{algorithm}
\DecMargin{1.2em} 	\vspace{-15pt}

	\caption{Query answering phase ($ \mathbcal{Step\ 1} $) of Algorithm~\Tree.}
	\label{pseudo:tree:filtering}
\end{figure}

However, by substituting the variable vertices with the generic ``?var'' requires us to keep information about the joining order of each edge $ e_i \in P_i $, as well as, how each $ P_i \in CP(Q_i) $ intersects with the rest of the paths in $ CP(Q_i) $. 
In order to calculate the subgraphs that satisfy each covering path $ P_i \in CP(Q_i) $,  each $ \mv[e_i]: e_i \in P_i $ must be joined. Each path $ P_i $ that is indexed under a trie path $ \{n_1 \rightarrow \ldots n_i \rightarrow \ldots n_k\} $  maintains the original ordering of its edges and vertices, while the order under which each edge of a node $ n_i $ is connected with its children nodes ($ \children(n_i) $), is determined as follows: the target vertex $ t \in e_i $ (where $ e_i $ is indexed under $ n_i $) is connected with the source node $ s \in e_{i+1}: e_{i+1} \in \children(n_i)$ of the parent node $ n_i $. 
Finally, for each covering path $ P_i \in CP(Q_i)$ \Tree\ maintains information about the vertices that intersected in the original query graph pattern $ Q_i $; this information is utilized during the query answering phase.

\subsection{Query Answering Phase}
\label{sec:tree:answering}

During the evolution of the graph, a constant stream of updates $ S = (u_1, u_2, \dots, u_k) $ arrives at the system. For each update $u_i \in S$ Algorithm~\Tree\ performs the following steps:
\begin{mylisting}
	\item 
	Determines which tries are affected by update $ u_i $ and proceeds in examining them.	
	\item
	While traversing the affected tries, performs the materialization and prunes sub-tries that are not affected by $ u_i $.
\end{mylisting}

In the following, we describe each step of the query answering phase of Algorithm~\Tree. The pseudocode for each step is provided in Figs.~\ref{pseudo:tree:filtering} and \ref{pseudo:tree:filtering:traversal}.

\ctitle{\underline{$ \mathbcal{Step\ 1:} $} Locate and Traverse Affected Tries.}
When an update $ u_i $ arrives at the system, Algorithm~\Tree\ utilizes the edge $ e_i \in u_i$ to locate the tries that are affected by $ u_i $. To achieve this, \Tree\ uses the hash table $ \edgeInd $ to obtain the list of tries that contain $ e_i $ in their children set. Thus, Algorithm~\Tree\ receives a list ($ affectedTries $) that contains all the tries that were affected by $ u_i $ and must be examined (Fig.~\ref{pseudo:tree:filtering}, line 1). Subsequently, Algorithm~\Tree\ proceeds into examining each trie $ T_i \in affectedTries $ by traversing each $T_i $ in order to locate the node $ n_i $ that indexes edge $ e_i \in u_i $. When node $ n_i $ is located, the algorithm proceeds in $ \mathcal{Step 2} $ of the query answering process described below (Fig.~\ref{pseudo:tree:filtering}, lines $ 3 - 7 $). 

\begin{figure}[!t]
	\centering
	\includegraphics[scale=.5]{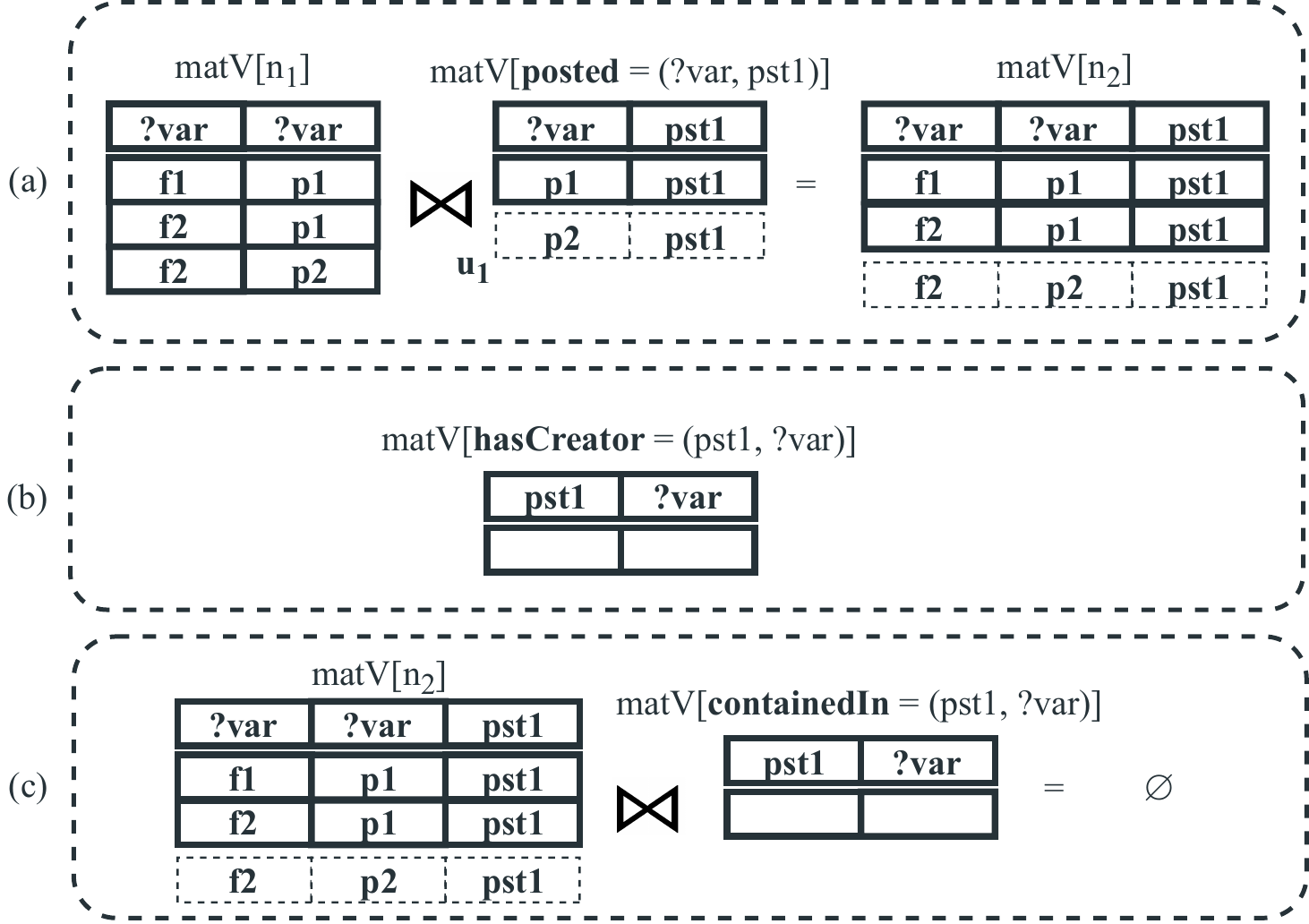}	
	
	\caption{\textcolor{black}{Updating materialized views.}}
	\label{fig:joining}
\end{figure}%

\color{black}
\begin{example}
	Let us consider the data structures presented in Fig.~\ref{fig:forest}, the materialized views in Fig.~\ref{fig:joining}, and an update $ u_1 = ( posted = (p2, pst1)) $ that arrives into the evolving graph (Fig.~\ref{fig:joining}(a)). Algorithm~\Tree\ prompts hash table $ \edgeInd $ and obtains list  $ \{\&T_1, \&T_3\} $. Subsequently, \Tree\ will traverse tries $ T_1 $ and $ T_3 $. When traversing trie $ T_1 $ \Tree\ locates node $ n_2 $ that matches update $ e_1 \in u_1 $ and proceeds in $ \mathcal{Step 2} $ (described below). Finally, when traversing $ T_3 $ \Tree\ will stop the traversal at root node $ n_6 $ as its materialized view is empty $ \mv[hasCreator = (pst1, ?var)] = \O$ (Fig.~\ref{fig:joining} (b)), thus all sub-tries will yield empty materialized views.
	\label{ex:trie:ans:upd}	
\end{example}
\color{black}

\ctitle{\underline{$ \mathbcal{Step\ 2:} $} Trie Traversal and Materialization.}
Intuitively, a trie path $ \{n_0 \rightarrow \ldots n_i \rightarrow \ldots n_k\} $ represents a series of joined materialized views $ \mv s = \{\mv_{1}$, $ \mv_{2} $ , $ \dots $, $ \mv_{k} \}$. Each materialized view $ \mv_i \in \mv s $ corresponds to a node $ n_i $ 
that stores edge $ e_i $ and the materialized view $ \mv_i $. The materialized view contains the results of the join operation between the $ \mv[e_i] $ and the materialized view of the parent node $ n_i $ ($ \mv (\parent(n_i)] $), i.e., $ \mv_i = \mv [\parent(n_i)] \Join \mv [e_i] $. 
Thus, when an update $ u_i $ affects a node $ n_i $ in this ``chain'' of joins, $ n_i $'s and its children's ($ \children(n_i) $) materialized views must be updated with $ u_i $. Based on this \Tree\ searches for and locates node $ n_i $ inside $ T_i $ that is affected by $ u_i $ and updates $ n_i $'s sub-trie.

After locating node $ n_i \in T_i $ that is affected by $ u_i $, Algorithm~\Tree\ continues the traversal of $ n_i $'s sub-trie and prunes the remaining sub-tries of $ T_i $ (Fig.~\ref{pseudo:tree:filtering}, line $ 7 $). Subsequently, \Tree\ updates the materialized view of $ n_i $ by performing a join operation between its parent's node materialized view $ \mv[\parent(n_i)] $  and the update $ u_i $, i.e., $ results = \mv[\parent(n_i)] \Join u_i $. Notice, that Algorithm~\Tree\ calculates the subgraphs formulated by the current update solely based on the update $ u_1 $ and does not perform a full join operation between $ \mv[\parent(n_i)] $ and $ \mv[\edge(n_i)] $, the updated results are then stored in the corresponding $ \mv[n_i] $.

\begin{figure}[!t]
	\vspace{-15pt}
\IncMargin{1.2em}
\begin{algorithm}[H]
\scriptsize
	
	\Function{Trie Traversal \& Materialization}
	\KwIn{Node $n_i$}
	\KwOut{Locate matched queries}
	\tcp{Update the current materialized view by joining the parent materialized view with the  materialized view of the edge in node $n_i$\hspace*{-10mm}}
	\tikzmark{left}\tikzmark{top1}$result \leftarrow \mv[\parent(n_i)] \Join \mv[\edge(n_i)]$\;

	\If{$result = \O$}{
		return\;
	}

	\tcp{Store the query identifiers of node $ n_i $}
	$ affectedQueries \leftarrow affectedQueries \cup qIDs(n_i)$\;

	\tcp{Recursively update the $ \mv $s of $ n_i $'s children}
	\ForEach{$n_{c} \in \children(n_i)$}{
		 Trie Traversal \& Materialization ($n_{c}$)\;
	}

	\tikzmark{bottom1}\AddNote{top1}{bottom1}{left}{$ \mathbcal{Step\ 2} $}return $ affectedQueries$ \tcp*[l]{Return the affected $ qIDs $}

\end{algorithm}
\DecMargin{1.2em} 	\vspace{-15pt}

	\caption{Query answering phase ($ \mathbcal{Step\ 2} $) of Algorithm~\Tree.}
	\label{pseudo:tree:filtering:traversal}
\end{figure}

For each child node $ n_j \in \children(n_i) $, \Tree\ updates its corresponding materialized view by joining its view $ \mv[n_j] $ that corresponds to the edge that it stores (given by $ \mv[\edge(n_j)] $) with its parent node materialized view $ \mv[n_i] $ (Fig.~\ref{pseudo:tree:filtering:traversal}, lines $ 1 - 7 $). 
If  at any point the process of joining the materialized views returns an empty result set the specific sub-trie is pruned, while, the traversal continues in a different sub-trie of $ T_i $ (Fig.~\ref{pseudo:tree:filtering:traversal}, lines $ 5 - 6 $). Subsequently, for each trie node $ n_j $ in the trie traversal when there is a successful join operation among $ \mv[e_j] : e_j \in n_j $ and $ \mv[n_i] $, the query identifiers indexed under $ n_j $ are stored in $ affectedQueries $ list  (Fig.~\ref{pseudo:tree:filtering:traversal}, lines $ 4 $ and $ 7 $). Note that similarly to before, only the updated part of a materialized view is utilized as the parent's materialized view, an approach applied on database-management system \cite{Gupta:1993}.

\begin{example}
	Let us consider the data structures presented in Fig.~\ref{fig:forest}, Fig.~\ref{fig:joining}, and an update $ u_1 = ( posted = (p2, pst1)) $ that arrives into the evolving graph. After locating the affected trie node $ n_2 $ (described in Example~\ref{ex:trie:ans:upd}) \Tree\ proceeds in updating the materialized view of $ n_2 $, i.e., $ \mv[n_2] $, by calculating the join operation between its parents materialized view, i.e., $ \mv[n_1] $ and the update $ u_1 $. Fig.~\ref{fig:joining}, demonstrates the operations of joining $ \mv[n_2] $ with update $ u_1 $, the result of the operation is tuple $ (f2, p2, pst1) $, which is added into $ \mv[n2] $, presented in Fig.~\ref{fig:joining}(a). While the query identifiers of $ n_2 $ (i.e., $ Q_1 $) are indexed in $ affectedQueries $. Finally, \Tree\ proceeds in updating the sub-trie of $ n_2 $, node $ n_3 $, where the updated tuple $ (f2, p2, pst1) $ is joined with $ \mv[\edge(n_3)] $ (i.e., $ \mv[containedIn = (pst1, ?var)] $). This operation yields an empty result (Fig.~\ref{fig:joining}(c)), thus terminating the traversal.
\end{example}

Finally, to complete the filtering phase Algorithm~\Tree\ iterates through the affected list of queries and performs the join operations among the paths that form a query, thus, yielding the final answer (Fig.~\ref{pseudo:tree:filtering}, lines $ 8 - 13 $).

The time complexity, of Algorithm \Tree\ when filtering an update $ u_i $, is calculated as follows: The traversal complexity is $\mathcal{O}(T * (P_{m} * B))$, where $ T $ denotes the number of tries that contain $ e_i \in u_i $, $ P_{m} $ denotes the size of the longest trie path, and $ B $ the branching factor. The time complexity of joining two materialized views $ \mv_1 $ and $ \mv_2 $, where $ |\mv_1| = N $ and $ |\mv_2| = M $, is $\mathcal{O}(N * M)$. Finally, the total time complexity is calculated as $ \mathcal{O}((T * (P_{m} * B)) * (N * M)) $.

\ctitle{Caching.} 
\textcolor{black}{
During $ \mathbcal{Step\ 2} $, two materialized views are joined using a typical hash join operation with a build and a probe phase. In the build phase, a hash table for the smallest (in the number of tuples) table is constructed, while in the probe phase the largest table is scanned and the hash table is probed to perform the join. Algorithm~\Tree\ discards all the data structures and intermediate results after the join operation commences. In order to enhance this resource intensive operation, we cache and reuse the data structures generated during the build and probe phases as well as the intermediate results whenever possible. This approach constitutes an extension of our proposed solution (Algorithm~\Tree) and it is coined \TreeCache.   
}

\section{Advanced Baselines}
\label{sec:competitors}
Since no prior work in the literature considers the problem of continuous multi-query evaluation, we designed and implemented Algorithms~\Inv\ and \InvIncr, two advanced baselines that utilize inverted index data structures. \textcolor{black}{Finally, we provide a third baseline that was based on the well-established graph database \Neo~\cite{neo4j}.}

\subsection{Algorithm INV}
\label{sec:algInv}

Algorithm~\Inv\ (\textsc{Inv}erted Index), utilizes inverted index data structures to index the query graph patterns. The inverted index data structure is able to capture and index common elements of the graph patterns at the edge level during indexing time. Subsequently, the inverted index is utilized during filtering time to determine which queries have been satisfied. In the following sections we describe the query indexing and answering phase of \Inv.

\ctitle{The Query Indexing Phase} of Algorithm \Inv, for each query graph pattern $ Q_i $, is performed in two steps:
	(1) Transforming the original query graph pattern $ Q_i $ into a set of path conjuncts, that cover all vertices and edges of $ Q_i $, and  when combined can effectively re-compose $ Q_i $, and finally, indexing those covering paths in a matrix along the unique query identifier,	
	(2) Indexing all edges $ e_i \in Q_i $ into an inverted index structure.
In the following, we present each step of the query indexing phase of \Inv\ and give details about the data structures utilized. 

\ctitle{\underline{$ \mathbcal{Step\ 1:} $} Extracting the Covering Paths.}
In the first step of the query indexing phase, Algorithm~\Inv\ decomposes a query graph $ Q_i $ into a set of paths $ CP $, a process described in detail in Section~\ref{sec:tree:indexing}.
 Thus, given the  query set presented in Fig.~\ref{fig:exampleQueries} (a), \Inv\ yields the same set of covering paths $ CP $ (Fig.~\ref{fig:exampleQueries} (b)). Finally, the covering path set $ CP $ is indexed into an array ($ \queryInd $) with the query identifier of $ Q_i $. 

\begin{figure}[t]
	\centering

	\includegraphics[width=\linewidth]{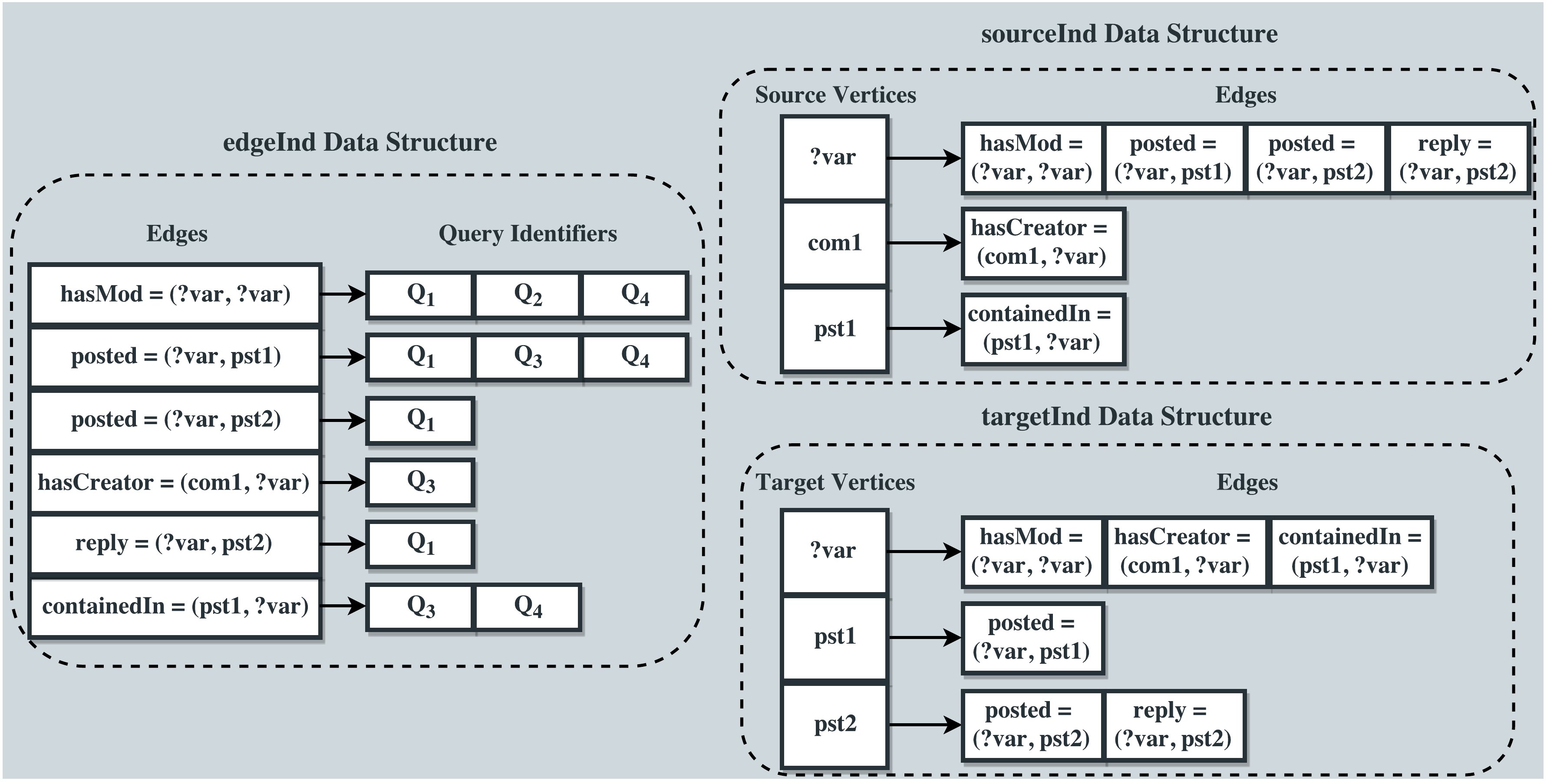}
	
	\caption{Data structures utilized by Algorithm~\Inv\ to index query graph patterns.}
	\label{fig:Inv:indexStructs}
\end{figure}

\ctitle{\underline{$ \mathbcal{Step\ 2:} $} Indexing the Query Graph.}
Algorithm \Inv\ builds three inverted indexes, 
where it stores the structural and attribute constrains of the query graph pattern $ Q_i $. 
Hash table $ \edgeInd $ indexes all edges $ e_i \in Q_{DB} $ (keys),  and the respective query identifiers as values,  
hash table $ \sourceInd $ indexes the source vertices of each edge (key), where the edges are indexed as values 
, and hash table $ \targetInd $ that indexes the target vertices of each edge (key), where the edges are indexed as values.  
In Fig.~\ref{fig:exampleQueries}(a) we present four query graph patterns, and in Fig.~\ref{fig:Inv:indexStructs} the data structures of \Inv\ when indexing those queries. Finally, \Inv\ applies the same techniques of handling variables as Algorithm~\Tree\ (Section~\ref{sec:tree:indexing}).

\ctitle{The Query Answering Phase} of Algorithm~\Inv, when a constant stream of updates $ S = (u_1, u_2, \dots, u_k) $ arrives at the system, is performed in three steps: 
	(1) Determines which queries are affected by update $ u_i $,
	(2) Prompts the inverted index data structure and determines which paths have been affected by update $ u_i $, 
	(3) Performs the materialization while querying the inverted index data structures.
In the following, we describe each step of the query answering phase: 

\ctitle{\underline{$ \mathbcal{Step\ 1:} $} Locate the Affected Queries.}
When a new update $ u_i $ arrives at the system, Algorithm~\Inv\ utilizes the edge $ e_i \in u_i $ to locate the queries that are affected, by querying the hash table $ \edgeInd $ to obtain the query identifier $ qIDs $ that contain $ e_i $. 
Subsequently, the algorithm iterates through the list of $ affectedQueries $ and checks each query $ Q_i \in qIDs $. For each query $ Q_i $ the algorithm checks $ \forall e_i \in Q_i  $ if  $ \mv [e_i] $ $\neq \O  $, i.e., each $ e_i $ should have a \emph{non empty} materialized view. The check is performed by iterating through the edge list that is provided by $ \queryInd $ and a hash table that keeps all materialized views present in the system. Intuitively, a query $ Q_i $ is candidate to match, as long as, all materialized views that correspond to its edges can be used in the query answering process.

\ctitle{\underline{$ \mathbcal{Step\ 2:} $} Locate the Affected Paths.}
Algorithm~\Inv\ proceeds to examine the inverted index structures $ \sourceInd $ and $ \targetInd $ by making use of $ e_i \in u_i $. \Inv\ queries $ \sourceInd $ and $ \targetInd $ to determine which edges are affected by the update, by utilizing the source and target vertices of update $ u_i $. 
\Inv\ examines each current edge $ e_c $ of the affected edge set and recursively visits all edges connected to $ e_c $, which are determined by querying the $ \sourceInd $ and $ \targetInd $. 
While examining the current edge $ e_c $, \Inv\ checks if $ e_c $ is part of $ affectedQueries $, if not, the examination of the specific path is pruned. 
For efficiency reasons, the examination is bound by the maximum length of a path present in $ affectedQueries $ which is calculated by utilizing the $ \queryInd $ data structure.

\ctitle{\underline{$ \mathbcal{Step\ 3:} $} Path Examination and Materialization.} 
While \Inv\ examines the paths affected by update $ u_i $ ($ \mathcal{Step} $ 2), it performs the materialization on the currently examined path. 
More specifically, while \Inv\ searches through the paths formulated by the visits of edge sets determined by $ \targetInd $ and $ \sourceInd $, it maintains a path $ P_c = \{ v_1 \xrightarrow{e_1} v_2 \xrightarrow{e_2} \dots v_k \xrightarrow{e_k} v_{k+1}\} $ that corresponds to the edges already visited. 

While, visiting each edge $ e_c $, \Inv\ accesses the materialized view that corresponds to it (i.e., $ \mv[e_c] $) and updates the set of materialized views $ \mv s = \{\mv_1, \mv_2, \dots, \mv_k\}  $ that correspond to the current path. For example, given an already visited path $ P =  \{v_1 \xrightarrow{e_1} v_2 \xrightarrow{e_2} v_3\}$ its materialized view $ \mv[P] $ will be generated, by $\mv[P] = \mv[e_1] \Join \mv[e_2] $. When visiting the next edge $ e_n $, a new path $ P' $ is generated and its materialized view $ \mv[P'] = \mv[P] \Join \mv[e_n]  $ will be generated.
If at any point, the process of joining the materialized views yields an empty result set 
 the examination of the edge is terminated (pruning).  
 This allows us to prune paths that are not going to satisfy any $ Q_i \in affectedQueries $. 
 If a path $  P_i $ yields a successful series of join operations (i.e., $ \mv [P_i]  \neq \O $), it is marked as matched. 

Finally, to produce the final answer subgraphs Algorithm~\Inv\ iterates through the affected list of queries $ qIDs \in affectedQueries $ and performs the final join operation among all the paths that comprise the query.

\ctitle{Caching.}
In the spirit of Algorithm~\TreeCache\ (Section~\ref{sec:tree:answering}), we developed an extension of Algorithm~\Inv, namely Algorithm~\InvCache, that caches and reuses the calculated data structures of the hash join phase.

\subsection{Algorithm INC}
Based on Algorithm~\Inv\ we developed an algorithmic extension, namely Algorithm~\InvIncr. Algorithm~\InvIncr\ utilizes the same inverted index data structures to index the covering paths, edges, source and target vertices as Algorithm~\Inv, while the examination of a path affected during query answering remains similar. The key difference lies in executing the joining operations between the materialized views that correspond to edges belonging to a path. More specifically, when Algorithm~\Inv\ executes a series of joins between the materialized views (that formulate a path) to determine which subgraphs match a path; it utilizes all tuples of each materialized view that participate in the joining process. On the other hand, Algorithm~\InvIncr\ makes use of only the update $ u_i $ and thus reduces the number of tuples examined through out the joining process of the paths.

\ctitle{Caching.}
In the spirit of Algorithm~\TreeCache\ (Section~\ref{sec:tree:answering}), we developed an extension of Algorithm~\InvIncr, namely Algorithm~\InvIncrCache, that caches and reuses the calculated data structures of the hash join phase. 

\color{black}
\subsection{\Neo\ }
To evaluate the efficiency of the proposed algorithm against a real-world approach, we implemented a solution based on the well-established graph database \Neo~\cite{neo4j}. In this approach, we extend \Neo's native functionality with auxiliary data structures to efficiently store the query set. They are used during the answering phase to located affected queries and execute them on \Neo. 

\ctitle{The Query Indexing Phase.} To address the continuous multi-query evaluation scenario, we designed main-memory data structures to facilitate indexing of query graph patterns.  
To this end, 
each query graph pattern $Q_i $ is indexed in three steps: 
(1) $Q_i $ is converted into \Neo's native query language Cypher\footnote{\url{https://neo4j.com/developer/cypher/}}, 
(2) indexing each query in a matrix ($ \queryInd $), and 
(3) indexing all edges $e_i \in Q_i $ by an inverted index structure ($\edgeInd $), where $e_i $ is used as key and a collection of query identifiers as values (similarly to Algorithms~\Inv/\InvIncr\  Fig.~\ref{fig:Inv:indexStructs}).

\ctitle{The Query Answering Phase.} Each update that is received as part of an incoming stream of updates  $S = (u_1, u_2, \ldots, u_k)$ is processed as follows: 
(1) an incoming update $ u_i $ is applied to \Neo\,%
(2) by querying the inverted index $ \edgeInd $ with $ e_i \in u_i $, it is determined which queries are affected, 
(3) all affected queries are retrieved from matrix $ \queryInd $ and the appropriate parameters are set, 
(4) the affected queries are executed. 

To enhance performance, the following configurations are applied: 
(1) the graph database builds indexes on all labels of the schema allowing for faster look up times of nodes,
(2) the execution of Cypher queries employs the \emph{parameters syntax}\footnote{\url{https://neo4j.com/docs/cypher-manual/current/syntax/parameters/}} as it enables the execution planner of \Neo\ to cache the query plans for future use, 
(3) the number of writes per transaction\footnote{\url{https://neo4j.com/docs/cypher-manual/current/introduction/transactions/}} in the database and the allocated memory were optimized based on the hardware configuration (see Section~\ref{sec:expSetup}).

\color{black}

\begin{figure*}[!ht]
	\begin{subfigure}[t]{0.32\textwidth}
		\centering
		\resizebox{\linewidth}{0.18\textheight}{
\begingroup
  \makeatletter
  \providecommand\color[2][]{%
    \GenericError{(gnuplot) \space\space\space\@spaces}{%
      Package color not loaded in conjunction with
      terminal option `colourtext'%
    }{See the gnuplot documentation for explanation.%
    }{Either use 'blacktext' in gnuplot or load the package
      color.sty in LaTeX.}%
    \renewcommand\color[2][]{}%
  }%
  \providecommand\includegraphics[2][]{%
    \GenericError{(gnuplot) \space\space\space\@spaces}{%
      Package graphicx or graphics not loaded%
    }{See the gnuplot documentation for explanation.%
    }{The gnuplot epslatex terminal needs graphicx.sty or graphics.sty.}%
    \renewcommand\includegraphics[2][]{}%
  }%
  \providecommand\rotatebox[2]{#2}%
  \@ifundefined{ifGPcolor}{%
    \newif\ifGPcolor
    \GPcolortrue
  }{}%
  \@ifundefined{ifGPblacktext}{%
    \newif\ifGPblacktext
    \GPblacktexttrue
  }{}%
  \let\gplgaddtomacro\g@addto@macro
  \gdef\gplbacktext{}%
  \gdef\gplfronttext{}%
  \makeatother
  \ifGPblacktext
    \def\colorrgb#1{}%
    \def\colorgray#1{}%
  \else
    \ifGPcolor
      \def\colorrgb#1{\color[rgb]{#1}}%
      \def\colorgray#1{\color[gray]{#1}}%
      \expandafter\def\csname LTw\endcsname{\color{white}}%
      \expandafter\def\csname LTb\endcsname{\color{black}}%
      \expandafter\def\csname LTa\endcsname{\color{black}}%
      \expandafter\def\csname LT0\endcsname{\color[rgb]{1,0,0}}%
      \expandafter\def\csname LT1\endcsname{\color[rgb]{0,1,0}}%
      \expandafter\def\csname LT2\endcsname{\color[rgb]{0,0,1}}%
      \expandafter\def\csname LT3\endcsname{\color[rgb]{1,0,1}}%
      \expandafter\def\csname LT4\endcsname{\color[rgb]{0,1,1}}%
      \expandafter\def\csname LT5\endcsname{\color[rgb]{1,1,0}}%
      \expandafter\def\csname LT6\endcsname{\color[rgb]{0,0,0}}%
      \expandafter\def\csname LT7\endcsname{\color[rgb]{1,0.3,0}}%
      \expandafter\def\csname LT8\endcsname{\color[rgb]{0.5,0.5,0.5}}%
    \else
      \def\colorrgb#1{\color{black}}%
      \def\colorgray#1{\color[gray]{#1}}%
      \expandafter\def\csname LTw\endcsname{\color{white}}%
      \expandafter\def\csname LTb\endcsname{\color{black}}%
      \expandafter\def\csname LTa\endcsname{\color{black}}%
      \expandafter\def\csname LT0\endcsname{\color{black}}%
      \expandafter\def\csname LT1\endcsname{\color{black}}%
      \expandafter\def\csname LT2\endcsname{\color{black}}%
      \expandafter\def\csname LT3\endcsname{\color{black}}%
      \expandafter\def\csname LT4\endcsname{\color{black}}%
      \expandafter\def\csname LT5\endcsname{\color{black}}%
      \expandafter\def\csname LT6\endcsname{\color{black}}%
      \expandafter\def\csname LT7\endcsname{\color{black}}%
      \expandafter\def\csname LT8\endcsname{\color{black}}%
    \fi
  \fi
    \setlength{\unitlength}{0.0500bp}%
    \ifx\gptboxheight\undefined%
      \newlength{\gptboxheight}%
      \newlength{\gptboxwidth}%
      \newsavebox{\gptboxtext}%
    \fi%
    \setlength{\fboxrule}{0.5pt}%
    \setlength{\fboxsep}{1pt}%
\begin{picture}(7200.00,5040.00)%
    \gplgaddtomacro\gplbacktext{%
      \csname LTb\endcsname%
      \put(732,756){\makebox(0,0)[r]{\strut{}0.0}}%
      \csname LTb\endcsname%
      \put(732,1134){\makebox(0,0)[r]{\strut{}0.5}}%
      \csname LTb\endcsname%
      \put(732,1512){\makebox(0,0)[r]{\strut{}1.0}}%
      \csname LTb\endcsname%
      \put(732,1889){\makebox(0,0)[r]{\strut{}1.5}}%
      \csname LTb\endcsname%
      \put(732,2267){\makebox(0,0)[r]{\strut{}2.0}}%
      \csname LTb\endcsname%
      \put(864,536){\makebox(0,0){\strut{}$10$}}%
      \csname LTb\endcsname%
      \put(1528,536){\makebox(0,0){\strut{}$20$}}%
      \csname LTb\endcsname%
      \put(2192,536){\makebox(0,0){\strut{}$30$}}%
      \csname LTb\endcsname%
      \put(2856,536){\makebox(0,0){\strut{}$40$}}%
      \csname LTb\endcsname%
      \put(3520,536){\makebox(0,0){\strut{}$50$}}%
      \csname LTb\endcsname%
      \put(4183,536){\makebox(0,0){\strut{}$60$}}%
      \csname LTb\endcsname%
      \put(4847,536){\makebox(0,0){\strut{}$70$}}%
      \csname LTb\endcsname%
      \put(5511,536){\makebox(0,0){\strut{}$80$}}%
      \csname LTb\endcsname%
      \put(6175,536){\makebox(0,0){\strut{}$90$}}%
      \csname LTb\endcsname%
      \put(6839,536){\makebox(0,0){\strut{}$100$}}%
    }%
    \gplgaddtomacro\gplfronttext{%
      \csname LTb\endcsname%
      \put(116,2831){\rotatebox{-270}{\makebox(0,0){\strut{}Answering time (msec/update)}}}%
      \put(3851,206){\makebox(0,0){\strut{}Graph size (Edges x$1000$)}}%
      \csname LTb\endcsname%
      \put(996,2072){\makebox(0,0)[l]{\strut{}\Tree}}%
      \csname LTb\endcsname%
      \put(996,1808){\makebox(0,0)[l]{\strut{}\TreeCache}}%
    }%
    \gplgaddtomacro\gplbacktext{%
      \csname LTb\endcsname%
      \put(732,2684){\makebox(0,0)[r]{\strut{}20}}%
      \csname LTb\endcsname%
      \put(732,2889){\makebox(0,0)[r]{\strut{}40}}%
      \csname LTb\endcsname%
      \put(732,3095){\makebox(0,0)[r]{\strut{}60}}%
      \csname LTb\endcsname%
      \put(732,3301){\makebox(0,0)[r]{\strut{}80}}%
      \csname LTb\endcsname%
      \put(732,3506){\makebox(0,0)[r]{\strut{}100}}%
      \csname LTb\endcsname%
      \put(732,3712){\makebox(0,0)[r]{\strut{}120}}%
      \csname LTb\endcsname%
      \put(732,3918){\makebox(0,0)[r]{\strut{}140}}%
      \csname LTb\endcsname%
      \put(732,4124){\makebox(0,0)[r]{\strut{}160}}%
      \csname LTb\endcsname%
      \put(732,4329){\makebox(0,0)[r]{\strut{}180}}%
      \csname LTb\endcsname%
      \put(732,4535){\makebox(0,0)[r]{\strut{}200}}%
      \csname LTb\endcsname%
      \put(864,4755){\makebox(0,0){\strut{}10}}%
      \csname LTb\endcsname%
      \put(1528,4755){\makebox(0,0){\strut{}16}}%
      \csname LTb\endcsname%
      \put(2192,4755){\makebox(0,0){\strut{}22}}%
      \csname LTb\endcsname%
      \put(2856,4755){\makebox(0,0){\strut{}28}}%
      \csname LTb\endcsname%
      \put(3520,4755){\makebox(0,0){\strut{}33}}%
      \csname LTb\endcsname%
      \put(4183,4755){\makebox(0,0){\strut{}38}}%
      \csname LTb\endcsname%
      \put(4847,4755){\makebox(0,0){\strut{}43}}%
      \csname LTb\endcsname%
      \put(5511,4755){\makebox(0,0){\strut{}48}}%
      \csname LTb\endcsname%
      \put(6175,4755){\makebox(0,0){\strut{}52}}%
      \csname LTb\endcsname%
      \put(6839,4755){\makebox(0,0){\strut{}57}}%
    }%
    \gplgaddtomacro\gplfronttext{%
      \csname LTb\endcsname%
      \put(3851,5085){\makebox(0,0){\strut{}Graph size (Vertices x$1000$)}}%
      \csname LTb\endcsname%
      \put(996,4340){\makebox(0,0)[l]{\strut{}\Inv}}%
      \csname LTb\endcsname%
      \put(996,4076){\makebox(0,0)[l]{\strut{}\InvCache}}%
      \csname LTb\endcsname%
      \put(996,3812){\makebox(0,0)[l]{\strut{}\InvIncr}}%
      \csname LTb\endcsname%
      \put(996,3548){\makebox(0,0)[l]{\strut{}\InvIncrCache}}%
      \csname LTb\endcsname%
      \put(996,3284){\makebox(0,0)[l]{\strut{}\Neo}}%
    }%
    \gplbacktext
    \put(0,0){\includegraphics{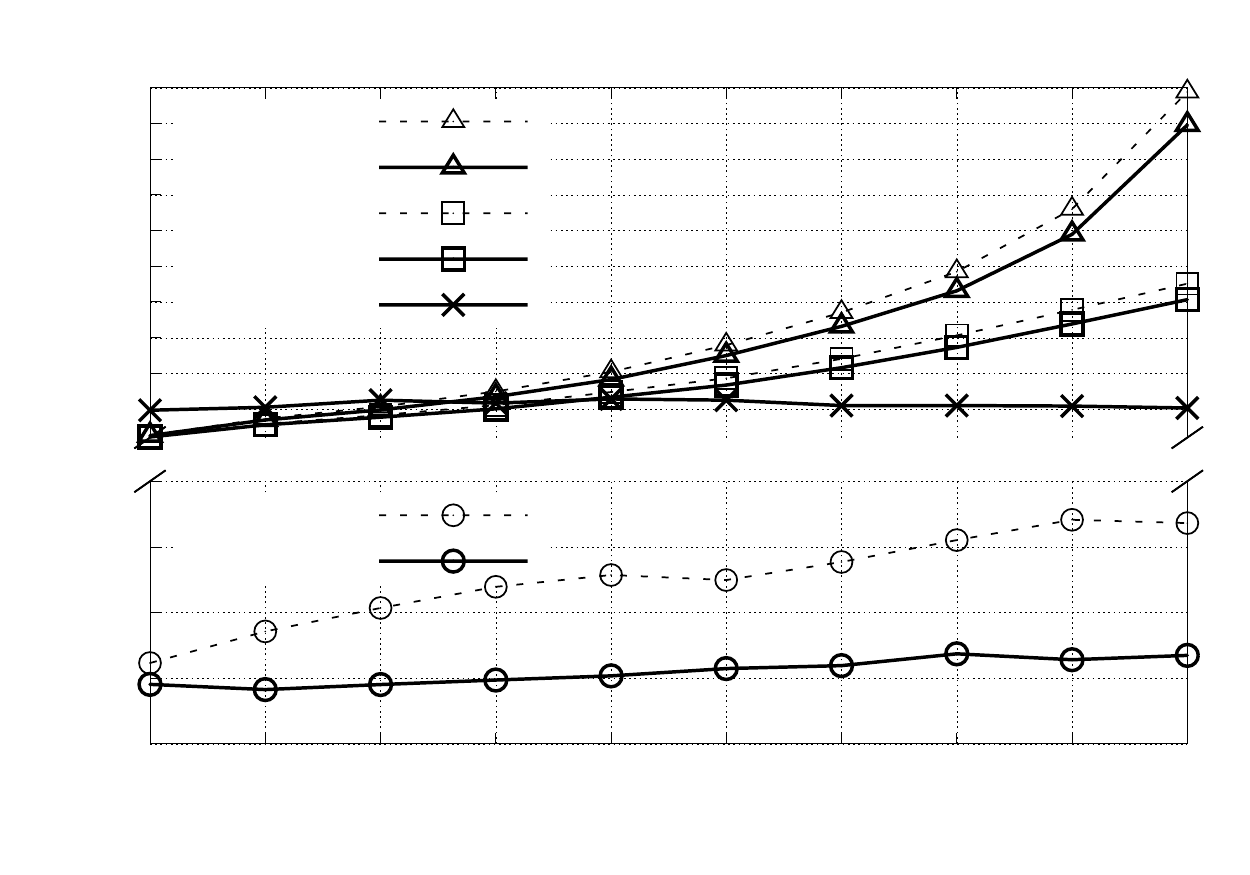}}%
    \gplfronttext
  \end{picture}%
\endgroup
 		}
		\caption{Query answering time for $ |Q_{DB}| = 5K $, $ \len = 5 $, $ \select = 25\% $, $ \overlap = 35\% $ and $ |G_E| = 10K  $ to $ |G_E| = 100K $.}
		\label{fig:exp:social-p100-sub5-S25:filteringTime}
	\end{subfigure}
	\hfill
	\begin{subfigure}[t]{0.32\textwidth}
		\centering
		\resizebox{\linewidth}{0.18\textheight}{
\begingroup
  \makeatletter
  \providecommand\color[2][]{%
    \GenericError{(gnuplot) \space\space\space\@spaces}{%
      Package color not loaded in conjunction with
      terminal option `colourtext'%
    }{See the gnuplot documentation for explanation.%
    }{Either use 'blacktext' in gnuplot or load the package
      color.sty in LaTeX.}%
    \renewcommand\color[2][]{}%
  }%
  \providecommand\includegraphics[2][]{%
    \GenericError{(gnuplot) \space\space\space\@spaces}{%
      Package graphicx or graphics not loaded%
    }{See the gnuplot documentation for explanation.%
    }{The gnuplot epslatex terminal needs graphicx.sty or graphics.sty.}%
    \renewcommand\includegraphics[2][]{}%
  }%
  \providecommand\rotatebox[2]{#2}%
  \@ifundefined{ifGPcolor}{%
    \newif\ifGPcolor
    \GPcolortrue
  }{}%
  \@ifundefined{ifGPblacktext}{%
    \newif\ifGPblacktext
    \GPblacktexttrue
  }{}%
  \let\gplgaddtomacro\g@addto@macro
  \gdef\gplbacktext{}%
  \gdef\gplfronttext{}%
  \makeatother
  \ifGPblacktext
    \def\colorrgb#1{}%
    \def\colorgray#1{}%
  \else
    \ifGPcolor
      \def\colorrgb#1{\color[rgb]{#1}}%
      \def\colorgray#1{\color[gray]{#1}}%
      \expandafter\def\csname LTw\endcsname{\color{white}}%
      \expandafter\def\csname LTb\endcsname{\color{black}}%
      \expandafter\def\csname LTa\endcsname{\color{black}}%
      \expandafter\def\csname LT0\endcsname{\color[rgb]{1,0,0}}%
      \expandafter\def\csname LT1\endcsname{\color[rgb]{0,1,0}}%
      \expandafter\def\csname LT2\endcsname{\color[rgb]{0,0,1}}%
      \expandafter\def\csname LT3\endcsname{\color[rgb]{1,0,1}}%
      \expandafter\def\csname LT4\endcsname{\color[rgb]{0,1,1}}%
      \expandafter\def\csname LT5\endcsname{\color[rgb]{1,1,0}}%
      \expandafter\def\csname LT6\endcsname{\color[rgb]{0,0,0}}%
      \expandafter\def\csname LT7\endcsname{\color[rgb]{1,0.3,0}}%
      \expandafter\def\csname LT8\endcsname{\color[rgb]{0.5,0.5,0.5}}%
    \else
      \def\colorrgb#1{\color{black}}%
      \def\colorgray#1{\color[gray]{#1}}%
      \expandafter\def\csname LTw\endcsname{\color{white}}%
      \expandafter\def\csname LTb\endcsname{\color{black}}%
      \expandafter\def\csname LTa\endcsname{\color{black}}%
      \expandafter\def\csname LT0\endcsname{\color{black}}%
      \expandafter\def\csname LT1\endcsname{\color{black}}%
      \expandafter\def\csname LT2\endcsname{\color{black}}%
      \expandafter\def\csname LT3\endcsname{\color{black}}%
      \expandafter\def\csname LT4\endcsname{\color{black}}%
      \expandafter\def\csname LT5\endcsname{\color{black}}%
      \expandafter\def\csname LT6\endcsname{\color{black}}%
      \expandafter\def\csname LT7\endcsname{\color{black}}%
      \expandafter\def\csname LT8\endcsname{\color{black}}%
    \fi
  \fi
    \setlength{\unitlength}{0.0500bp}%
    \ifx\gptboxheight\undefined%
      \newlength{\gptboxheight}%
      \newlength{\gptboxwidth}%
      \newsavebox{\gptboxtext}%
    \fi%
    \setlength{\fboxrule}{0.5pt}%
    \setlength{\fboxsep}{1pt}%
\begin{picture}(7200.00,5040.00)%
    \gplgaddtomacro\gplbacktext{%
      \csname LTb\endcsname%
      \put(732,756){\makebox(0,0)[r]{\strut{}0.0}}%
      \csname LTb\endcsname%
      \put(732,972){\makebox(0,0)[r]{\strut{}0.5}}%
      \csname LTb\endcsname%
      \put(732,1188){\makebox(0,0)[r]{\strut{}1.0}}%
      \csname LTb\endcsname%
      \put(732,1404){\makebox(0,0)[r]{\strut{}1.5}}%
      \csname LTb\endcsname%
      \put(732,1619){\makebox(0,0)[r]{\strut{}2.0}}%
      \csname LTb\endcsname%
      \put(732,1835){\makebox(0,0)[r]{\strut{}2.5}}%
      \csname LTb\endcsname%
      \put(732,2051){\makebox(0,0)[r]{\strut{}3.0}}%
      \csname LTb\endcsname%
      \put(732,2267){\makebox(0,0)[r]{\strut{}3.5}}%
      \csname LTb\endcsname%
      \put(864,536){\makebox(0,0){\strut{}10\%}}%
      \csname LTb\endcsname%
      \put(2358,536){\makebox(0,0){\strut{}15\%}}%
      \csname LTb\endcsname%
      \put(3852,536){\makebox(0,0){\strut{}20\%}}%
      \csname LTb\endcsname%
      \put(5345,536){\makebox(0,0){\strut{}25\%}}%
      \csname LTb\endcsname%
      \put(6839,536){\makebox(0,0){\strut{}30\%}}%
    }%
    \gplgaddtomacro\gplfronttext{%
      \csname LTb\endcsname%
      \put(116,2831){\rotatebox{-270}{\makebox(0,0){\strut{}Answering time (msec/update)}}}%
      \put(3851,206){\makebox(0,0){\strut{}Varying $\select$}}%
      \csname LTb\endcsname%
      \put(996,2072){\makebox(0,0)[l]{\strut{}\Tree}}%
      \csname LTb\endcsname%
      \put(996,1808){\makebox(0,0)[l]{\strut{}\TreeCache}}%
    }%
    \gplgaddtomacro\gplbacktext{%
      \csname LTb\endcsname%
      \put(732,2531){\makebox(0,0)[r]{\strut{}20}}%
      \csname LTb\endcsname%
      \put(732,2713){\makebox(0,0)[r]{\strut{}50}}%
      \csname LTb\endcsname%
      \put(732,3017){\makebox(0,0)[r]{\strut{}100}}%
      \csname LTb\endcsname%
      \put(732,3321){\makebox(0,0)[r]{\strut{}150}}%
      \csname LTb\endcsname%
      \put(732,3624){\makebox(0,0)[r]{\strut{}200}}%
      \csname LTb\endcsname%
      \put(732,3928){\makebox(0,0)[r]{\strut{}250}}%
      \csname LTb\endcsname%
      \put(732,4231){\makebox(0,0)[r]{\strut{}300}}%
      \csname LTb\endcsname%
      \put(732,4535){\makebox(0,0)[r]{\strut{}350}}%
      \csname LTb\endcsname%
      \put(864,2299){\makebox(0,0){\strut{}}}%
      \csname LTb\endcsname%
      \put(2358,2299){\makebox(0,0){\strut{}}}%
      \csname LTb\endcsname%
      \put(3852,2299){\makebox(0,0){\strut{}}}%
      \csname LTb\endcsname%
      \put(5345,2299){\makebox(0,0){\strut{}}}%
      \csname LTb\endcsname%
      \put(6839,2299){\makebox(0,0){\strut{}}}%
    }%
    \gplgaddtomacro\gplfronttext{%
      \csname LTb\endcsname%
      \put(996,4340){\makebox(0,0)[l]{\strut{}\Inv}}%
      \csname LTb\endcsname%
      \put(996,4076){\makebox(0,0)[l]{\strut{}\InvCache}}%
      \csname LTb\endcsname%
      \put(996,3812){\makebox(0,0)[l]{\strut{}\InvIncr}}%
      \csname LTb\endcsname%
      \put(996,3548){\makebox(0,0)[l]{\strut{}\InvIncrCache}}%
      \csname LTb\endcsname%
      \put(996,3284){\makebox(0,0)[l]{\strut{}\Neo}}%
    }%
    \gplbacktext
    \put(0,0){\includegraphics{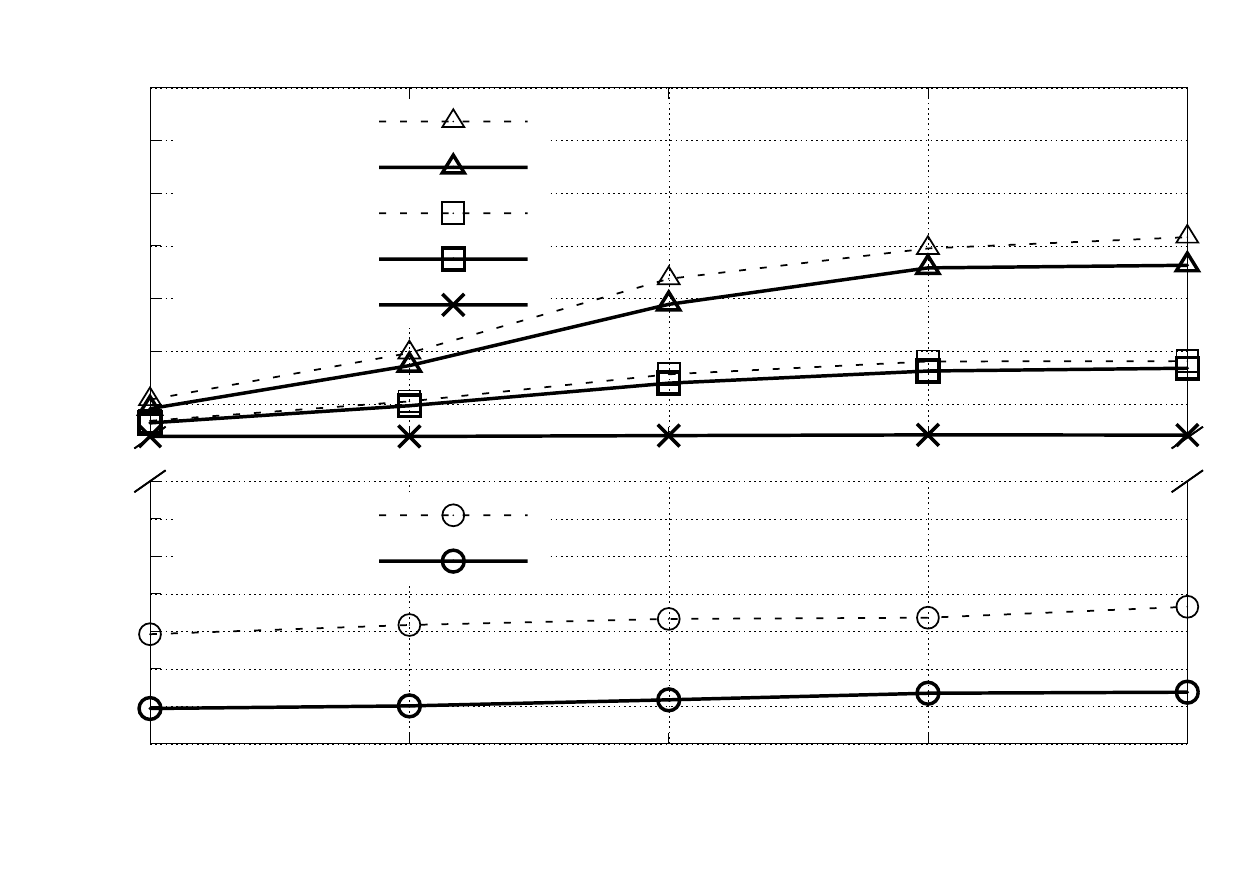}}%
    \gplfronttext
  \end{picture}%
\endgroup
 		}
		\caption{Query answering time for $ |Q_{DB}| = 5K $, $ \len = 5 $,  $ \overlap = 35\% $,  $ |G_E| = 100K$ and $ \select = 10\% $ to $ \select = 30\% $.}
		\label{fig:exp:social-p100-sub5-S10:VaryingSelec}
	\end{subfigure}
	\hfill
	\begin{subfigure}[t]{0.32\textwidth}
		\centering
		\resizebox{\linewidth}{0.18\textheight}{
			\small{
\begingroup
  \makeatletter
  \providecommand\color[2][]{%
    \GenericError{(gnuplot) \space\space\space\@spaces}{%
      Package color not loaded in conjunction with
      terminal option `colourtext'%
    }{See the gnuplot documentation for explanation.%
    }{Either use 'blacktext' in gnuplot or load the package
      color.sty in LaTeX.}%
    \renewcommand\color[2][]{}%
  }%
  \providecommand\includegraphics[2][]{%
    \GenericError{(gnuplot) \space\space\space\@spaces}{%
      Package graphicx or graphics not loaded%
    }{See the gnuplot documentation for explanation.%
    }{The gnuplot epslatex terminal needs graphicx.sty or graphics.sty.}%
    \renewcommand\includegraphics[2][]{}%
  }%
  \providecommand\rotatebox[2]{#2}%
  \@ifundefined{ifGPcolor}{%
    \newif\ifGPcolor
    \GPcolortrue
  }{}%
  \@ifundefined{ifGPblacktext}{%
    \newif\ifGPblacktext
    \GPblacktexttrue
  }{}%
  \let\gplgaddtomacro\g@addto@macro
  \gdef\gplbacktext{}%
  \gdef\gplfronttext{}%
  \makeatother
  \ifGPblacktext
    \def\colorrgb#1{}%
    \def\colorgray#1{}%
  \else
    \ifGPcolor
      \def\colorrgb#1{\color[rgb]{#1}}%
      \def\colorgray#1{\color[gray]{#1}}%
      \expandafter\def\csname LTw\endcsname{\color{white}}%
      \expandafter\def\csname LTb\endcsname{\color{black}}%
      \expandafter\def\csname LTa\endcsname{\color{black}}%
      \expandafter\def\csname LT0\endcsname{\color[rgb]{1,0,0}}%
      \expandafter\def\csname LT1\endcsname{\color[rgb]{0,1,0}}%
      \expandafter\def\csname LT2\endcsname{\color[rgb]{0,0,1}}%
      \expandafter\def\csname LT3\endcsname{\color[rgb]{1,0,1}}%
      \expandafter\def\csname LT4\endcsname{\color[rgb]{0,1,1}}%
      \expandafter\def\csname LT5\endcsname{\color[rgb]{1,1,0}}%
      \expandafter\def\csname LT6\endcsname{\color[rgb]{0,0,0}}%
      \expandafter\def\csname LT7\endcsname{\color[rgb]{1,0.3,0}}%
      \expandafter\def\csname LT8\endcsname{\color[rgb]{0.5,0.5,0.5}}%
    \else
      \def\colorrgb#1{\color{black}}%
      \def\colorgray#1{\color[gray]{#1}}%
      \expandafter\def\csname LTw\endcsname{\color{white}}%
      \expandafter\def\csname LTb\endcsname{\color{black}}%
      \expandafter\def\csname LTa\endcsname{\color{black}}%
      \expandafter\def\csname LT0\endcsname{\color{black}}%
      \expandafter\def\csname LT1\endcsname{\color{black}}%
      \expandafter\def\csname LT2\endcsname{\color{black}}%
      \expandafter\def\csname LT3\endcsname{\color{black}}%
      \expandafter\def\csname LT4\endcsname{\color{black}}%
      \expandafter\def\csname LT5\endcsname{\color{black}}%
      \expandafter\def\csname LT6\endcsname{\color{black}}%
      \expandafter\def\csname LT7\endcsname{\color{black}}%
      \expandafter\def\csname LT8\endcsname{\color{black}}%
    \fi
  \fi
    \setlength{\unitlength}{0.0500bp}%
    \ifx\gptboxheight\undefined%
      \newlength{\gptboxheight}%
      \newlength{\gptboxwidth}%
      \newsavebox{\gptboxtext}%
    \fi%
    \setlength{\fboxrule}{0.5pt}%
    \setlength{\fboxsep}{1pt}%
\begin{picture}(7200.00,5040.00)%
    \gplgaddtomacro\gplbacktext{%
      \csname LTb\endcsname%
      \put(1078,704){\makebox(0,0)[r]{\strut{}0}}%
      \csname LTb\endcsname%
      \put(1078,1733){\makebox(0,0)[r]{\strut{}1}}%
      \csname LTb\endcsname%
      \put(1078,2762){\makebox(0,0)[r]{\strut{}10}}%
      \csname LTb\endcsname%
      \put(1078,3790){\makebox(0,0)[r]{\strut{}100}}%
      \csname LTb\endcsname%
      \put(1078,4819){\makebox(0,0)[r]{\strut{}1000}}%
      \csname LTb\endcsname%
      \put(2032,484){\makebox(0,0){\strut{}1000}}%
      \csname LTb\endcsname%
      \put(3941,484){\makebox(0,0){\strut{}3000}}%
      \csname LTb\endcsname%
      \put(5849,484){\makebox(0,0){\strut{}5000}}%
    }%
    \gplgaddtomacro\gplfronttext{%
      \csname LTb\endcsname%
      \put(198,2761){\rotatebox{-270}{\makebox(0,0){\strut{}Answering time (msec/update)}}}%
      \put(3940,154){\makebox(0,0){\strut{}Varying $ |Q_{DB}| $}}%
      \csname LTb\endcsname%
      \put(1210,4646){\makebox(0,0)[l]{\strut{}\TreeCache}}%
      \csname LTb\endcsname%
      \put(1210,4426){\makebox(0,0)[l]{\strut{}\Tree}}%
      \csname LTb\endcsname%
      \put(1210,4206){\makebox(0,0)[l]{\strut{}\Neo}}%
      \csname LTb\endcsname%
      \put(1210,3986){\makebox(0,0)[l]{\strut{}\InvIncrCache}}%
      \csname LTb\endcsname%
      \put(1210,3766){\makebox(0,0)[l]{\strut{}\InvIncr}}%
      \csname LTb\endcsname%
      \put(1210,3546){\makebox(0,0)[l]{\strut{}\InvCache}}%
      \csname LTb\endcsname%
      \put(1210,3326){\makebox(0,0)[l]{\strut{}\Inv}}%
    }%
    \gplbacktext
    \put(0,0){\includegraphics{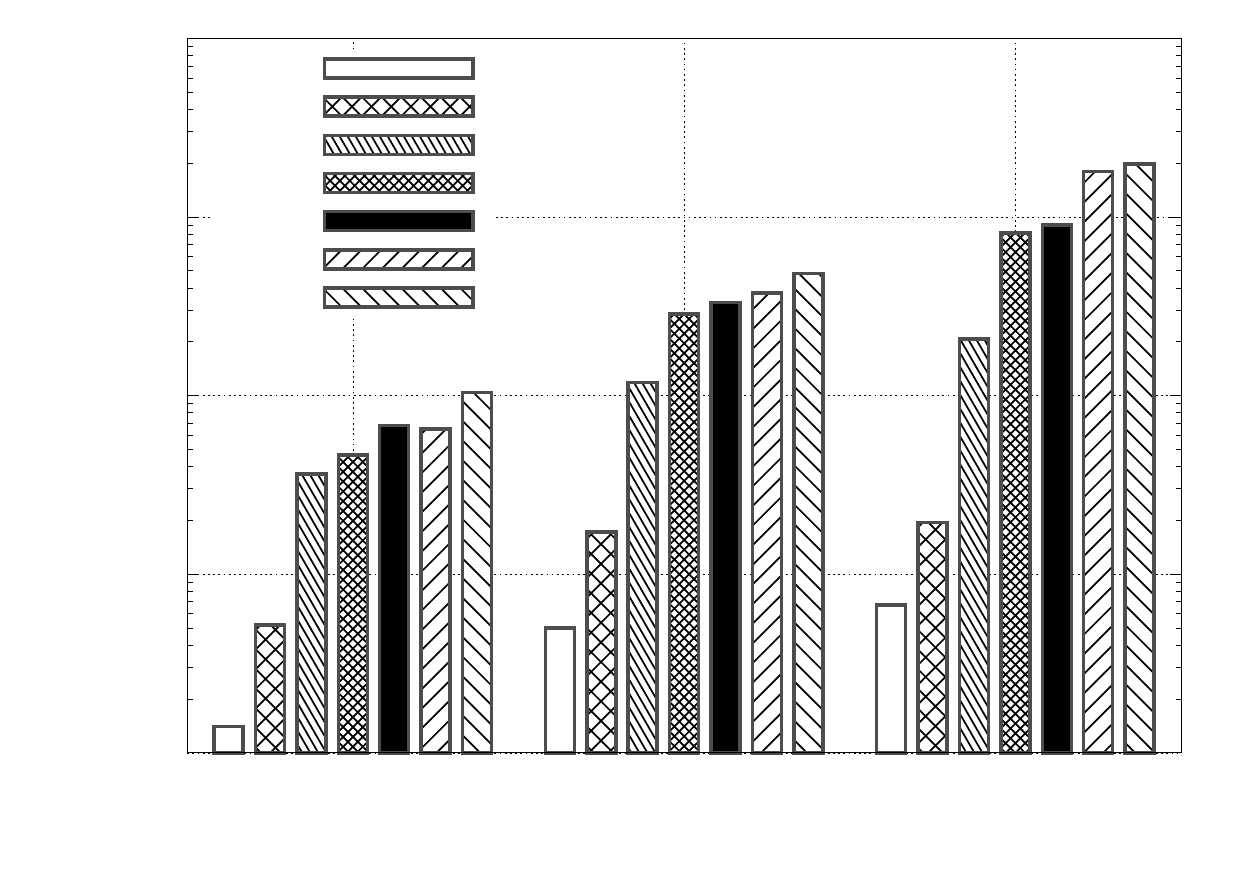}}%
    \gplfronttext
  \end{picture}%
\endgroup
 			}
		}
		\caption{Query answering time for $ |G_E| = 100K$, $ \len = 5 $,  $ \overlap = 35\% $,   $ \select = 25\% $, and $ |Q_{DB}| = 1K $ to $ |Q_{DB}| = 5K $.}
		\label{fig:exp:social-p100-sub5-S10:VaryingQDBsize}
	\end{subfigure}
	\vspace{-0.3cm}
	\begin{subfigure}[t]{0.32\textwidth}
		\centering
		\resizebox{\linewidth}{0.18\textheight}{
\begingroup
  \makeatletter
  \providecommand\color[2][]{%
    \GenericError{(gnuplot) \space\space\space\@spaces}{%
      Package color not loaded in conjunction with
      terminal option `colourtext'%
    }{See the gnuplot documentation for explanation.%
    }{Either use 'blacktext' in gnuplot or load the package
      color.sty in LaTeX.}%
    \renewcommand\color[2][]{}%
  }%
  \providecommand\includegraphics[2][]{%
    \GenericError{(gnuplot) \space\space\space\@spaces}{%
      Package graphicx or graphics not loaded%
    }{See the gnuplot documentation for explanation.%
    }{The gnuplot epslatex terminal needs graphicx.sty or graphics.sty.}%
    \renewcommand\includegraphics[2][]{}%
  }%
  \providecommand\rotatebox[2]{#2}%
  \@ifundefined{ifGPcolor}{%
    \newif\ifGPcolor
    \GPcolortrue
  }{}%
  \@ifundefined{ifGPblacktext}{%
    \newif\ifGPblacktext
    \GPblacktexttrue
  }{}%
  \let\gplgaddtomacro\g@addto@macro
  \gdef\gplbacktext{}%
  \gdef\gplfronttext{}%
  \makeatother
  \ifGPblacktext
    \def\colorrgb#1{}%
    \def\colorgray#1{}%
  \else
    \ifGPcolor
      \def\colorrgb#1{\color[rgb]{#1}}%
      \def\colorgray#1{\color[gray]{#1}}%
      \expandafter\def\csname LTw\endcsname{\color{white}}%
      \expandafter\def\csname LTb\endcsname{\color{black}}%
      \expandafter\def\csname LTa\endcsname{\color{black}}%
      \expandafter\def\csname LT0\endcsname{\color[rgb]{1,0,0}}%
      \expandafter\def\csname LT1\endcsname{\color[rgb]{0,1,0}}%
      \expandafter\def\csname LT2\endcsname{\color[rgb]{0,0,1}}%
      \expandafter\def\csname LT3\endcsname{\color[rgb]{1,0,1}}%
      \expandafter\def\csname LT4\endcsname{\color[rgb]{0,1,1}}%
      \expandafter\def\csname LT5\endcsname{\color[rgb]{1,1,0}}%
      \expandafter\def\csname LT6\endcsname{\color[rgb]{0,0,0}}%
      \expandafter\def\csname LT7\endcsname{\color[rgb]{1,0.3,0}}%
      \expandafter\def\csname LT8\endcsname{\color[rgb]{0.5,0.5,0.5}}%
    \else
      \def\colorrgb#1{\color{black}}%
      \def\colorgray#1{\color[gray]{#1}}%
      \expandafter\def\csname LTw\endcsname{\color{white}}%
      \expandafter\def\csname LTb\endcsname{\color{black}}%
      \expandafter\def\csname LTa\endcsname{\color{black}}%
      \expandafter\def\csname LT0\endcsname{\color{black}}%
      \expandafter\def\csname LT1\endcsname{\color{black}}%
      \expandafter\def\csname LT2\endcsname{\color{black}}%
      \expandafter\def\csname LT3\endcsname{\color{black}}%
      \expandafter\def\csname LT4\endcsname{\color{black}}%
      \expandafter\def\csname LT5\endcsname{\color{black}}%
      \expandafter\def\csname LT6\endcsname{\color{black}}%
      \expandafter\def\csname LT7\endcsname{\color{black}}%
      \expandafter\def\csname LT8\endcsname{\color{black}}%
    \fi
  \fi
    \setlength{\unitlength}{0.0500bp}%
    \ifx\gptboxheight\undefined%
      \newlength{\gptboxheight}%
      \newlength{\gptboxwidth}%
      \newsavebox{\gptboxtext}%
    \fi%
    \setlength{\fboxrule}{0.5pt}%
    \setlength{\fboxsep}{1pt}%
\begin{picture}(7200.00,5040.00)%
    \gplgaddtomacro\gplbacktext{%
      \csname LTb\endcsname%
      \put(732,756){\makebox(0,0)[r]{\strut{}0.0}}%
      \csname LTb\endcsname%
      \put(732,1008){\makebox(0,0)[r]{\strut{}1.0}}%
      \csname LTb\endcsname%
      \put(732,1260){\makebox(0,0)[r]{\strut{}2.0}}%
      \csname LTb\endcsname%
      \put(732,1512){\makebox(0,0)[r]{\strut{}3.0}}%
      \csname LTb\endcsname%
      \put(732,1763){\makebox(0,0)[r]{\strut{}4.0}}%
      \csname LTb\endcsname%
      \put(732,2015){\makebox(0,0)[r]{\strut{}5.0}}%
      \csname LTb\endcsname%
      \put(732,2267){\makebox(0,0)[r]{\strut{}6.0}}%
      \csname LTb\endcsname%
      \put(864,536){\makebox(0,0){\strut{}3}}%
      \csname LTb\endcsname%
      \put(2856,536){\makebox(0,0){\strut{}5}}%
      \csname LTb\endcsname%
      \put(4847,536){\makebox(0,0){\strut{}7}}%
      \csname LTb\endcsname%
      \put(6839,536){\makebox(0,0){\strut{}9}}%
    }%
    \gplgaddtomacro\gplfronttext{%
      \csname LTb\endcsname%
      \put(116,2611){\rotatebox{-270}{\makebox(0,0){\strut{}Answering time (msec/update)}}}%
      \put(3851,206){\makebox(0,0){\strut{}Varying $ \len $}}%
      \csname LTb\endcsname%
      \put(996,2072){\makebox(0,0)[l]{\strut{}\Tree}}%
      \csname LTb\endcsname%
      \put(996,1808){\makebox(0,0)[l]{\strut{}\TreeCache}}%
    }%
    \gplgaddtomacro\gplbacktext{%
      \csname LTb\endcsname%
      \put(732,2519){\makebox(0,0)[r]{\strut{}15}}%
      \csname LTb\endcsname%
      \put(732,2800){\makebox(0,0)[r]{\strut{}500}}%
      \csname LTb\endcsname%
      \put(732,3089){\makebox(0,0)[r]{\strut{}1000}}%
      \csname LTb\endcsname%
      \put(732,3378){\makebox(0,0)[r]{\strut{}1500}}%
      \csname LTb\endcsname%
      \put(732,3667){\makebox(0,0)[r]{\strut{}2000}}%
      \csname LTb\endcsname%
      \put(732,3957){\makebox(0,0)[r]{\strut{}2500}}%
      \csname LTb\endcsname%
      \put(732,4246){\makebox(0,0)[r]{\strut{}3000}}%
      \csname LTb\endcsname%
      \put(732,4535){\makebox(0,0)[r]{\strut{}3500}}%
      \csname LTb\endcsname%
      \put(864,2299){\makebox(0,0){\strut{}}}%
      \csname LTb\endcsname%
      \put(2856,2299){\makebox(0,0){\strut{}}}%
      \csname LTb\endcsname%
      \put(4847,2299){\makebox(0,0){\strut{}}}%
      \csname LTb\endcsname%
      \put(6839,2299){\makebox(0,0){\strut{}}}%
    }%
    \gplgaddtomacro\gplfronttext{%
      \csname LTb\endcsname%
      \put(996,4340){\makebox(0,0)[l]{\strut{}\Neo}}%
      \csname LTb\endcsname%
      \put(996,4076){\makebox(0,0)[l]{\strut{}\Inv}}%
      \csname LTb\endcsname%
      \put(996,3812){\makebox(0,0)[l]{\strut{}\InvCache}}%
      \csname LTb\endcsname%
      \put(996,3548){\makebox(0,0)[l]{\strut{}\InvIncr}}%
      \csname LTb\endcsname%
      \put(996,3284){\makebox(0,0)[l]{\strut{}\InvIncrCache}}%
    }%
    \gplbacktext
    \put(0,0){\includegraphics{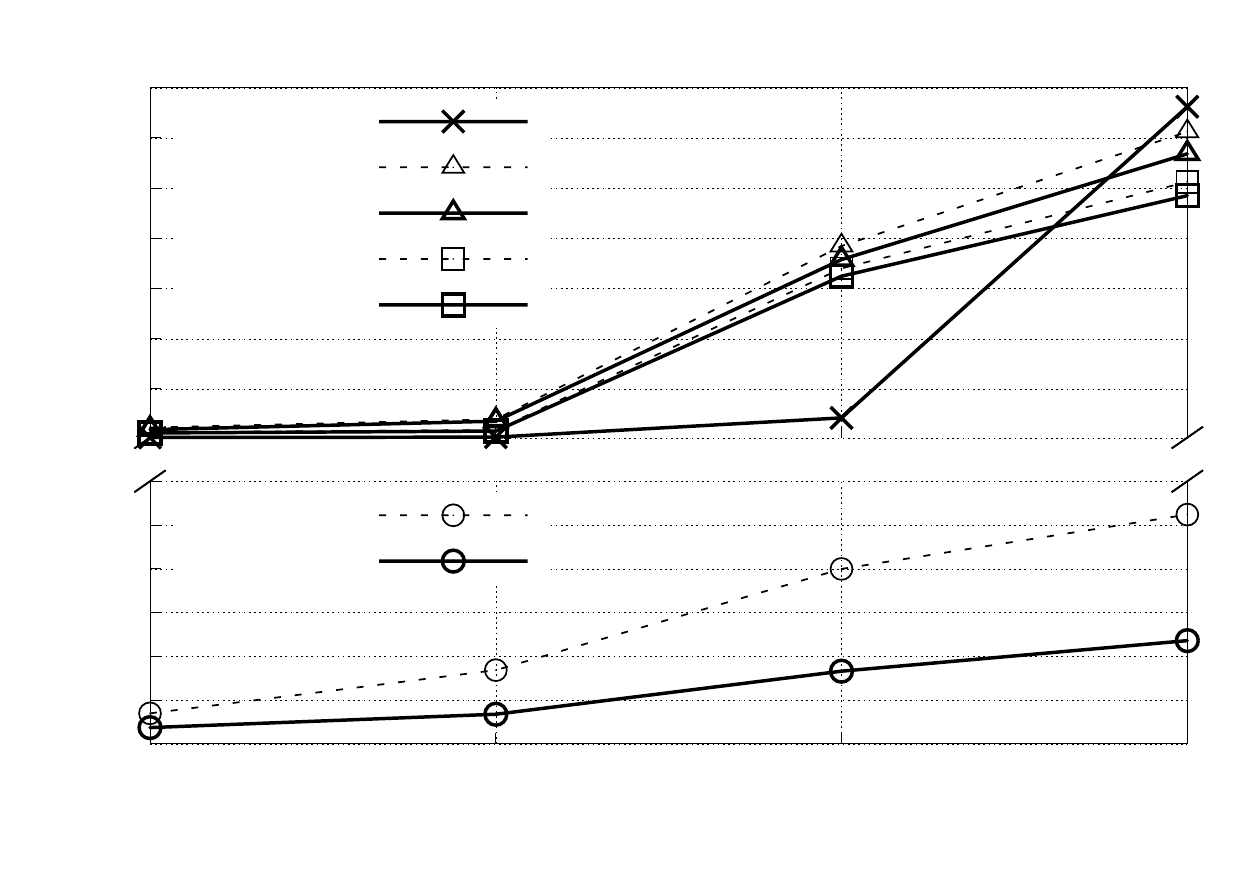}}%
    \gplfronttext
  \end{picture}%
\endgroup
 		}
		\caption{Query answering time for $ |Q_{DB}| = 5K $, $ \overlap = 35\% $,   $ \select = 25\% $, $ |G_E| = 100K $, and $ \len = 3 $ to $ \len = 9 $.}
		\label{fig:exp:social-p10M-sub5-S25:VaryingLength}
	\end{subfigure}
	\hfill
	\begin{subfigure}[t]{0.32\textwidth}
		\centering
		\resizebox{\linewidth}{0.18\textheight}{
\begingroup
  \makeatletter
  \providecommand\color[2][]{%
    \GenericError{(gnuplot) \space\space\space\@spaces}{%
      Package color not loaded in conjunction with
      terminal option `colourtext'%
    }{See the gnuplot documentation for explanation.%
    }{Either use 'blacktext' in gnuplot or load the package
      color.sty in LaTeX.}%
    \renewcommand\color[2][]{}%
  }%
  \providecommand\includegraphics[2][]{%
    \GenericError{(gnuplot) \space\space\space\@spaces}{%
      Package graphicx or graphics not loaded%
    }{See the gnuplot documentation for explanation.%
    }{The gnuplot epslatex terminal needs graphicx.sty or graphics.sty.}%
    \renewcommand\includegraphics[2][]{}%
  }%
  \providecommand\rotatebox[2]{#2}%
  \@ifundefined{ifGPcolor}{%
    \newif\ifGPcolor
    \GPcolortrue
  }{}%
  \@ifundefined{ifGPblacktext}{%
    \newif\ifGPblacktext
    \GPblacktexttrue
  }{}%
  \let\gplgaddtomacro\g@addto@macro
  \gdef\gplbacktext{}%
  \gdef\gplfronttext{}%
  \makeatother
  \ifGPblacktext
    \def\colorrgb#1{}%
    \def\colorgray#1{}%
  \else
    \ifGPcolor
      \def\colorrgb#1{\color[rgb]{#1}}%
      \def\colorgray#1{\color[gray]{#1}}%
      \expandafter\def\csname LTw\endcsname{\color{white}}%
      \expandafter\def\csname LTb\endcsname{\color{black}}%
      \expandafter\def\csname LTa\endcsname{\color{black}}%
      \expandafter\def\csname LT0\endcsname{\color[rgb]{1,0,0}}%
      \expandafter\def\csname LT1\endcsname{\color[rgb]{0,1,0}}%
      \expandafter\def\csname LT2\endcsname{\color[rgb]{0,0,1}}%
      \expandafter\def\csname LT3\endcsname{\color[rgb]{1,0,1}}%
      \expandafter\def\csname LT4\endcsname{\color[rgb]{0,1,1}}%
      \expandafter\def\csname LT5\endcsname{\color[rgb]{1,1,0}}%
      \expandafter\def\csname LT6\endcsname{\color[rgb]{0,0,0}}%
      \expandafter\def\csname LT7\endcsname{\color[rgb]{1,0.3,0}}%
      \expandafter\def\csname LT8\endcsname{\color[rgb]{0.5,0.5,0.5}}%
    \else
      \def\colorrgb#1{\color{black}}%
      \def\colorgray#1{\color[gray]{#1}}%
      \expandafter\def\csname LTw\endcsname{\color{white}}%
      \expandafter\def\csname LTb\endcsname{\color{black}}%
      \expandafter\def\csname LTa\endcsname{\color{black}}%
      \expandafter\def\csname LT0\endcsname{\color{black}}%
      \expandafter\def\csname LT1\endcsname{\color{black}}%
      \expandafter\def\csname LT2\endcsname{\color{black}}%
      \expandafter\def\csname LT3\endcsname{\color{black}}%
      \expandafter\def\csname LT4\endcsname{\color{black}}%
      \expandafter\def\csname LT5\endcsname{\color{black}}%
      \expandafter\def\csname LT6\endcsname{\color{black}}%
      \expandafter\def\csname LT7\endcsname{\color{black}}%
      \expandafter\def\csname LT8\endcsname{\color{black}}%
    \fi
  \fi
    \setlength{\unitlength}{0.0500bp}%
    \ifx\gptboxheight\undefined%
      \newlength{\gptboxheight}%
      \newlength{\gptboxwidth}%
      \newsavebox{\gptboxtext}%
    \fi%
    \setlength{\fboxrule}{0.5pt}%
    \setlength{\fboxsep}{1pt}%
\begin{picture}(7200.00,5040.00)%
    \gplgaddtomacro\gplbacktext{%
      \csname LTb\endcsname%
      \put(732,756){\makebox(0,0)[r]{\strut{}0.0}}%
      \csname LTb\endcsname%
      \put(732,1008){\makebox(0,0)[r]{\strut{}0.5}}%
      \csname LTb\endcsname%
      \put(732,1260){\makebox(0,0)[r]{\strut{}1.0}}%
      \csname LTb\endcsname%
      \put(732,1512){\makebox(0,0)[r]{\strut{}1.5}}%
      \csname LTb\endcsname%
      \put(732,1763){\makebox(0,0)[r]{\strut{}2.0}}%
      \csname LTb\endcsname%
      \put(732,2015){\makebox(0,0)[r]{\strut{}2.5}}%
      \csname LTb\endcsname%
      \put(732,2267){\makebox(0,0)[r]{\strut{}3.0}}%
      \csname LTb\endcsname%
      \put(864,536){\makebox(0,0){\strut{}25\%}}%
      \csname LTb\endcsname%
      \put(2358,536){\makebox(0,0){\strut{}35\%}}%
      \csname LTb\endcsname%
      \put(3852,536){\makebox(0,0){\strut{}45\%}}%
      \csname LTb\endcsname%
      \put(5345,536){\makebox(0,0){\strut{}55\%}}%
      \csname LTb\endcsname%
      \put(6839,536){\makebox(0,0){\strut{}65\%}}%
    }%
    \gplgaddtomacro\gplfronttext{%
      \csname LTb\endcsname%
      \put(116,2611){\rotatebox{-270}{\makebox(0,0){\strut{}Answering time (msec/update)}}}%
      \put(3851,206){\makebox(0,0){\strut{}Varying $ \overlap $}}%
      \csname LTb\endcsname%
      \put(4532,2072){\makebox(0,0)[l]{\strut{}\Tree}}%
      \csname LTb\endcsname%
      \put(4532,1808){\makebox(0,0)[l]{\strut{}\TreeCache}}%
    }%
    \gplgaddtomacro\gplbacktext{%
      \csname LTb\endcsname%
      \put(732,2562){\makebox(0,0)[r]{\strut{}20}}%
      \csname LTb\endcsname%
      \put(732,2819){\makebox(0,0)[r]{\strut{}50}}%
      \csname LTb\endcsname%
      \put(732,3248){\makebox(0,0)[r]{\strut{}100}}%
      \csname LTb\endcsname%
      \put(732,3677){\makebox(0,0)[r]{\strut{}150}}%
      \csname LTb\endcsname%
      \put(732,4106){\makebox(0,0)[r]{\strut{}200}}%
      \csname LTb\endcsname%
      \put(732,4535){\makebox(0,0)[r]{\strut{}250}}%
      \csname LTb\endcsname%
      \put(864,2299){\makebox(0,0){\strut{}}}%
      \csname LTb\endcsname%
      \put(2358,2299){\makebox(0,0){\strut{}}}%
      \csname LTb\endcsname%
      \put(3852,2299){\makebox(0,0){\strut{}}}%
      \csname LTb\endcsname%
      \put(5345,2299){\makebox(0,0){\strut{}}}%
      \csname LTb\endcsname%
      \put(6839,2299){\makebox(0,0){\strut{}}}%
    }%
    \gplgaddtomacro\gplfronttext{%
      \csname LTb\endcsname%
      \put(4532,4340){\makebox(0,0)[l]{\strut{}\Inv}}%
      \csname LTb\endcsname%
      \put(4532,4076){\makebox(0,0)[l]{\strut{}\InvCache}}%
      \csname LTb\endcsname%
      \put(4532,3812){\makebox(0,0)[l]{\strut{}\InvIncr}}%
      \csname LTb\endcsname%
      \put(4532,3548){\makebox(0,0)[l]{\strut{}\InvIncrCache}}%
      \csname LTb\endcsname%
      \put(4532,3284){\makebox(0,0)[l]{\strut{}\Neo}}%
    }%
    \gplbacktext
    \put(0,0){\includegraphics{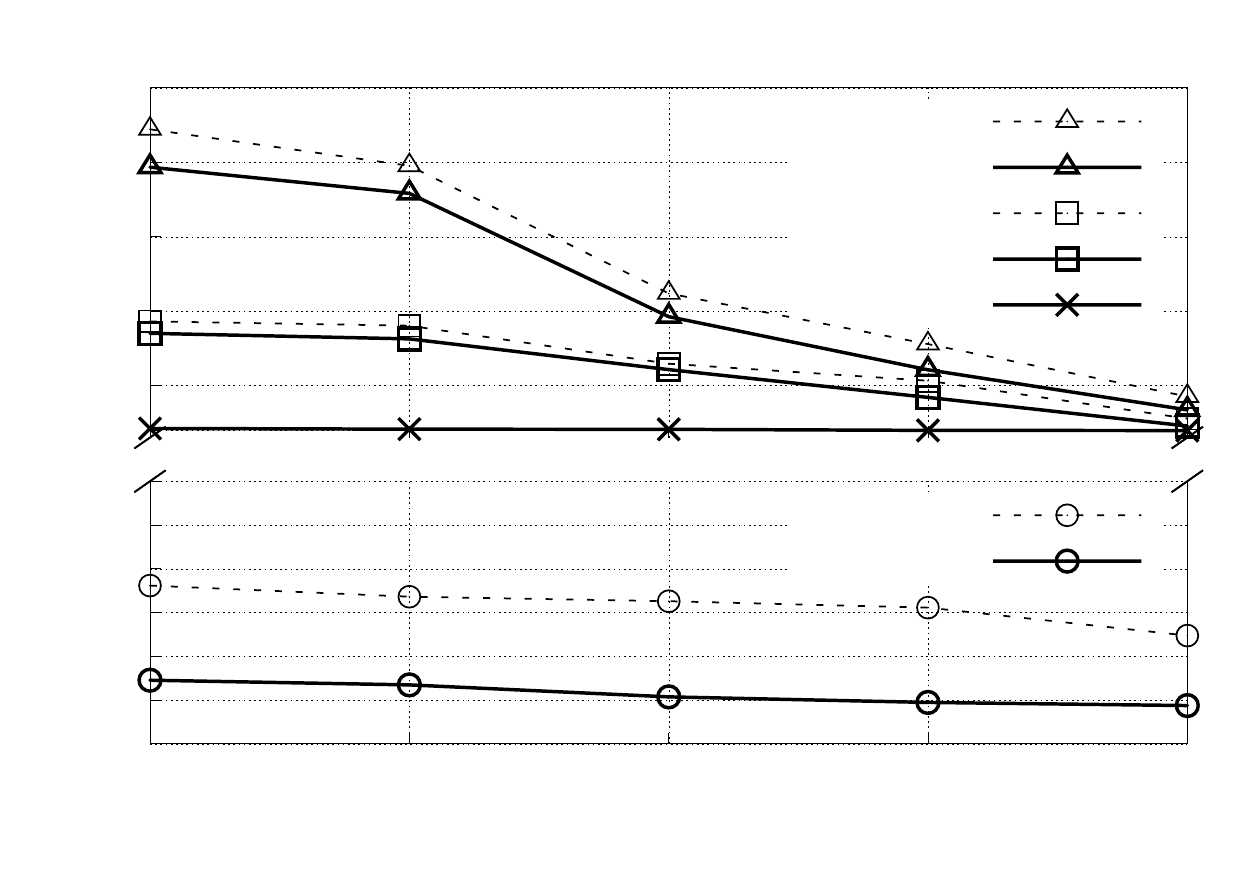}}%
    \gplfronttext
  \end{picture}%
\endgroup
 		}
		\caption{Query answering time for $ |Q_{DB}| = 5K $,  $ \len = 5 $, $ \select = 25\% $, $ |G_E| = 100K $, and  $ \overlap = 25\% $ to  $ \overlap = 65\% $.}
		\label{fig:exp:social-p10M-sub5-S25:VaryingOverlap}
	\end{subfigure}
	\hfill
	\begin{subfigure}[t]{0.32\textwidth}
		\centering
		\resizebox{\linewidth}{0.18\textheight}{
\begingroup
  \makeatletter
  \providecommand\color[2][]{%
    \GenericError{(gnuplot) \space\space\space\@spaces}{%
      Package color not loaded in conjunction with
      terminal option `colourtext'%
    }{See the gnuplot documentation for explanation.%
    }{Either use 'blacktext' in gnuplot or load the package
      color.sty in LaTeX.}%
    \renewcommand\color[2][]{}%
  }%
  \providecommand\includegraphics[2][]{%
    \GenericError{(gnuplot) \space\space\space\@spaces}{%
      Package graphicx or graphics not loaded%
    }{See the gnuplot documentation for explanation.%
    }{The gnuplot epslatex terminal needs graphicx.sty or graphics.sty.}%
    \renewcommand\includegraphics[2][]{}%
  }%
  \providecommand\rotatebox[2]{#2}%
  \@ifundefined{ifGPcolor}{%
    \newif\ifGPcolor
    \GPcolortrue
  }{}%
  \@ifundefined{ifGPblacktext}{%
    \newif\ifGPblacktext
    \GPblacktexttrue
  }{}%
  \let\gplgaddtomacro\g@addto@macro
  \gdef\gplbacktext{}%
  \gdef\gplfronttext{}%
  \makeatother
  \ifGPblacktext
    \def\colorrgb#1{}%
    \def\colorgray#1{}%
  \else
    \ifGPcolor
      \def\colorrgb#1{\color[rgb]{#1}}%
      \def\colorgray#1{\color[gray]{#1}}%
      \expandafter\def\csname LTw\endcsname{\color{white}}%
      \expandafter\def\csname LTb\endcsname{\color{black}}%
      \expandafter\def\csname LTa\endcsname{\color{black}}%
      \expandafter\def\csname LT0\endcsname{\color[rgb]{1,0,0}}%
      \expandafter\def\csname LT1\endcsname{\color[rgb]{0,1,0}}%
      \expandafter\def\csname LT2\endcsname{\color[rgb]{0,0,1}}%
      \expandafter\def\csname LT3\endcsname{\color[rgb]{1,0,1}}%
      \expandafter\def\csname LT4\endcsname{\color[rgb]{0,1,1}}%
      \expandafter\def\csname LT5\endcsname{\color[rgb]{1,1,0}}%
      \expandafter\def\csname LT6\endcsname{\color[rgb]{0,0,0}}%
      \expandafter\def\csname LT7\endcsname{\color[rgb]{1,0.3,0}}%
      \expandafter\def\csname LT8\endcsname{\color[rgb]{0.5,0.5,0.5}}%
    \else
      \def\colorrgb#1{\color{black}}%
      \def\colorgray#1{\color[gray]{#1}}%
      \expandafter\def\csname LTw\endcsname{\color{white}}%
      \expandafter\def\csname LTb\endcsname{\color{black}}%
      \expandafter\def\csname LTa\endcsname{\color{black}}%
      \expandafter\def\csname LT0\endcsname{\color{black}}%
      \expandafter\def\csname LT1\endcsname{\color{black}}%
      \expandafter\def\csname LT2\endcsname{\color{black}}%
      \expandafter\def\csname LT3\endcsname{\color{black}}%
      \expandafter\def\csname LT4\endcsname{\color{black}}%
      \expandafter\def\csname LT5\endcsname{\color{black}}%
      \expandafter\def\csname LT6\endcsname{\color{black}}%
      \expandafter\def\csname LT7\endcsname{\color{black}}%
      \expandafter\def\csname LT8\endcsname{\color{black}}%
    \fi
  \fi
    \setlength{\unitlength}{0.0500bp}%
    \ifx\gptboxheight\undefined%
      \newlength{\gptboxheight}%
      \newlength{\gptboxwidth}%
      \newsavebox{\gptboxtext}%
    \fi%
    \setlength{\fboxrule}{0.5pt}%
    \setlength{\fboxsep}{1pt}%
\begin{picture}(7200.00,5040.00)%
    \gplgaddtomacro\gplbacktext{%
      \csname LTb\endcsname%
      \put(732,756){\makebox(0,0)[r]{\strut{}0}}%
      \csname LTb\endcsname%
      \put(732,1058){\makebox(0,0)[r]{\strut{}2}}%
      \csname LTb\endcsname%
      \put(732,1360){\makebox(0,0)[r]{\strut{}4}}%
      \csname LTb\endcsname%
      \put(732,1663){\makebox(0,0)[r]{\strut{}6}}%
      \csname LTb\endcsname%
      \put(732,1965){\makebox(0,0)[r]{\strut{}8}}%
      \csname LTb\endcsname%
      \put(732,2267){\makebox(0,0)[r]{\strut{}10}}%
      \csname LTb\endcsname%
      \put(864,536){\makebox(0,0){\strut{}$100$}}%
      \csname LTb\endcsname%
      \put(1528,536){\makebox(0,0){\strut{}$200$}}%
      \csname LTb\endcsname%
      \put(2192,536){\makebox(0,0){\strut{}$300$}}%
      \csname LTb\endcsname%
      \put(2856,536){\makebox(0,0){\strut{}$400$}}%
      \csname LTb\endcsname%
      \put(3520,536){\makebox(0,0){\strut{}$500$}}%
      \csname LTb\endcsname%
      \put(4183,536){\makebox(0,0){\strut{}$600$}}%
      \csname LTb\endcsname%
      \put(4847,536){\makebox(0,0){\strut{}$700$}}%
      \csname LTb\endcsname%
      \put(5511,536){\makebox(0,0){\strut{}$800$}}%
      \csname LTb\endcsname%
      \put(6175,536){\makebox(0,0){\strut{}$900$}}%
      \csname LTb\endcsname%
      \put(6839,536){\makebox(0,0){\strut{}$1000$}}%
    }%
    \gplgaddtomacro\gplfronttext{%
      \csname LTb\endcsname%
      \put(248,2611){\rotatebox{-270}{\makebox(0,0){\strut{}Answering time (msec/update)}}}%
      \put(3851,206){\makebox(0,0){\strut{}Graph size (Edges x$1000$)}}%
      \csname LTb\endcsname%
      \put(4532,2072){\makebox(0,0)[l]{\strut{}\Tree}}%
      \csname LTb\endcsname%
      \put(4532,1808){\makebox(0,0)[l]{\strut{}\TreeCache}}%
    }%
    \gplgaddtomacro\gplbacktext{%
      \csname LTb\endcsname%
      \put(732,2519){\makebox(0,0)[r]{\strut{}20}}%
      \csname LTb\endcsname%
      \put(732,2963){\makebox(0,0)[r]{\strut{}500}}%
      \csname LTb\endcsname%
      \put(732,3425){\makebox(0,0)[r]{\strut{}1000}}%
      \csname LTb\endcsname%
      \put(732,3888){\makebox(0,0)[r]{\strut{}1500}}%
      \csname LTb\endcsname%
      \put(732,4350){\makebox(0,0)[r]{\strut{}2000}}%
      \csname LTb\endcsname%
      \put(864,4755){\makebox(0,0){\strut{}57}}%
      \csname LTb\endcsname%
      \put(1528,4755){\makebox(0,0){\strut{}106}}%
      \csname LTb\endcsname%
      \put(2192,4755){\makebox(0,0){\strut{}149}}%
      \csname LTb\endcsname%
      \put(2856,4755){\makebox(0,0){\strut{}188}}%
      \csname LTb\endcsname%
      \put(3520,4755){\makebox(0,0){\strut{}226}}%
      \csname LTb\endcsname%
      \put(4183,4755){\makebox(0,0){\strut{}270}}%
      \csname LTb\endcsname%
      \put(4847,4755){\makebox(0,0){\strut{}318}}%
      \csname LTb\endcsname%
      \put(5511,4755){\makebox(0,0){\strut{}367}}%
      \csname LTb\endcsname%
      \put(6175,4755){\makebox(0,0){\strut{}415}}%
      \csname LTb\endcsname%
      \put(6839,4755){\makebox(0,0){\strut{}463}}%
      \put(1661,4381){\makebox(0,0)[l]{\strut{}\emph{*}}}%
      \put(2325,3272){\makebox(0,0)[l]{\strut{}\emph{*}}}%
      \put(2325,3332){\makebox(0,0)[l]{\strut{}\emph{*}}}%
    }%
    \gplgaddtomacro\gplfronttext{%
      \csname LTb\endcsname%
      \put(3851,4975){\makebox(0,0){\strut{}Graph size (Vertices x$1000$)}}%
      \csname LTb\endcsname%
      \put(4532,4340){\makebox(0,0)[l]{\strut{}\Inv}}%
      \csname LTb\endcsname%
      \put(4532,4076){\makebox(0,0)[l]{\strut{}\InvCache}}%
      \csname LTb\endcsname%
      \put(4532,3812){\makebox(0,0)[l]{\strut{}\InvIncr}}%
      \csname LTb\endcsname%
      \put(4532,3548){\makebox(0,0)[l]{\strut{}\InvIncrCache}}%
      \csname LTb\endcsname%
      \put(4532,3284){\makebox(0,0)[l]{\strut{}\Neo}}%
    }%
    \gplbacktext
    \put(0,0){\includegraphics{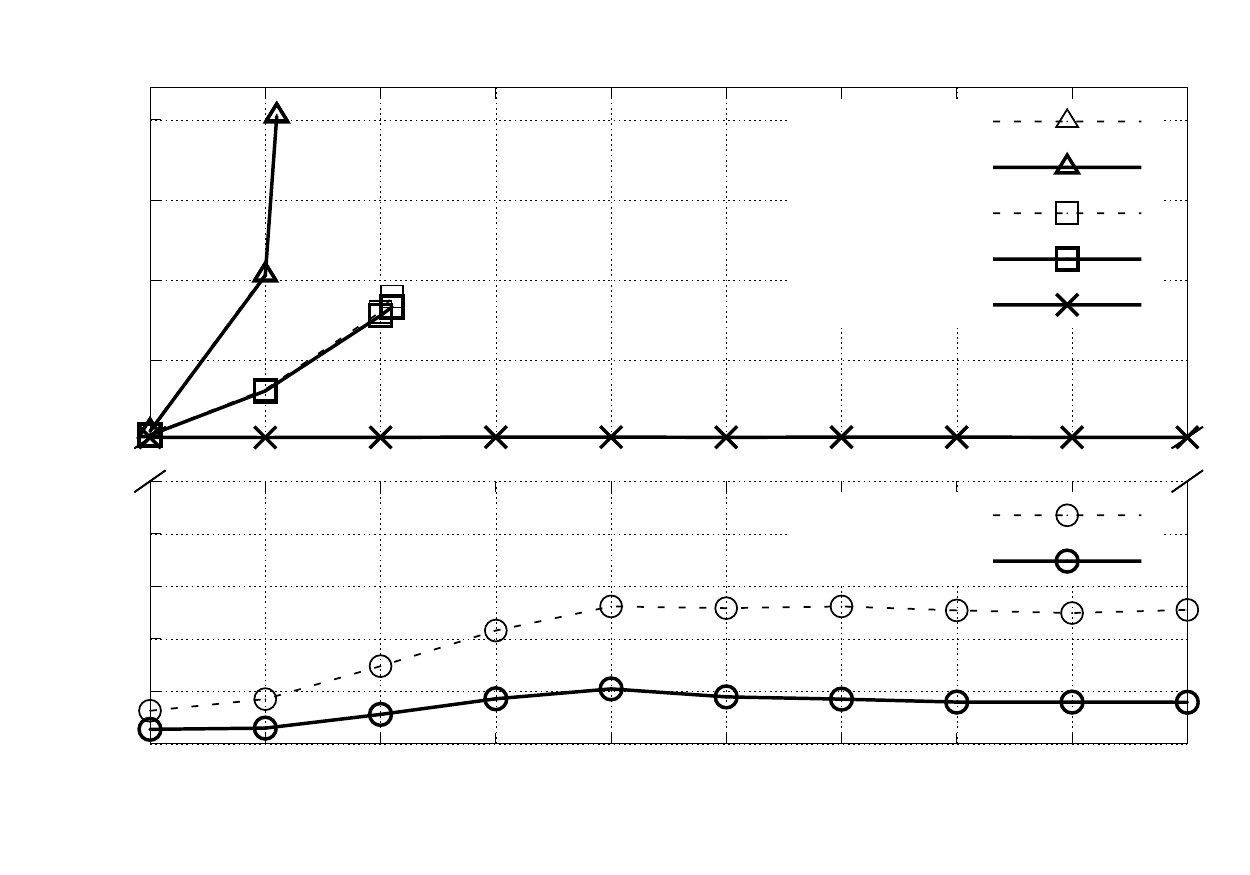}}%
    \gplfronttext
  \end{picture}%
\endgroup
 		}
		\caption{Query answering time for $ |Q_{DB}| = 5K $, $ \len = 5 $, $ \select = 25\% $, $ \overlap = 35\% $ and $ |G_E| = 100K  $ to $ |G_E| = 1M $.}
		\label{fig:exp:social-p1M-sub5-S25:filteringTime}
	\end{subfigure}	
	\caption{Results for the $ SNB $ dataset.}
	\label{fig:exp:social:all}
\end{figure*}

\section{Experimental Evaluation}
\label{sec:exps}

In this section we present the data and query sets, the algorithmic and technical configuration, the metrics employed and finally, present and extensively discuss the experimental evaluation results.

\subsection{Experimental Setup}
\label{sec:expSetup}

\ctitle{Data and Query Sets.} 
For the experimental evaluation we used a synthetic and two real-world datasets.

\cititle{The SNB Dataset.} 
The first dataset we utilized is the LDBC Social Network Benchmark ($ SNB $) \cite{ErlingALCGPPB15}. 
$ SNB $ is a synthetic benchmark designed to accurately simulate the evolution of a social network through time (i.e, vertex and edge sets labels, event distribution etc). 
This evolution is modeled using activities that occur inside a social network (i.e. user account creation, friendship linking, content creation, user interactions etc). 
\textcolor{black}{Based on the $ SNB $ generator we simulated the evolution of a graph consisting of user activities over a time period of $ 2 $ years. From this dataset we derived $ 3 $ query loads and configurations: (i) a set with a graph size of $ |G_E| = 100K $ edges and $ |G_V| = 57K $ vertices, (ii) a set with a graph size of $ |G_E| = 1M $ edges and $ |G_V| = 463K $ vertices, and (iii) a set with a graph size of $ |G_E| = 10M $ edges and $ |G_V| = 3.5M$.}

\cititle{The NYC Dataset.} 
The second dataset we utilized is a real world set of taxi rides performed in New York City ($ TAXI $) in 2013 \cite{web:taxiTripData} utilized in \emph{DEBS 2015 Grand Challenge} \cite{debs2015}. $ TAXI $ contains more that $ 160M $ entries of taxi rides with information about the license, pickup and drop-off location, the trip distance, the date and duration of the trip, and the fare. \textcolor{black}{We utilized the available data to generate a stream of updates that result in a graph of $ |G_E| = 1M $ edges and $ |G_V| = 280K  $, accompanied by a set of $ 5K $ query graph patterns.}

\cititle{The BioGRID Dataset.}
The third dataset we utilized is BioGRID~\cite{biogrid}, a real world dataset that represents protein to protein interactions. 
\textcolor{black}{This dataset is used as a stress test for our algorithms since it contains one type of edge (interacts) and vertex (protein), and thus every update affects the whole query database. We used BioGRID to generate a stream of updates that result in a graph size of $ |G_E| = 1M $ edges and $ |G_V| = 63K$ vertices, with a set of $ 5K $ query graph patterns.}

\ctitle{Query Set Configuration.}
In order to construct the set of query graph patterns $ Q_{DB} $ we identified three distinct query classes that are typical in the relevant literature: chains, stars, and cycles \cite{GillaniPL16, MondalD16}.  Each type of query graph pattern was chosen equiprobably during the generation of the query set. The baseline values for the query set are: \textcolor{black}{(i) an average size $ \len $ of 5 edges/query graph pattern, a value derived from the query workloads presented in $ SNB $ \cite{ErlingALCGPPB15}, (ii) a query database $ |Q_{DB}| $ size of $ 5K $  graph patterns, (iii) a factor that denotes the percentage of the query set $ Q_{DB}  $ that will ultimately be satisfied, denoted as selectivity $ \select = 25 \% $, and (iv) a factor that denotes the percentage of overlap between the queries in the set, $ \overlap = 35 \% $.}

\ctitle{Metrics.}
In our evaluation, we present and discuss the filtering and indexing time of each algorithm, along with the total memory requirements.

\color{black}
\ctitle{Technical Configuration.} All algorithms were implemented in Java $ 8 $ while for the materialization implementation the Stream API was employed. The \Neo-based approach was implemented using the embedded version of \Neo\ $ 3.4.7 $. Extensive experimentation evaluation concluded that a transaction\footnote{\url{https://neo4j.com/docs/cypher-manual/current/introduction/transactions/}} can perform up to $ 20K $ writes in the database without degrading \Neo's performance, while in order to ensure indexes are cached in main memory $ 55GB $ of main memory were allocated. A machine with Intel(R) Xeon(R) Processor E5-2650 at $ 2.00GHz $, $ 64GB $ RAM, and Ubuntu Linux $ 14.04 $ was used. The time shown in the graphs is wall-clock time and the results of each experiment are averaged over $ 10 $ runs to eliminate fluctuations in measurements.
\color{black}

\subsection{Results for the SNB Dataset}
In this section, we present the evaluation for the $ SNB $ benchmark and highlight the most significant findings.

\color{black}

\ctitle{Query Answering Time.}
Fig.~\ref{fig:exp:social-p100-sub5-S25:filteringTime} presents the results regarding the query answering time, i.e., the average time in milliseconds needed to determine which queries are satisfied by an incoming update, against a query set of $  Q_{DB} =  5K $. Please notice that the y-axis is split due to the high
differences in the performance of \Tree/\TreeCache and its competitors. We observe that the answering time increases for all algorithms as the graph size increases. Algorithms~\Tree/\TreeCache\ achieve the lowest answering times, suggesting better performance. Contrary, the competitors are more sensitive in graph size changes, with Algorithm~\Inv\ performing the worst (highest query answering time).
When comparing Algorithm~\Tree\ to \Inv, \InvIncr\ and \Neo\ the query answering time is improved by $ 99.15\% $, $ 98.14\% $ and $ 91.86\% $ respectively, while the improvement between \InvIncr\ and \Inv\ is $ 54.33 \% $. Finally,  comparing Algorithm~\TreeCache\ to \InvCache, \InvIncrCache\ and \Neo\ demonstrates a performance improvement of $ 99.62 \% $, $ 99.17 \% $ and $ 96.74 \% $ respectively, while the difference of \InvIncrCache\ and \InvCache\ is $ 54.6 \% $. 

\color{black}

The results (Fig.~\ref{fig:exp:social-p100-sub5-S25:filteringTime}) suggest that all solutions that implement caching are faster compared to the versions without it. In more detail, Algorithms~\TreeCache/\InvCache/\InvIncrCache\ are consistently faster than their non-caching counterparts, by  $ 59.95\% $, $ 9.36\% $ and $  9.91\% $ respectively. This is attributed to the fact that Algorithms~\Tree/\Inv/\InvIncr, have to recalculate the probe and build structures required for the joining process, in contrast to Algorithms \TreeCache/\InvCache/\InvIncrCache\ that store these structures and incrementally update them, thus providing better performance. 

In Fig.~\ref{fig:exp:social-p100-sub5-S10:VaryingSelec} we present the results when varying the parameter $ \sigma $, for $ 10\% $, $ 15\% $, $ 20\% $, $ 25\% $ and $ 30\% $ of a query set for $ |Q_{DB}|  = 5K $ and $ |G_E| = 100K$. 
In this setup the algorithms  are  evaluated for a varying percentage of queries that match. 
A higher number of queries satisfied,  increases the number of calculations performed by each algorithm. The results show that all algorithms behave in a similar manner as previously described. In more detail, Algorithm \TreeCache\ is the most efficient of all, and thus the fastest among the extensions that utilize caching, while \Tree\ is the most efficient solution among the solutions that do not employ a caching strategy. Finally, the percentage differences, between the algorithmic solutions remain the same as before in most cases. 

In Fig.~\ref{fig:exp:social-p100-sub5-S10:VaryingQDBsize} we give the results of the experimental evaluation when varying the size of the query database $ |Q_{DB}| $. More specifically, we present the answering time per triple when $ |Q_{DB}| = 1K$, $ 3K $ and $ 5K $, and $ |G_E|  = 100K $. Please notice the y-axis is in logarithmic scale. The results demonstrate that all algorithm's behavior is aligned with our previous observations. More specifically, Algorithms~\TreeCache\ and \Tree\ exhibit the highest performance (i.e., lowest answering time), throughout the increase of $ |Q_{DB}| $s, and thus determine faster which queries of $ |Q_{DB}| $ have matched given an update $ u_i $. Similarly to the previous setups, the competitors have the lowest performance, while Algorithms~\InvIncr\ and \InvIncrCache\ perform better compared to \Inv\ and \InvCache.

\color{black}
In Fig.~\ref{fig:exp:social-p10M-sub5-S25:VaryingLength} we give the results of the experimental evaluation when varying the average query size $ \len $. More specifically, we present the answering time per triple when $ \len = 3 $, $ 5 $, $ 7 $ and $ 9 $ of a query set for $ |Q_{DB}|  = 5K $ and $ |G_E| = 100K$. We observe that the answering time increases for all algorithms as the average query length increases.  More specifically, Algorithms~\TreeCache\ and \Tree\ exhibit the highest performance (i.e., lowest answering time), throughout the increase of $ \len $s, and thus determine faster which queries have  been satisfied. Similarly to the previous evaluation setups, the Algorithms~\Inv/\InvCache/\InvIncr/\InvIncrCache/\Neo\ have the lowest performance, and increase significantly their answering time when $ \len $ increases, while Algorithms~\InvIncr\ and \InvIncrCache\ perform better compared to \Inv, \InvCache\ and \Neo\ when $ \len = 9 $.
\color{black}

\color{black}
In Fig.~\ref{fig:exp:social-p10M-sub5-S25:VaryingOverlap} we give the results of the experimental evaluation when varying the parameter $ \overlap $, for $ 25\% $, $ 35\% $, $ 45\% $, $ 55\% $ and $ 65\% $ of a query set for $ |Q_{DB}|  = 5K $ and $ |G_E| = 100K$. In this setup the algorithms are evaluated for varying percentage of query overlap. A higher number of query overlap, should decrease the number of calculations performed by algorithms designed to exploit commonalities among the query set. The results show that all algorithm behave in a similar manner as previously described, while Algorithms\Inv/\InvCache/\InvIncr/\InvIncrCache\ observe higher performance gains. Algorithm~\TreeCache\ is the most efficient of all, and thus the fastest among the extensions that utilize caching techniques, while \Tree\ is the most efficient solution among the solutions that do not employ caching.
\color{black}

\begin{figure}
	\centering
	\resizebox{!}{0.18\textheight}{
\begingroup
  \makeatletter
  \providecommand\color[2][]{%
    \GenericError{(gnuplot) \space\space\space\@spaces}{%
      Package color not loaded in conjunction with
      terminal option `colourtext'%
    }{See the gnuplot documentation for explanation.%
    }{Either use 'blacktext' in gnuplot or load the package
      color.sty in LaTeX.}%
    \renewcommand\color[2][]{}%
  }%
  \providecommand\includegraphics[2][]{%
    \GenericError{(gnuplot) \space\space\space\@spaces}{%
      Package graphicx or graphics not loaded%
    }{See the gnuplot documentation for explanation.%
    }{The gnuplot epslatex terminal needs graphicx.sty or graphics.sty.}%
    \renewcommand\includegraphics[2][]{}%
  }%
  \providecommand\rotatebox[2]{#2}%
  \@ifundefined{ifGPcolor}{%
    \newif\ifGPcolor
    \GPcolortrue
  }{}%
  \@ifundefined{ifGPblacktext}{%
    \newif\ifGPblacktext
    \GPblacktexttrue
  }{}%
  \let\gplgaddtomacro\g@addto@macro
  \gdef\gplbacktext{}%
  \gdef\gplfronttext{}%
  \makeatother
  \ifGPblacktext
    \def\colorrgb#1{}%
    \def\colorgray#1{}%
  \else
    \ifGPcolor
      \def\colorrgb#1{\color[rgb]{#1}}%
      \def\colorgray#1{\color[gray]{#1}}%
      \expandafter\def\csname LTw\endcsname{\color{white}}%
      \expandafter\def\csname LTb\endcsname{\color{black}}%
      \expandafter\def\csname LTa\endcsname{\color{black}}%
      \expandafter\def\csname LT0\endcsname{\color[rgb]{1,0,0}}%
      \expandafter\def\csname LT1\endcsname{\color[rgb]{0,1,0}}%
      \expandafter\def\csname LT2\endcsname{\color[rgb]{0,0,1}}%
      \expandafter\def\csname LT3\endcsname{\color[rgb]{1,0,1}}%
      \expandafter\def\csname LT4\endcsname{\color[rgb]{0,1,1}}%
      \expandafter\def\csname LT5\endcsname{\color[rgb]{1,1,0}}%
      \expandafter\def\csname LT6\endcsname{\color[rgb]{0,0,0}}%
      \expandafter\def\csname LT7\endcsname{\color[rgb]{1,0.3,0}}%
      \expandafter\def\csname LT8\endcsname{\color[rgb]{0.5,0.5,0.5}}%
    \else
      \def\colorrgb#1{\color{black}}%
      \def\colorgray#1{\color[gray]{#1}}%
      \expandafter\def\csname LTw\endcsname{\color{white}}%
      \expandafter\def\csname LTb\endcsname{\color{black}}%
      \expandafter\def\csname LTa\endcsname{\color{black}}%
      \expandafter\def\csname LT0\endcsname{\color{black}}%
      \expandafter\def\csname LT1\endcsname{\color{black}}%
      \expandafter\def\csname LT2\endcsname{\color{black}}%
      \expandafter\def\csname LT3\endcsname{\color{black}}%
      \expandafter\def\csname LT4\endcsname{\color{black}}%
      \expandafter\def\csname LT5\endcsname{\color{black}}%
      \expandafter\def\csname LT6\endcsname{\color{black}}%
      \expandafter\def\csname LT7\endcsname{\color{black}}%
      \expandafter\def\csname LT8\endcsname{\color{black}}%
    \fi
  \fi
    \setlength{\unitlength}{0.0500bp}%
    \ifx\gptboxheight\undefined%
      \newlength{\gptboxheight}%
      \newlength{\gptboxwidth}%
      \newsavebox{\gptboxtext}%
    \fi%
    \setlength{\fboxrule}{0.5pt}%
    \setlength{\fboxsep}{1pt}%
\begin{picture}(7200.00,5040.00)%
    \gplgaddtomacro\gplbacktext{%
      \csname LTb\endcsname%
      \put(682,704){\makebox(0,0)[r]{\strut{}0}}%
      \csname LTb\endcsname%
      \put(682,1278){\makebox(0,0)[r]{\strut{}5}}%
      \csname LTb\endcsname%
      \put(682,1852){\makebox(0,0)[r]{\strut{}10}}%
      \csname LTb\endcsname%
      \put(682,2427){\makebox(0,0)[r]{\strut{}15}}%
      \csname LTb\endcsname%
      \put(682,3001){\makebox(0,0)[r]{\strut{}20}}%
      \csname LTb\endcsname%
      \put(682,3575){\makebox(0,0)[r]{\strut{}25}}%
      \csname LTb\endcsname%
      \put(682,4149){\makebox(0,0)[r]{\strut{}30}}%
      \csname LTb\endcsname%
      \put(814,484){\makebox(0,0){\strut{}$1$}}%
      \csname LTb\endcsname%
      \put(1479,484){\makebox(0,0){\strut{}$2$}}%
      \csname LTb\endcsname%
      \put(2145,484){\makebox(0,0){\strut{}$3$}}%
      \csname LTb\endcsname%
      \put(2810,484){\makebox(0,0){\strut{}$4$}}%
      \csname LTb\endcsname%
      \put(3476,484){\makebox(0,0){\strut{}$5$}}%
      \csname LTb\endcsname%
      \put(4141,484){\makebox(0,0){\strut{}$6$}}%
      \csname LTb\endcsname%
      \put(4807,484){\makebox(0,0){\strut{}$7$}}%
      \csname LTb\endcsname%
      \put(5472,484){\makebox(0,0){\strut{}$8$}}%
      \csname LTb\endcsname%
      \put(6138,484){\makebox(0,0){\strut{}$9$}}%
      \csname LTb\endcsname%
      \put(6803,484){\makebox(0,0){\strut{}$10$}}%
      \put(814,4599){\makebox(0,0)[l]{\strut{}463}}%
      \put(1479,4599){\makebox(0,0)[l]{\strut{}829}}%
      \put(2145,4599){\makebox(0,0)[l]{\strut{}1128}}%
      \put(2810,4599){\makebox(0,0)[l]{\strut{}1444}}%
      \put(3476,4599){\makebox(0,0)[l]{\strut{}1878}}%
      \put(4141,4599){\makebox(0,0)[l]{\strut{}2159}}%
      \put(4807,4599){\makebox(0,0)[l]{\strut{}2465}}%
      \put(5472,4599){\makebox(0,0)[l]{\strut{}2820}}%
      \put(6138,4599){\makebox(0,0)[l]{\strut{}3146}}%
      \put(6803,4599){\makebox(0,0)[l]{\strut{}3531}}%
      \put(3875,4126){\makebox(0,0)[l]{\strut{}\emph{*}}}%
      \put(3077,3231){\makebox(0,0)[l]{\strut{}\emph{*}}}%
    }%
    \gplgaddtomacro\gplfronttext{%
      \csname LTb\endcsname%
      \put(198,2541){\rotatebox{-270}{\makebox(0,0){\strut{}Answering time (msec/update)}}}%
      \put(3808,154){\makebox(0,0){\strut{}Graph size (Edges x$10^{6}$)}}%
      \put(3808,4929){\makebox(0,0){\strut{}Graph size (Vertices x$1000$)}}%
      \csname LTb\endcsname%
      \put(4496,2805){\makebox(0,0)[l]{\strut{}\Tree}}%
      \csname LTb\endcsname%
      \put(4496,2541){\makebox(0,0)[l]{\strut{}\Neo}}%
      \csname LTb\endcsname%
      \put(4496,2277){\makebox(0,0)[l]{\strut{}\TreeCache}}%
    }%
    \gplbacktext
    \put(0,0){\includegraphics{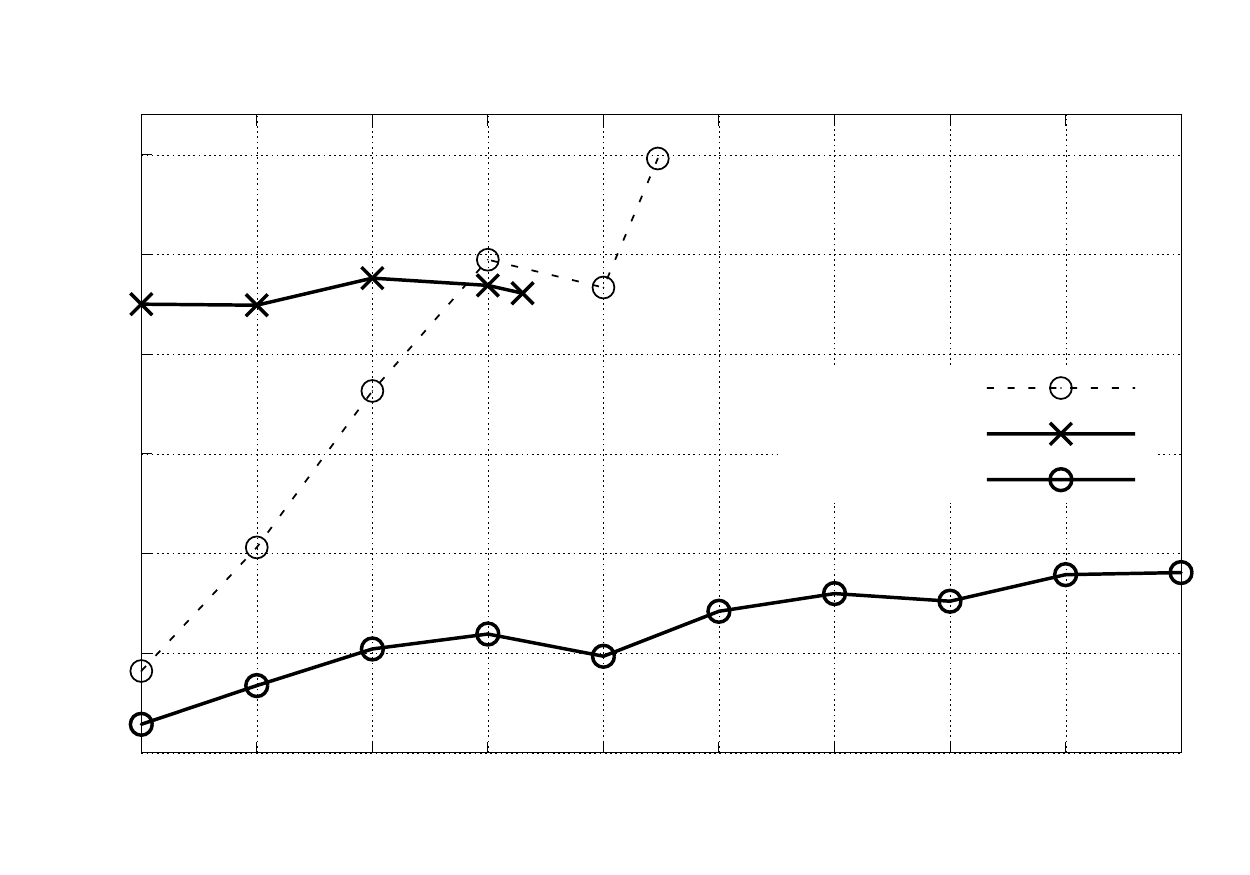}}%
    \gplfronttext
  \end{picture}%
\endgroup
 	}
	\caption{Query answering time for $ |Q_{DB}| = 5K $, $ \len = 5 $, $ \select = 25\% $, $ \overlap = 35\% $ and $ |G_E| = 1M  $ to $ |G_E| = 10M $.}
	\label{fig:exp:social-p10M-sub5-S25:filteringTime}
\end{figure}

\color{black}
Fig.~\ref{fig:exp:social-p1M-sub5-S25:filteringTime} presents the results regarding the query answering time, for all algorithms when indexing a query set of $ |Q_{DB}| = 5K $ and a final graph size of $ |G_E| = 1M $ and $ |G_V| = 463K $. Given the extremely slow performance of some algorithms we have set a \emph{execution time threshold} of 24 hours, for all algorithms under evaluation, thus, when the threshold was crossed the evaluation was terminated. 
We again observe that the answering time increases for all algorithms as the graph size increases. Algorithms~\Tree/\TreeCache\ achieve the lowest answering times, suggesting better performance. Contrary, the Algorithms~\Inv/\InvCache/\InvIncr/\InvIncrCache\ are more sensitive in graph size changes and thus fail to terminate within the time threshold. More specifically, Algorithms~\Inv/\InvCache\  \emph{time out} at $ |G_E| = 210K $, while Algorithms~\InvIncr/\InvIncrCache\ \emph{time out} at $ |G_E| = 310K $ as denoted by the asterisks in the plot. When comparing Algorithms~\Tree\ and \TreeCache\ to \Neo\ the query answering is improved by $ 77.01\% $ and $ 92.86\% $ respectively. 
\color{black}

\color{black}
Fig.~\ref{fig:exp:social-p10M-sub5-S25:filteringTime} presents the results regarding the query answering time, for Algorithms~\Tree, \TreeCache\ and \Neo\ when indexing a query set of $ Q_{DB} = 5K $ and a final graph size of $ |G_E| = 10M $ and $ |G_V| = 3.5M $. Again, we have set an \emph{execution time threshold} of 24 hours, for all the algorithms under evaluation. In this experimental setup, we observe that the answering time increases for all algorithms as the graph size increases. Algorithm~\TreeCache\ achieves the lowest answering times, suggesting better performance, while Algorithms~\Tree\ and \Neo\ fail to terminate within the given time threshold. More specifically, Algorithm~\Tree\ \emph{times out} at $ |G_E| = 5.47M $, while Algorithm~\Neo\ times out at $ |G_E| = 4.3M $ as denoted by the asterisks in the plot. 
\color{black}

\color{black}
Overall, Algorithms~\TreeCache\ and \Tree, the two solutions that utilize trie structures to capture and index the common structural and attribute restrictions of query graphs achieve the lowest query answering times, compared to Algorithms~\Inv/\InvCache/\InvIncr/\InvIncrCache\ that employ no clustering techniques, as well as when compared with commercial solutions such as \Neo.
Additionally, adopting the incremental joining techniques (found in Algorithm~\Tree) into Algorithm~\InvIncr\ does not seem to significantly improve its performance when compared to Algorithm~\Inv.
In the same manner, adopting caching techniques that store the data structures generated during the join operations, change significantly the performance of the algorithms applied on, i.e., Algorithms~\TreeCache/\InvCache/\InvIncrCache. 
Taking all the above into consideration, we conclude that the algorithms that utilize trie-based indexing are able to achieve low query answering times compared to their competitors. 
\color{black}

\begin{figure}[!t]
	\centering
	\resizebox{!}{0.18\textheight}{
\begingroup
  \makeatletter
  \providecommand\color[2][]{%
    \GenericError{(gnuplot) \space\space\space\@spaces}{%
      Package color not loaded in conjunction with
      terminal option `colourtext'%
    }{See the gnuplot documentation for explanation.%
    }{Either use 'blacktext' in gnuplot or load the package
      color.sty in LaTeX.}%
    \renewcommand\color[2][]{}%
  }%
  \providecommand\includegraphics[2][]{%
    \GenericError{(gnuplot) \space\space\space\@spaces}{%
      Package graphicx or graphics not loaded%
    }{See the gnuplot documentation for explanation.%
    }{The gnuplot epslatex terminal needs graphicx.sty or graphics.sty.}%
    \renewcommand\includegraphics[2][]{}%
  }%
  \providecommand\rotatebox[2]{#2}%
  \@ifundefined{ifGPcolor}{%
    \newif\ifGPcolor
    \GPcolortrue
  }{}%
  \@ifundefined{ifGPblacktext}{%
    \newif\ifGPblacktext
    \GPblacktexttrue
  }{}%
  \let\gplgaddtomacro\g@addto@macro
  \gdef\gplbacktext{}%
  \gdef\gplfronttext{}%
  \makeatother
  \ifGPblacktext
    \def\colorrgb#1{}%
    \def\colorgray#1{}%
  \else
    \ifGPcolor
      \def\colorrgb#1{\color[rgb]{#1}}%
      \def\colorgray#1{\color[gray]{#1}}%
      \expandafter\def\csname LTw\endcsname{\color{white}}%
      \expandafter\def\csname LTb\endcsname{\color{black}}%
      \expandafter\def\csname LTa\endcsname{\color{black}}%
      \expandafter\def\csname LT0\endcsname{\color[rgb]{1,0,0}}%
      \expandafter\def\csname LT1\endcsname{\color[rgb]{0,1,0}}%
      \expandafter\def\csname LT2\endcsname{\color[rgb]{0,0,1}}%
      \expandafter\def\csname LT3\endcsname{\color[rgb]{1,0,1}}%
      \expandafter\def\csname LT4\endcsname{\color[rgb]{0,1,1}}%
      \expandafter\def\csname LT5\endcsname{\color[rgb]{1,1,0}}%
      \expandafter\def\csname LT6\endcsname{\color[rgb]{0,0,0}}%
      \expandafter\def\csname LT7\endcsname{\color[rgb]{1,0.3,0}}%
      \expandafter\def\csname LT8\endcsname{\color[rgb]{0.5,0.5,0.5}}%
    \else
      \def\colorrgb#1{\color{black}}%
      \def\colorgray#1{\color[gray]{#1}}%
      \expandafter\def\csname LTw\endcsname{\color{white}}%
      \expandafter\def\csname LTb\endcsname{\color{black}}%
      \expandafter\def\csname LTa\endcsname{\color{black}}%
      \expandafter\def\csname LT0\endcsname{\color{black}}%
      \expandafter\def\csname LT1\endcsname{\color{black}}%
      \expandafter\def\csname LT2\endcsname{\color{black}}%
      \expandafter\def\csname LT3\endcsname{\color{black}}%
      \expandafter\def\csname LT4\endcsname{\color{black}}%
      \expandafter\def\csname LT5\endcsname{\color{black}}%
      \expandafter\def\csname LT6\endcsname{\color{black}}%
      \expandafter\def\csname LT7\endcsname{\color{black}}%
      \expandafter\def\csname LT8\endcsname{\color{black}}%
    \fi
  \fi
    \setlength{\unitlength}{0.0500bp}%
    \ifx\gptboxheight\undefined%
      \newlength{\gptboxheight}%
      \newlength{\gptboxwidth}%
      \newsavebox{\gptboxtext}%
    \fi%
    \setlength{\fboxrule}{0.5pt}%
    \setlength{\fboxsep}{1pt}%
\begin{picture}(7200.00,5040.00)%
    \gplgaddtomacro\gplbacktext{%
      \csname LTb\endcsname%
      \put(946,704){\makebox(0,0)[r]{\strut{}0.00}}%
      \csname LTb\endcsname%
      \put(946,2076){\makebox(0,0)[r]{\strut{}0.01}}%
      \csname LTb\endcsname%
      \put(946,3447){\makebox(0,0)[r]{\strut{}0.10}}%
      \csname LTb\endcsname%
      \put(946,4819){\makebox(0,0)[r]{\strut{}1.00}}%
      \csname LTb\endcsname%
      \put(1078,484){\makebox(0,0){\strut{}$1$}}%
      \csname LTb\endcsname%
      \put(2509,484){\makebox(0,0){\strut{}$2$}}%
      \csname LTb\endcsname%
      \put(3941,484){\makebox(0,0){\strut{}$3$}}%
      \csname LTb\endcsname%
      \put(5372,484){\makebox(0,0){\strut{}$4$}}%
      \csname LTb\endcsname%
      \put(6803,484){\makebox(0,0){\strut{}$5$}}%
    }%
    \gplgaddtomacro\gplfronttext{%
      \csname LTb\endcsname%
      \put(198,2761){\rotatebox{-270}{\makebox(0,0){\strut{}Indexing time (msec/query)}}}%
      \put(3940,154){\makebox(0,0){\strut{}Varying $Q_{DB}$ (x$1000$)}}%
      \csname LTb\endcsname%
      \put(2849,4624){\makebox(0,0)[l]{\strut{}\InvIncrCache}}%
      \csname LTb\endcsname%
      \put(2849,4360){\makebox(0,0)[l]{\strut{}\InvIncr}}%
      \csname LTb\endcsname%
      \put(2849,4096){\makebox(0,0)[l]{\strut{}\TreeCache}}%
      \csname LTb\endcsname%
      \put(2849,3832){\makebox(0,0)[l]{\strut{}\Tree}}%
      \csname LTb\endcsname%
      \put(4760,4624){\makebox(0,0)[l]{\strut{}\InvCache}}%
      \csname LTb\endcsname%
      \put(4760,4360){\makebox(0,0)[l]{\strut{}\Inv}}%
      \csname LTb\endcsname%
      \put(4760,4096){\makebox(0,0)[l]{\strut{}\Neo}}%
    }%
    \gplbacktext
    \put(0,0){\includegraphics{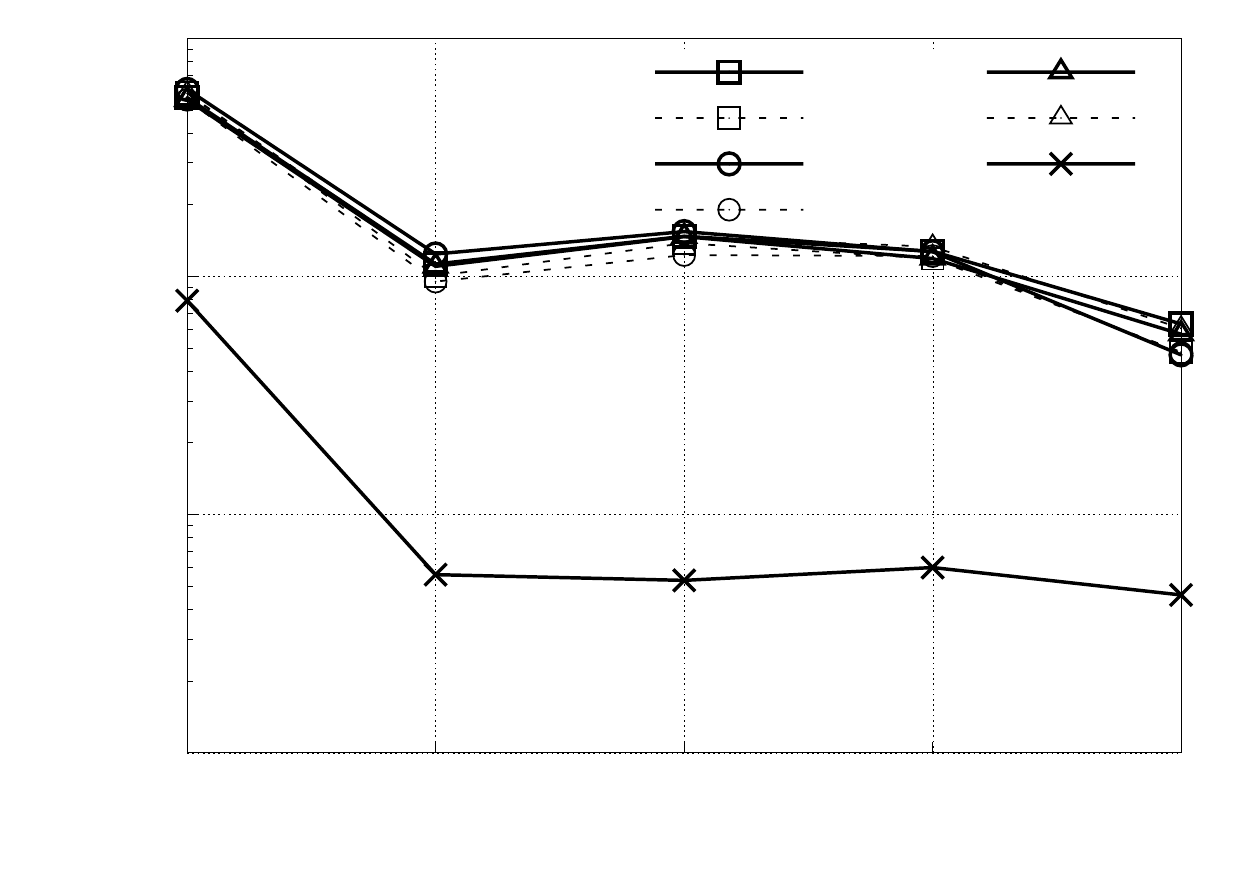}}%
    \gplfronttext
  \end{picture}%
\endgroup
 	}
	\caption{Query insertion time for $ \len = 5 $, $ \select = 25\% $, $ \overlap = 35\% $ and $ |Q_{DB}| = 1K $ to $ |Q_{DB}| = 5K $.}
	\label{fig:exp:social-p100-sub5-S25:indexing}
\end{figure}

\color{black}
\ctitle{Indexing Time.}
Fig.~\ref{fig:exp:social-p100-sub5-S25:indexing} presents the indexing time in milliseconds required to insert $ 1,000 $ query graph patterns when the query database size increases. We observe that the time required to go from an empty query database to a query database of size $ 1,000 $ is higher compared to the time required for the next iterations of the query database. Please notice the y-axis is in logarithmic scale. This can be explained as follows, all algorithms utilize data structures that need to be initialized during the initial stages of query indexing phase, i.e. when inserting queries in an empty database, additionally as the queries share common restrictions less time is required for creating new entries in the data structures of algorithms. Additionally, the time required to index a query graph pattern in the database does not vary significantly for all algorithms. Notice that query indexing time is not a critical performance parameter in the proposed paradigm, since the most important dimension is query answering time. 
\color{black}

\subsection{Results for the NYC and BioGRID Dataset}
In this section, we present the evaluation for the $ NYC $ and $ BioGRID $ dataset and highlight the most significant findings.

\begin{figure*}[!ht]
	\begin{subfigure}[t]{0.32\textwidth}
		\centering
		\resizebox{\linewidth}{0.18\textheight}{
\begingroup
  \makeatletter
  \providecommand\color[2][]{%
    \GenericError{(gnuplot) \space\space\space\@spaces}{%
      Package color not loaded in conjunction with
      terminal option `colourtext'%
    }{See the gnuplot documentation for explanation.%
    }{Either use 'blacktext' in gnuplot or load the package
      color.sty in LaTeX.}%
    \renewcommand\color[2][]{}%
  }%
  \providecommand\includegraphics[2][]{%
    \GenericError{(gnuplot) \space\space\space\@spaces}{%
      Package graphicx or graphics not loaded%
    }{See the gnuplot documentation for explanation.%
    }{The gnuplot epslatex terminal needs graphicx.sty or graphics.sty.}%
    \renewcommand\includegraphics[2][]{}%
  }%
  \providecommand\rotatebox[2]{#2}%
  \@ifundefined{ifGPcolor}{%
    \newif\ifGPcolor
    \GPcolortrue
  }{}%
  \@ifundefined{ifGPblacktext}{%
    \newif\ifGPblacktext
    \GPblacktexttrue
  }{}%
  \let\gplgaddtomacro\g@addto@macro
  \gdef\gplbacktext{}%
  \gdef\gplfronttext{}%
  \makeatother
  \ifGPblacktext
    \def\colorrgb#1{}%
    \def\colorgray#1{}%
  \else
    \ifGPcolor
      \def\colorrgb#1{\color[rgb]{#1}}%
      \def\colorgray#1{\color[gray]{#1}}%
      \expandafter\def\csname LTw\endcsname{\color{white}}%
      \expandafter\def\csname LTb\endcsname{\color{black}}%
      \expandafter\def\csname LTa\endcsname{\color{black}}%
      \expandafter\def\csname LT0\endcsname{\color[rgb]{1,0,0}}%
      \expandafter\def\csname LT1\endcsname{\color[rgb]{0,1,0}}%
      \expandafter\def\csname LT2\endcsname{\color[rgb]{0,0,1}}%
      \expandafter\def\csname LT3\endcsname{\color[rgb]{1,0,1}}%
      \expandafter\def\csname LT4\endcsname{\color[rgb]{0,1,1}}%
      \expandafter\def\csname LT5\endcsname{\color[rgb]{1,1,0}}%
      \expandafter\def\csname LT6\endcsname{\color[rgb]{0,0,0}}%
      \expandafter\def\csname LT7\endcsname{\color[rgb]{1,0.3,0}}%
      \expandafter\def\csname LT8\endcsname{\color[rgb]{0.5,0.5,0.5}}%
    \else
      \def\colorrgb#1{\color{black}}%
      \def\colorgray#1{\color[gray]{#1}}%
      \expandafter\def\csname LTw\endcsname{\color{white}}%
      \expandafter\def\csname LTb\endcsname{\color{black}}%
      \expandafter\def\csname LTa\endcsname{\color{black}}%
      \expandafter\def\csname LT0\endcsname{\color{black}}%
      \expandafter\def\csname LT1\endcsname{\color{black}}%
      \expandafter\def\csname LT2\endcsname{\color{black}}%
      \expandafter\def\csname LT3\endcsname{\color{black}}%
      \expandafter\def\csname LT4\endcsname{\color{black}}%
      \expandafter\def\csname LT5\endcsname{\color{black}}%
      \expandafter\def\csname LT6\endcsname{\color{black}}%
      \expandafter\def\csname LT7\endcsname{\color{black}}%
      \expandafter\def\csname LT8\endcsname{\color{black}}%
    \fi
  \fi
    \setlength{\unitlength}{0.0500bp}%
    \ifx\gptboxheight\undefined%
      \newlength{\gptboxheight}%
      \newlength{\gptboxwidth}%
      \newsavebox{\gptboxtext}%
    \fi%
    \setlength{\fboxrule}{0.5pt}%
    \setlength{\fboxsep}{1pt}%
\begin{picture}(7200.00,5040.00)%
    \gplgaddtomacro\gplbacktext{%
      \csname LTb\endcsname%
      \put(732,756){\makebox(0,0)[r]{\strut{}0}}%
      \csname LTb\endcsname%
      \put(732,1008){\makebox(0,0)[r]{\strut{}2}}%
      \csname LTb\endcsname%
      \put(732,1260){\makebox(0,0)[r]{\strut{}4}}%
      \csname LTb\endcsname%
      \put(732,1512){\makebox(0,0)[r]{\strut{}6}}%
      \csname LTb\endcsname%
      \put(732,1763){\makebox(0,0)[r]{\strut{}8}}%
      \csname LTb\endcsname%
      \put(732,2015){\makebox(0,0)[r]{\strut{}10}}%
      \csname LTb\endcsname%
      \put(732,2267){\makebox(0,0)[r]{\strut{}12}}%
      \csname LTb\endcsname%
      \put(864,536){\makebox(0,0){\strut{}$100$}}%
      \csname LTb\endcsname%
      \put(1528,536){\makebox(0,0){\strut{}$200$}}%
      \csname LTb\endcsname%
      \put(2192,536){\makebox(0,0){\strut{}$300$}}%
      \csname LTb\endcsname%
      \put(2856,536){\makebox(0,0){\strut{}$400$}}%
      \csname LTb\endcsname%
      \put(3520,536){\makebox(0,0){\strut{}$500$}}%
      \csname LTb\endcsname%
      \put(4183,536){\makebox(0,0){\strut{}$600$}}%
      \csname LTb\endcsname%
      \put(4847,536){\makebox(0,0){\strut{}$700$}}%
      \csname LTb\endcsname%
      \put(5511,536){\makebox(0,0){\strut{}$800$}}%
      \csname LTb\endcsname%
      \put(6175,536){\makebox(0,0){\strut{}$900$}}%
      \csname LTb\endcsname%
      \put(6839,536){\makebox(0,0){\strut{}$1000$}}%
    }%
    \gplgaddtomacro\gplfronttext{%
      \csname LTb\endcsname%
      \put(248,2611){\rotatebox{-270}{\makebox(0,0){\strut{}Answering time (msec/update)}}}%
      \put(3851,206){\makebox(0,0){\strut{}Graph size (Edges x$1000$)}}%
      \csname LTb\endcsname%
      \put(996,2072){\makebox(0,0)[l]{\strut{}\TreeCache}}%
      \csname LTb\endcsname%
      \put(996,1808){\makebox(0,0)[l]{\strut{}\Tree}}%
    }%
    \gplgaddtomacro\gplbacktext{%
      \csname LTb\endcsname%
      \put(732,2519){\makebox(0,0)[r]{\strut{}25}}%
      \csname LTb\endcsname%
      \put(732,2714){\makebox(0,0)[r]{\strut{}100}}%
      \csname LTb\endcsname%
      \put(732,2974){\makebox(0,0)[r]{\strut{}200}}%
      \csname LTb\endcsname%
      \put(732,3234){\makebox(0,0)[r]{\strut{}300}}%
      \csname LTb\endcsname%
      \put(732,3494){\makebox(0,0)[r]{\strut{}400}}%
      \csname LTb\endcsname%
      \put(732,3755){\makebox(0,0)[r]{\strut{}500}}%
      \csname LTb\endcsname%
      \put(732,4015){\makebox(0,0)[r]{\strut{}600}}%
      \csname LTb\endcsname%
      \put(732,4275){\makebox(0,0)[r]{\strut{}700}}%
      \csname LTb\endcsname%
      \put(732,4535){\makebox(0,0)[r]{\strut{}800}}%
      \csname LTb\endcsname%
      \put(864,4755){\makebox(0,0)[l]{\strut{}44}}%
      \csname LTb\endcsname%
      \put(1528,4755){\makebox(0,0)[l]{\strut{}70}}%
      \csname LTb\endcsname%
      \put(2192,4755){\makebox(0,0)[l]{\strut{}96}}%
      \csname LTb\endcsname%
      \put(2856,4755){\makebox(0,0)[l]{\strut{}121}}%
      \csname LTb\endcsname%
      \put(3520,4755){\makebox(0,0)[l]{\strut{}147}}%
      \csname LTb\endcsname%
      \put(4183,4755){\makebox(0,0)[l]{\strut{}174}}%
      \csname LTb\endcsname%
      \put(4847,4755){\makebox(0,0)[l]{\strut{}202}}%
      \csname LTb\endcsname%
      \put(5511,4755){\makebox(0,0)[l]{\strut{}228}}%
      \csname LTb\endcsname%
      \put(6175,4755){\makebox(0,0)[l]{\strut{}254}}%
      \csname LTb\endcsname%
      \put(6839,4755){\makebox(0,0)[l]{\strut{}280}}%
      \put(1694,4260){\makebox(0,0)[l]{\strut{}\emph{*}}}%
      \put(1760,4160){\makebox(0,0)[l]{\strut{}\emph{*}}}%
      \put(2291,3412){\makebox(0,0)[l]{\strut{}\emph{*}}}%
      \put(2690,3440){\makebox(0,0)[l]{\strut{}\emph{*}}}%
    }%
    \gplgaddtomacro\gplfronttext{%
      \csname LTb\endcsname%
      \put(3851,5085){\makebox(0,0){\strut{}Graph size (Vertices x$1000$)}}%
      \csname LTb\endcsname%
      \put(4532,4340){\makebox(0,0)[l]{\strut{}\Inv}}%
      \csname LTb\endcsname%
      \put(4532,4076){\makebox(0,0)[l]{\strut{}\InvCache}}%
      \csname LTb\endcsname%
      \put(4532,3812){\makebox(0,0)[l]{\strut{}\InvIncr}}%
      \csname LTb\endcsname%
      \put(4532,3548){\makebox(0,0)[l]{\strut{}\InvIncrCache}}%
      \csname LTb\endcsname%
      \put(4532,3284){\makebox(0,0)[l]{\strut{}\Neo}}%
    }%
    \gplbacktext
    \put(0,0){\includegraphics{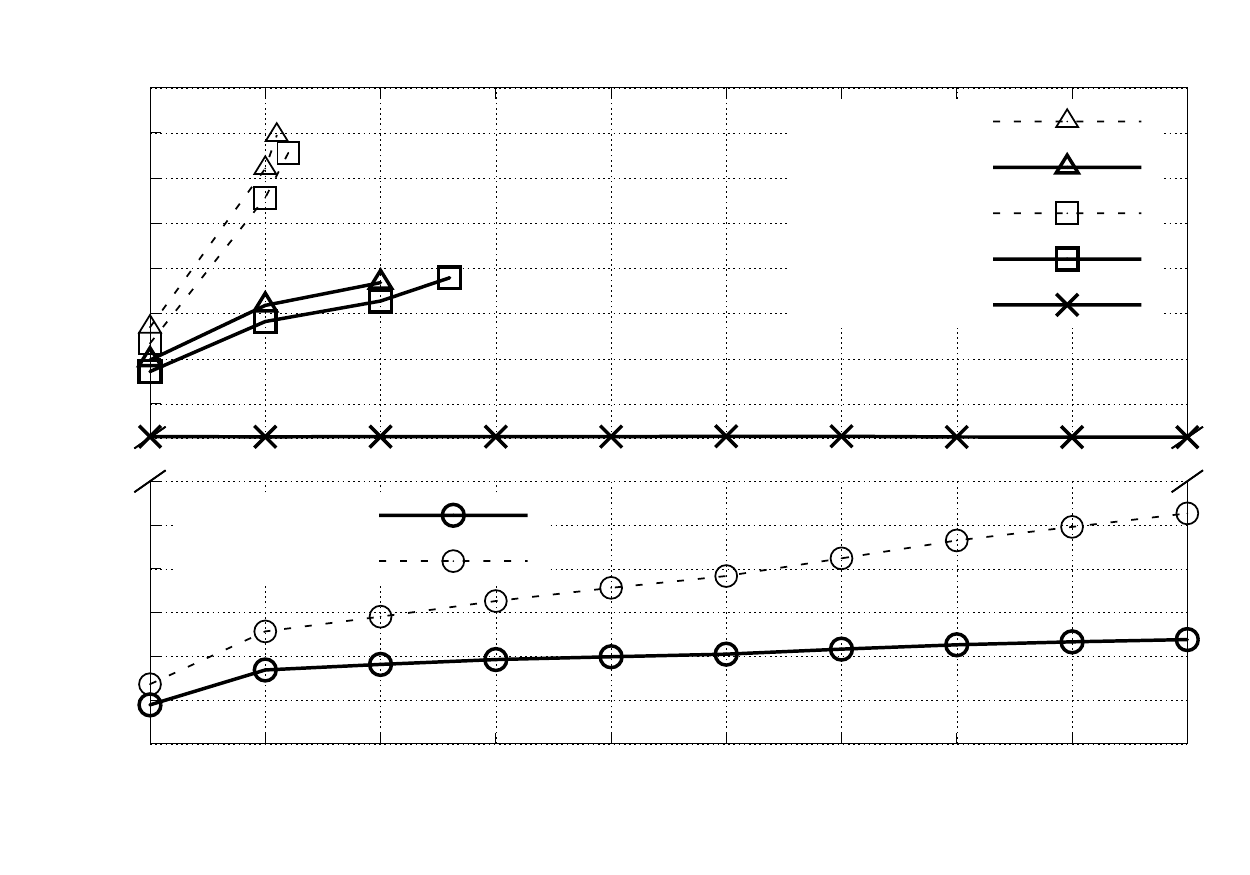}}%
    \gplfronttext
  \end{picture}%
\endgroup
 		}
		\caption{Query answering time for $ Q_{DB} = 5K $, $ \select = 25\% $ and $ G_E = 100K $ to $ 1M $.}
		\label{fig:exp:nyc-p100-sub5-s25:filteringTime}
	\end{subfigure}
	\hfil
	\begin{subfigure}[t]{0.32\textwidth}
		\centering
		\resizebox{\linewidth}{0.18\textheight}{
\begingroup
  \makeatletter
  \providecommand\color[2][]{%
    \GenericError{(gnuplot) \space\space\space\@spaces}{%
      Package color not loaded in conjunction with
      terminal option `colourtext'%
    }{See the gnuplot documentation for explanation.%
    }{Either use 'blacktext' in gnuplot or load the package
      color.sty in LaTeX.}%
    \renewcommand\color[2][]{}%
  }%
  \providecommand\includegraphics[2][]{%
    \GenericError{(gnuplot) \space\space\space\@spaces}{%
      Package graphicx or graphics not loaded%
    }{See the gnuplot documentation for explanation.%
    }{The gnuplot epslatex terminal needs graphicx.sty or graphics.sty.}%
    \renewcommand\includegraphics[2][]{}%
  }%
  \providecommand\rotatebox[2]{#2}%
  \@ifundefined{ifGPcolor}{%
    \newif\ifGPcolor
    \GPcolortrue
  }{}%
  \@ifundefined{ifGPblacktext}{%
    \newif\ifGPblacktext
    \GPblacktexttrue
  }{}%
  \let\gplgaddtomacro\g@addto@macro
  \gdef\gplbacktext{}%
  \gdef\gplfronttext{}%
  \makeatother
  \ifGPblacktext
    \def\colorrgb#1{}%
    \def\colorgray#1{}%
  \else
    \ifGPcolor
      \def\colorrgb#1{\color[rgb]{#1}}%
      \def\colorgray#1{\color[gray]{#1}}%
      \expandafter\def\csname LTw\endcsname{\color{white}}%
      \expandafter\def\csname LTb\endcsname{\color{black}}%
      \expandafter\def\csname LTa\endcsname{\color{black}}%
      \expandafter\def\csname LT0\endcsname{\color[rgb]{1,0,0}}%
      \expandafter\def\csname LT1\endcsname{\color[rgb]{0,1,0}}%
      \expandafter\def\csname LT2\endcsname{\color[rgb]{0,0,1}}%
      \expandafter\def\csname LT3\endcsname{\color[rgb]{1,0,1}}%
      \expandafter\def\csname LT4\endcsname{\color[rgb]{0,1,1}}%
      \expandafter\def\csname LT5\endcsname{\color[rgb]{1,1,0}}%
      \expandafter\def\csname LT6\endcsname{\color[rgb]{0,0,0}}%
      \expandafter\def\csname LT7\endcsname{\color[rgb]{1,0.3,0}}%
      \expandafter\def\csname LT8\endcsname{\color[rgb]{0.5,0.5,0.5}}%
    \else
      \def\colorrgb#1{\color{black}}%
      \def\colorgray#1{\color[gray]{#1}}%
      \expandafter\def\csname LTw\endcsname{\color{white}}%
      \expandafter\def\csname LTb\endcsname{\color{black}}%
      \expandafter\def\csname LTa\endcsname{\color{black}}%
      \expandafter\def\csname LT0\endcsname{\color{black}}%
      \expandafter\def\csname LT1\endcsname{\color{black}}%
      \expandafter\def\csname LT2\endcsname{\color{black}}%
      \expandafter\def\csname LT3\endcsname{\color{black}}%
      \expandafter\def\csname LT4\endcsname{\color{black}}%
      \expandafter\def\csname LT5\endcsname{\color{black}}%
      \expandafter\def\csname LT6\endcsname{\color{black}}%
      \expandafter\def\csname LT7\endcsname{\color{black}}%
      \expandafter\def\csname LT8\endcsname{\color{black}}%
    \fi
  \fi
    \setlength{\unitlength}{0.0500bp}%
    \ifx\gptboxheight\undefined%
      \newlength{\gptboxheight}%
      \newlength{\gptboxwidth}%
      \newsavebox{\gptboxtext}%
    \fi%
    \setlength{\fboxrule}{0.5pt}%
    \setlength{\fboxsep}{1pt}%
\begin{picture}(7200.00,5040.00)%
    \gplgaddtomacro\gplbacktext{%
      \csname LTb\endcsname%
      \put(732,756){\makebox(0,0)[r]{\strut{}0}}%
      \csname LTb\endcsname%
      \put(732,957){\makebox(0,0)[r]{\strut{}2}}%
      \csname LTb\endcsname%
      \put(732,1159){\makebox(0,0)[r]{\strut{}4}}%
      \csname LTb\endcsname%
      \put(732,1360){\makebox(0,0)[r]{\strut{}6}}%
      \csname LTb\endcsname%
      \put(732,1562){\makebox(0,0)[r]{\strut{}8}}%
      \csname LTb\endcsname%
      \put(732,1763){\makebox(0,0)[r]{\strut{}10}}%
      \csname LTb\endcsname%
      \put(732,1965){\makebox(0,0)[r]{\strut{}12}}%
      \csname LTb\endcsname%
      \put(732,2166){\makebox(0,0)[r]{\strut{}14}}%
      \csname LTb\endcsname%
      \put(864,536){\makebox(0,0){\strut{}$10$}}%
      \csname LTb\endcsname%
      \put(1528,536){\makebox(0,0){\strut{}$20$}}%
      \csname LTb\endcsname%
      \put(2192,536){\makebox(0,0){\strut{}$30$}}%
      \csname LTb\endcsname%
      \put(2856,536){\makebox(0,0){\strut{}$40$}}%
      \csname LTb\endcsname%
      \put(3520,536){\makebox(0,0){\strut{}$50$}}%
      \csname LTb\endcsname%
      \put(4183,536){\makebox(0,0){\strut{}$60$}}%
      \csname LTb\endcsname%
      \put(4847,536){\makebox(0,0){\strut{}$70$}}%
      \csname LTb\endcsname%
      \put(5511,536){\makebox(0,0){\strut{}$80$}}%
      \csname LTb\endcsname%
      \put(6175,536){\makebox(0,0){\strut{}$90$}}%
      \csname LTb\endcsname%
      \put(6839,536){\makebox(0,0){\strut{}$100$}}%
    }%
    \gplgaddtomacro\gplfronttext{%
      \csname LTb\endcsname%
      \put(248,2611){\rotatebox{-270}{\makebox(0,0){\strut{}Answering time (msec/update)}}}%
      \put(3851,206){\makebox(0,0){\strut{}Graph size (Edges x$1000$)}}%
      \csname LTb\endcsname%
      \put(996,2072){\makebox(0,0)[l]{\strut{}\Tree}}%
      \csname LTb\endcsname%
      \put(996,1808){\makebox(0,0)[l]{\strut{}\TreeCache}}%
    }%
    \gplgaddtomacro\gplbacktext{%
      \csname LTb\endcsname%
      \put(732,2519){\makebox(0,0)[r]{\strut{}100}}%
      \csname LTb\endcsname%
      \put(732,2855){\makebox(0,0)[r]{\strut{}1000}}%
      \csname LTb\endcsname%
      \put(732,3228){\makebox(0,0)[r]{\strut{}2000}}%
      \csname LTb\endcsname%
      \put(732,3602){\makebox(0,0)[r]{\strut{}3000}}%
      \csname LTb\endcsname%
      \put(732,3975){\makebox(0,0)[r]{\strut{}4000}}%
      \csname LTb\endcsname%
      \put(732,4348){\makebox(0,0)[r]{\strut{}5000}}%
      \csname LTb\endcsname%
      \put(864,4755){\makebox(0,0)[l]{\strut{}6.4}}%
      \csname LTb\endcsname%
      \put(1528,4755){\makebox(0,0)[l]{\strut{}7.6}}%
      \csname LTb\endcsname%
      \put(2192,4755){\makebox(0,0)[l]{\strut{}11.9}}%
      \csname LTb\endcsname%
      \put(2856,4755){\makebox(0,0)[l]{\strut{}15.0}}%
      \csname LTb\endcsname%
      \put(3520,4755){\makebox(0,0)[l]{\strut{}15.7}}%
      \csname LTb\endcsname%
      \put(4183,4755){\makebox(0,0)[l]{\strut{}16.2}}%
      \csname LTb\endcsname%
      \put(4847,4755){\makebox(0,0)[l]{\strut{}16.5}}%
      \csname LTb\endcsname%
      \put(5511,4755){\makebox(0,0)[l]{\strut{}16.8}}%
      \csname LTb\endcsname%
      \put(6175,4755){\makebox(0,0)[l]{\strut{}17.1}}%
      \csname LTb\endcsname%
      \put(6839,4755){\makebox(0,0)[l]{\strut{}17.2}}%
      \put(3619,4294){\makebox(0,0)[l]{\strut{}\emph{*}}}%
      \put(3619,4076){\makebox(0,0)[l]{\strut{}\emph{*}}}%
      \put(3619,3895){\makebox(0,0)[l]{\strut{}\emph{*}}}%
      \put(4283,4140){\makebox(0,0)[l]{\strut{}\emph{*}}}%
    }%
    \gplgaddtomacro\gplfronttext{%
      \csname LTb\endcsname%
      \put(3851,5085){\makebox(0,0){\strut{}Graph size (Vertices x$1000$)}}%
      \csname LTb\endcsname%
      \put(4532,4340){\makebox(0,0)[l]{\strut{}\Inv}}%
      \csname LTb\endcsname%
      \put(4532,4076){\makebox(0,0)[l]{\strut{}\InvCache}}%
      \csname LTb\endcsname%
      \put(4532,3812){\makebox(0,0)[l]{\strut{}\InvIncr}}%
      \csname LTb\endcsname%
      \put(4532,3548){\makebox(0,0)[l]{\strut{}\InvIncrCache}}%
      \csname LTb\endcsname%
      \put(4532,3284){\makebox(0,0)[l]{\strut{}\Neo}}%
    }%
    \gplbacktext
    \put(0,0){\includegraphics{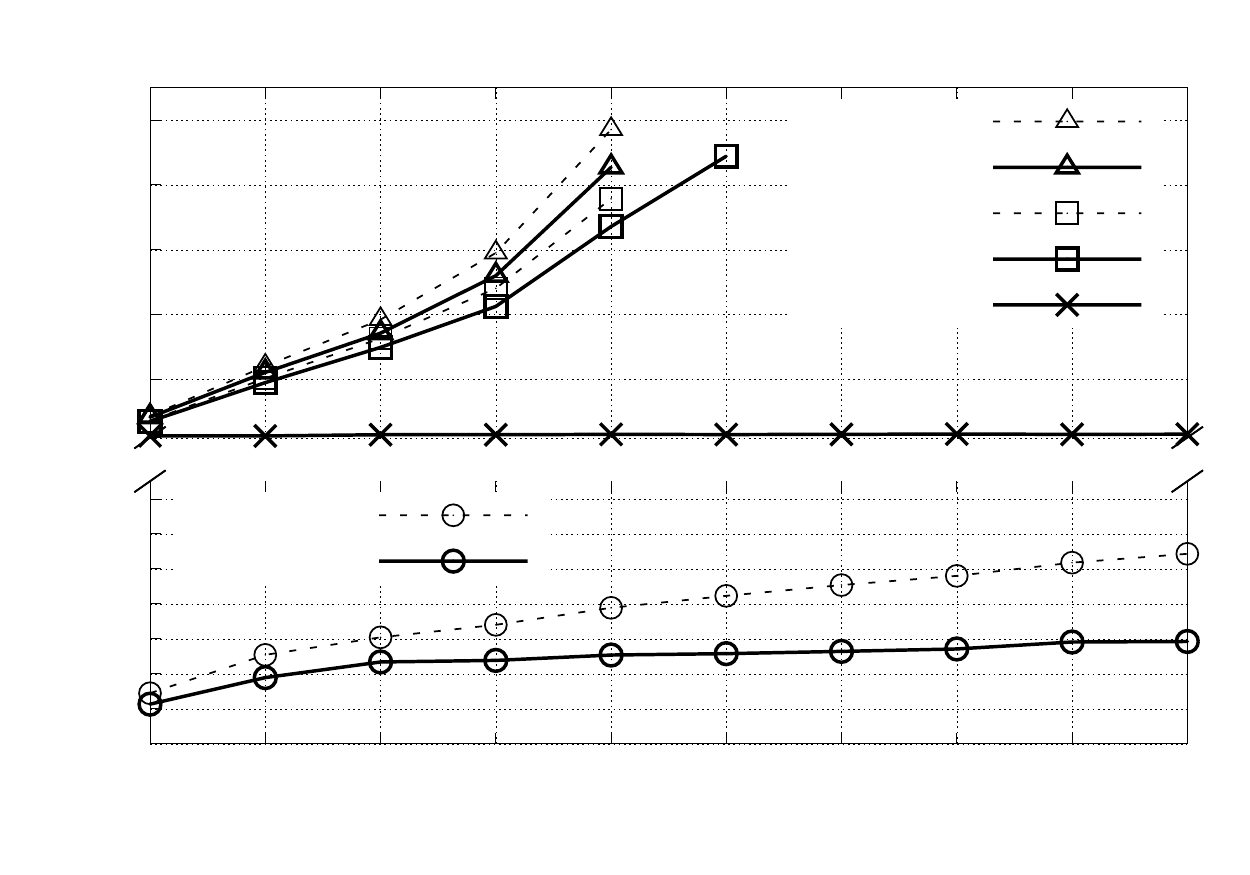}}%
    \gplfronttext
  \end{picture}%
\endgroup
 		}
		\caption{Query answering time for $ Q_{DB} = 5K $, $ \select = 25\% $ and $ G_E = 10K $ to $ 100K $.}
		\label{fig:biogrid:bio-pub100-sub5-s25:filteringTime}
	\end{subfigure}
	\hfill
	\begin{subfigure}[t]{0.32\textwidth}
		\centering
		\resizebox{\linewidth}{0.18\textheight}{
\begingroup
  \makeatletter
  \providecommand\color[2][]{%
    \GenericError{(gnuplot) \space\space\space\@spaces}{%
      Package color not loaded in conjunction with
      terminal option `colourtext'%
    }{See the gnuplot documentation for explanation.%
    }{Either use 'blacktext' in gnuplot or load the package
      color.sty in LaTeX.}%
    \renewcommand\color[2][]{}%
  }%
  \providecommand\includegraphics[2][]{%
    \GenericError{(gnuplot) \space\space\space\@spaces}{%
      Package graphicx or graphics not loaded%
    }{See the gnuplot documentation for explanation.%
    }{The gnuplot epslatex terminal needs graphicx.sty or graphics.sty.}%
    \renewcommand\includegraphics[2][]{}%
  }%
  \providecommand\rotatebox[2]{#2}%
  \@ifundefined{ifGPcolor}{%
    \newif\ifGPcolor
    \GPcolortrue
  }{}%
  \@ifundefined{ifGPblacktext}{%
    \newif\ifGPblacktext
    \GPblacktexttrue
  }{}%
  \let\gplgaddtomacro\g@addto@macro
  \gdef\gplbacktext{}%
  \gdef\gplfronttext{}%
  \makeatother
  \ifGPblacktext
    \def\colorrgb#1{}%
    \def\colorgray#1{}%
  \else
    \ifGPcolor
      \def\colorrgb#1{\color[rgb]{#1}}%
      \def\colorgray#1{\color[gray]{#1}}%
      \expandafter\def\csname LTw\endcsname{\color{white}}%
      \expandafter\def\csname LTb\endcsname{\color{black}}%
      \expandafter\def\csname LTa\endcsname{\color{black}}%
      \expandafter\def\csname LT0\endcsname{\color[rgb]{1,0,0}}%
      \expandafter\def\csname LT1\endcsname{\color[rgb]{0,1,0}}%
      \expandafter\def\csname LT2\endcsname{\color[rgb]{0,0,1}}%
      \expandafter\def\csname LT3\endcsname{\color[rgb]{1,0,1}}%
      \expandafter\def\csname LT4\endcsname{\color[rgb]{0,1,1}}%
      \expandafter\def\csname LT5\endcsname{\color[rgb]{1,1,0}}%
      \expandafter\def\csname LT6\endcsname{\color[rgb]{0,0,0}}%
      \expandafter\def\csname LT7\endcsname{\color[rgb]{1,0.3,0}}%
      \expandafter\def\csname LT8\endcsname{\color[rgb]{0.5,0.5,0.5}}%
    \else
      \def\colorrgb#1{\color{black}}%
      \def\colorgray#1{\color[gray]{#1}}%
      \expandafter\def\csname LTw\endcsname{\color{white}}%
      \expandafter\def\csname LTb\endcsname{\color{black}}%
      \expandafter\def\csname LTa\endcsname{\color{black}}%
      \expandafter\def\csname LT0\endcsname{\color{black}}%
      \expandafter\def\csname LT1\endcsname{\color{black}}%
      \expandafter\def\csname LT2\endcsname{\color{black}}%
      \expandafter\def\csname LT3\endcsname{\color{black}}%
      \expandafter\def\csname LT4\endcsname{\color{black}}%
      \expandafter\def\csname LT5\endcsname{\color{black}}%
      \expandafter\def\csname LT6\endcsname{\color{black}}%
      \expandafter\def\csname LT7\endcsname{\color{black}}%
      \expandafter\def\csname LT8\endcsname{\color{black}}%
    \fi
  \fi
    \setlength{\unitlength}{0.0500bp}%
    \ifx\gptboxheight\undefined%
      \newlength{\gptboxheight}%
      \newlength{\gptboxwidth}%
      \newsavebox{\gptboxtext}%
    \fi%
    \setlength{\fboxrule}{0.5pt}%
    \setlength{\fboxsep}{1pt}%
\begin{picture}(7200.00,5040.00)%
    \gplgaddtomacro\gplbacktext{%
      \csname LTb\endcsname%
      \put(814,704){\makebox(0,0)[r]{\strut{}0}}%
      \csname LTb\endcsname%
      \put(814,1112){\makebox(0,0)[r]{\strut{}20}}%
      \csname LTb\endcsname%
      \put(814,1521){\makebox(0,0)[r]{\strut{}40}}%
      \csname LTb\endcsname%
      \put(814,1929){\makebox(0,0)[r]{\strut{}60}}%
      \csname LTb\endcsname%
      \put(814,2337){\makebox(0,0)[r]{\strut{}80}}%
      \csname LTb\endcsname%
      \put(814,2746){\makebox(0,0)[r]{\strut{}100}}%
      \csname LTb\endcsname%
      \put(814,3154){\makebox(0,0)[r]{\strut{}120}}%
      \csname LTb\endcsname%
      \put(814,3562){\makebox(0,0)[r]{\strut{}140}}%
      \csname LTb\endcsname%
      \put(814,3971){\makebox(0,0)[r]{\strut{}160}}%
      \csname LTb\endcsname%
      \put(814,4379){\makebox(0,0)[r]{\strut{}180}}%
      \csname LTb\endcsname%
      \put(946,484){\makebox(0,0){\strut{}$100$}}%
      \csname LTb\endcsname%
      \put(1597,484){\makebox(0,0){\strut{}$200$}}%
      \csname LTb\endcsname%
      \put(2248,484){\makebox(0,0){\strut{}$300$}}%
      \csname LTb\endcsname%
      \put(2898,484){\makebox(0,0){\strut{}$400$}}%
      \csname LTb\endcsname%
      \put(3549,484){\makebox(0,0){\strut{}$500$}}%
      \csname LTb\endcsname%
      \put(4200,484){\makebox(0,0){\strut{}$600$}}%
      \csname LTb\endcsname%
      \put(4851,484){\makebox(0,0){\strut{}$700$}}%
      \csname LTb\endcsname%
      \put(5501,484){\makebox(0,0){\strut{}$800$}}%
      \csname LTb\endcsname%
      \put(6152,484){\makebox(0,0){\strut{}$900$}}%
      \csname LTb\endcsname%
      \put(6803,484){\makebox(0,0){\strut{}$1000$}}%
      \put(946,4599){\makebox(0,0)[l]{\strut{}17}}%
      \put(1597,4599){\makebox(0,0)[l]{\strut{}27}}%
      \put(2248,4599){\makebox(0,0)[l]{\strut{}29}}%
      \put(2898,4599){\makebox(0,0)[l]{\strut{}35}}%
      \put(3549,4599){\makebox(0,0)[l]{\strut{}44}}%
      \put(4200,4599){\makebox(0,0)[l]{\strut{}48}}%
      \put(4851,4599){\makebox(0,0)[l]{\strut{}54}}%
      \put(5501,4599){\makebox(0,0)[l]{\strut{}58}}%
      \put(6152,4599){\makebox(0,0)[l]{\strut{}61}}%
      \put(6803,4599){\makebox(0,0)[l]{\strut{}63}}%
      \put(3884,3991){\makebox(0,0)[l]{\strut{}\emph{*}}}%
    }%
    \gplgaddtomacro\gplfronttext{%
      \csname LTb\endcsname%
      \put(198,2541){\rotatebox{-270}{\makebox(0,0){\strut{}Answering time (msec/update)}}}%
      \put(3874,154){\makebox(0,0){\strut{}Graph size (Edges x$1000$)}}%
      \put(3874,4929){\makebox(0,0){\strut{}Graph size (Vertices x$1000$)}}%
      \csname LTb\endcsname%
      \put(4496,4184){\makebox(0,0)[l]{\strut{}\Neo}}%
      \csname LTb\endcsname%
      \put(4496,3920){\makebox(0,0)[l]{\strut{}\Tree}}%
      \csname LTb\endcsname%
      \put(4496,3656){\makebox(0,0)[l]{\strut{}\TreeCache}}%
    }%
    \gplbacktext
    \put(0,0){\includegraphics{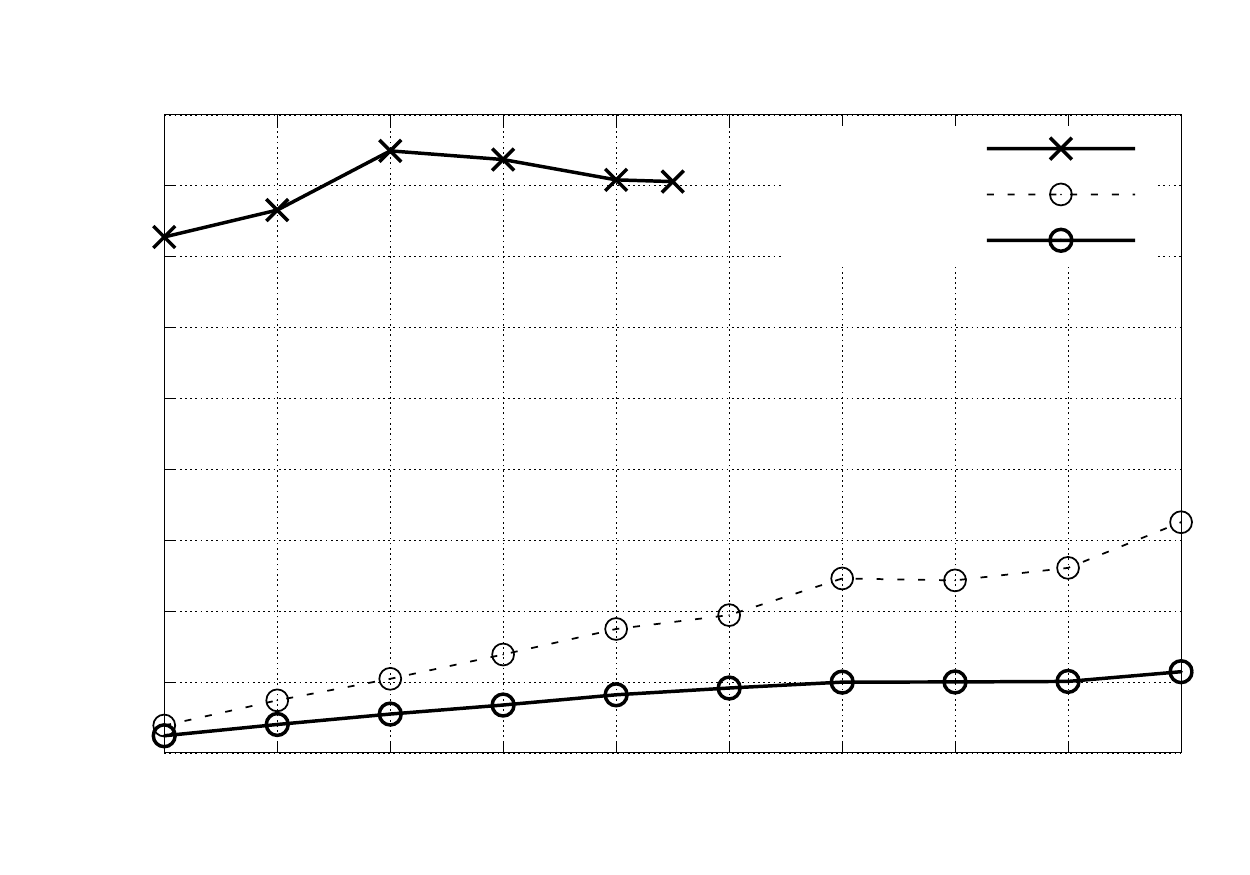}}%
    \gplfronttext
  \end{picture}%
\endgroup
 		}
		\caption{Query answering time for $ Q_{DB} = 5K $, $ \select = 25\% $ and $ G_E = 100K $ to $ 1M $.}
		\label{fig:biogrid:bio-pub100-sub5-s25:filteringTime:1M}
	\end{subfigure}
	
	\caption{Results for (a) the $ TAXI $ dataset, (b) \& (c) the $ BioGRID $ dataset.}
	\label{fig:exp:nyc:biogrid}
\end{figure*}

\color{black}
\ctitle{The $ NYC $ Dataset.}
Fig.~\ref{fig:exp:nyc-p100-sub5-s25:filteringTime} presents the results from the evaluation of the algorithms for the $ NYC $ dataset. More specifically, we present the results regarding the query answering performance of all algorithms when $ Q_{DB} = 5K $, $ \len = 5 $, $ \overlap = 35\% $, $ \select = 25\% $ and an execution time threshold of 24 hours.  Please notice that the y-axis is split due to high differences in the performance of the algorithms. Algorithms~\Inv\ and \InvCache\ fail to terminate within the time threshold and \emph{time out} at $ |G_E| = 210K $ and $ |G_E| = 300K $ respectively. Similarly, Algorithms~\InvIncr\ and \InvIncrCache\ \emph{time out} at $ |G_E| = 220K $ and $ 360K $ respectively. 
When comparing Algorithms~\Tree\ and \TreeCache\ to \Neo\ the query answering is improved by $ 59.68\% $ and $ 81.76\% $ respectively. These results indicate that again an algorithmic solution that exploits and indexes together the common parts of query graphs (i.e., Algorithms~\Tree/\TreeCache) achieves significantly lower query answering time compared to approaches that do not apply any clustering techniques (i.e., Algorithms\Inv/\InvCache\InvIncr/\InvIncrCache/\Neo). 
\color{black}

\color{black}
\ctitle{The $ BioGRID $ Dataset.}
Figs.~\ref{fig:biogrid:bio-pub100-sub5-s25:filteringTime} and \ref{fig:biogrid:bio-pub100-sub5-s25:filteringTime:1M} present the results from the evaluation of the algorithms for the $ BioGRID $ dataset. In Fig.~\ref{fig:biogrid:bio-pub100-sub5-s25:filteringTime} we present the results regarding the query answering performance of the algorithms, when $ Q_{DB} = 5K $, $ \select = 25\% $ for a final graph size of $ |G_E| = 100K $ and $ |G_V| = 17.2K $. Additionally, we set an execution time threshold of 24 hours due to the high differences in the performance of the algorithms. 
The $ BioGRID $ dataset serves as a stress test for our algorithms, since it contains only one type of edge and vertex, thus each incoming update will affect (but not necessarily satisfy) the entire query database. To this end, Algorithms~\Inv/\InvCache/\InvIncr\ exceed the time threshold and time out at $ |G_E| = 50K $, while \InvIncrCache\ times out at $ |G_E| = 60K $ as denoted by the asterisks in the plot. Finally, Fig.~\ref{fig:biogrid:bio-pub100-sub5-s25:filteringTime:1M} presents the results for the $ BioGRID $ dataset for a final graph size of $ |G_E| = 1M $ and $ |G_V| = 63K $. We again observe that Algorithms~\Tree\ and \TreeCache\ achieve the lowest answering time, while \Neo\ exceeds the time threshold and times out at $ |G_E| = 550K $. As it is demonstrated from the results yielded by the evaluation, Algorithms~\Tree\ and \TreeCache\ are the most efficient of all; this is attributed to the fact that both algorithms create a combined representation of the query graph patterns that can efficiently be utilized during query answering time.
\color{black}

 \begin{table}[!t]
 	\renewcommand{\arraystretch}{1.2}
 	\caption{\textcolor{black}{Memory usage for $ |Q_{DB}| = 5K $, $ \len = 5 $, $ \select = 25\% $, $ \overlap = 35\% $ and $ |G_E| = 100K  $.}}
 	\label{tab:exp:memory}
 	\centering
 	\begin{tabular}{lcrcrcr}
 		\hline
 		\multirow{2}{*}{Algorithm} & & \multicolumn{5}{c}{Dataset} \\
 		\cline{2-7}
 		& & $ SNB $ & & $ NYC $ & & $ BioGRID $ \\
 		\hline
 		\Tree &  & $ 201\textit{MB} $ & & $ 257 \textit{MB} $ & & $ 233\textit{MB} $\\
 		\TreeCache &  & $ 248\textit{MB} $ & & $ 273\textit{MB} $ & & $ 262\textit{MB} $\\
 		\Inv &  & $  205\textit{MB} $ & & $ 273\textit{MB} $ & & $  271\textit{MB} $\textsuperscript{50K} \\
 		\InvCache &  & $ 228\textit{MB} $ & & $ 381\textit{MB} $ & & $ 301\textit{MB} $\textsuperscript{50K} \\
 		\InvIncr &  & $ 206 \textit{MB} $ & & $ 273\textit{MB} $ & & $ 270\textit{MB} $\textsuperscript{50K}\\
 		\InvIncrCache &  & $ 228\textit{MB} $ & & $ 378\textit{MB} $ & & $ 310\textit{MB} $\textsuperscript{60K}\\
 		\Neo &  & $ 443\textit{MB} $ & & $ 590\textit{MB} $ & & $ 314\textit{MB} $\\
 		\hline
 	\end{tabular}
 \end{table}

\color{black}
\ctitle{Comparing Memory Requirements.}
Table~\ref{tab:exp:memory} presents the memory requirements of each algorithm, for the $ SNB $, $ NYC $ and $ BioGRID $ datasets when indexing $ |Q_{DB} = 5K| $ and a graph of $ |G_E| = 100K $. We observe, that across all datasets,  Algorithms~\Tree, \Inv\ and \InvIncr\ have the lowest main memory requirements, 
while, Algorithms~\TreeCache, \InvCache, \InvIncrCache\ and \Neo\  exhibit higher memory requirements. 
The higher memory requirements of algorithms that employ a caching strategy,  (i.e., Algorithms~\TreeCache/\InvCache/\InvIncrCache) is attributed to the fact that all structures calculated during the materialization phase are kept in memory for future usage; this results in higher memory requirements compared to algorithms that do not apply this caching technique (i.e., Algorithms~\Tree/\Inv/\InvIncr). Finally, \Neo\ is a full fledged database management system, thus it occupies more memory to support the required specifications. 

\color{black}

\section{Applications}
\label{sec:applications}

In this section, we briefly discuss additional application scenarios for 
continuous query evaluation over graph streams.

\ctitle{Social Networks.}
Social network graphs emerge naturally from the evolving social interactions and activities of the users. Many applications such as advertising, recommendation systems, and information discovery can benefit from continuous pattern matching. Prompt identification of influential users and active monitoring of content propagation inside the network could increase the effectiveness in those applications. In such scenarios, applications may leverage on sub-graph matching where patterns already observed in social networks can be utilized \cite{LeskovecSK06, LeskovecAH07, abs-1708-02377}. Finally, real-time reporting of influential users could be achieved through monitoring of the dissemination of posts inside the network \cite{ChaHBG10, ChoudhuryHCAF15}.

\ctitle{Protein Interaction Graphs.}
\emph{Protein-protein interaction} (PPI) graphs are important data repositories in which proteins are represented as vertices and identified interactions between them as edges. 
 PPI graphs are typically stored in central repositories \cite{biogrid, UniProt}, where they are constantly updated due to new protein interaction additions and invalidation of existing interactions. 
Scientists are typically forced to manually query these repositories on a regular basis to discover new patterns they are interested in, since the existing tools are unable to capture new patterns in the evolving graphs. Therefore, there is a clear need for an efficient solution that provides the continuous subgraph matching functionality over PPI graphs.

\ctitle{Other Domains.}
The techniques proposed in this paper can also be applied in a wide range of domains such as cybersecurity, knowledge graphs, road network monitoring, and co-authorship graphs. In cybersecurity, subgraph pattern matching could be applied to monitor the network traffic and capture denial of service and exfiltration attacks \cite{JoslynCHHNO13}. In road network monitoring, subgraph pattern matching could be applied to capture traffic congestion events, and taxi route pricing  \cite{web:taxiTripData}. In the domain of co-authorship graphs, users may utilize the continuous query evaluation algorithms in services similar to Google Scholar Alerts, when requesting to be notified about newly published content, by making use of appropriate graphical user interface tools.
\section{Conclusions and Outlook}
\label{sec:outlook}

\color{black}
In this work, we proposed a new paradigm to efficiently capture the evolving nature of graphs through query graph patterns. We proposed a novel method that indexes and continuously  evaluates queries over graph streams, by leveraging on the shared restrictions present in query sets. We evaluated our solution using three different datasets from social networks, transportation and biological interactions domains, and demonstrated that our approach is up to two orders of magnitude faster when compared to typical join-and-explore inverted index solutions, and the well-established graph database \Neo. We plan on extending our methods to support graph deletions and increase expressiveness through query classes that aim at clustering coefficient, shortest path, and betweenness centrality.
\color{black}

	\section*{Acknowledgments}
	This research was partially funded by the Danish Council for Independent Research (DFF) under grant agreement No. DFF-4093-00301.
	
	\bibliographystyle{IEEEtran}

\end{document}